\documentclass[11pt,times]{article}
\usepackage{natbib}

\usepackage{color}
\usepackage{graphicx}
\usepackage{sectsty}
\usepackage{boxedminipage}
\usepackage{amsmath}
\usepackage{epsfig}

\def\HI{{\rm H}{\sc i}}
\def\HII{{\rm H}{\sc ii}}
\def\HeII{{\rm He}{\sc ii}}

\sectionfont{\LARGE}
\subsectionfont{\Large}
\subsubsectionfont{\large}

\def\spose#1{\hbox to 0pt{#1\hss}}
\def\lta{\mathrel{\spose{\lower 3pt\hbox{$\mathchar"218$}}
     \raise 2.0pt\hbox{$\mathchar"13C$}}}
\def\gta{\mathrel{\spose{\lower 3pt\hbox{$\mathchar"218$}}
     \raise 2.0pt\hbox{$\mathchar"13E$}}}

\begin{document}
\thispagestyle{empty}
\begin{center}

\resizebox{!}{1.0cm}{\bf LOFAR-UK}
\vspace*{0.7cm}

\resizebox{!}{0.48cm}{\bf{\em A science case for UK involvement in LOFAR}}
\vspace*{0.5cm}

\begin{figure}[!h]
\centerline{
\psfig{file=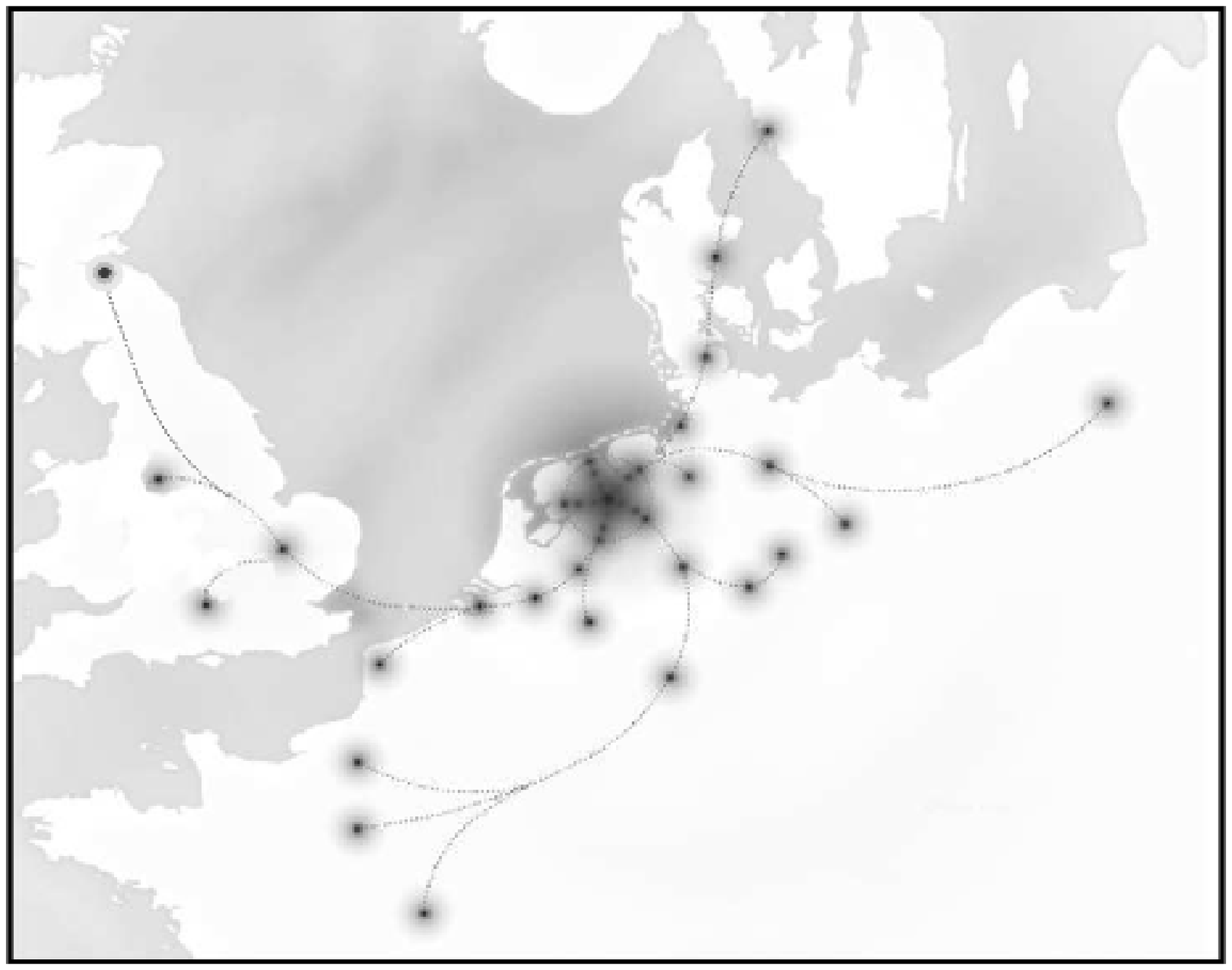,width=13cm,clip=}
}
\end{figure}
\vspace*{0.5cm}

\begin{large}
Assembled and edited on behalf of the LOFAR-UK consortium by
\end{large}
\smallskip

\begin{Large}
{\em Philip Best}
\vspace*{0.3cm}

{\em December 2007}
\vspace*{0.5cm}
\end{Large}
\end{center}

Contributors: 
Paul Alexander (Cambridge),
David Bacon (Edinburgh/Portsmouth),
David Bersier (Liverpool JMU),
Philip Best (Edinburgh), 
Rob Beswick (Manchester),
Andy Breen (Aberystwyth), 
Elias Brinks (Hertfordshire),
Catherine Brocksopp (UCL/MSSL),
Sandra Chapman (Warwick),
Michele Cirasuolo (Edinburgh),
Judith Croston (Hertfordshire),
Owain Davies (RAL),
Tom Dwelly (Southampton),
Steve Eales (Cardiff),
Alastair Edge (Durham),
Brian Ellison (RAL),
Rob Fender (Southampton),
Lyndsay Fletcher (Glasgow),
Martin F{\"u}llekrug (Bath),
Simon Garrington (Manchester),
Dave Green (Cambridge), 
Martin Haehnelt (Cambridge),
Martin Hardcastle (Hertfordshire),
Richard Harrison (RAL),
Faridey Honary (Lancaster),
Rob Ivison (ATC),
Neal Jackson (Manchester),
Matt Jarvis (Hertfordshire),
Christian Kaiser (Southampton),
Joe Khan (Glasgow),
Hans-Rainer Kl{\"o}ckner (Oxford),
Eduard Kontar (Glasgow),
Michael Kramer (Manchester),
Cedric Lacey (Durham),
Mark Lancaster (UCL),
Tom Maccarone (Southampton),
Alec MacKinnon (Glasgow),
Ross McLure (Edinburgh),
Avery Meiksin (Edinburgh), 
Cathryn Mitchell (Bath),
Bob Nichol (Portsmouth),
Will Percival (Portsmouth),
Robert Priddey (Hertfordshire),
Steve Rawlings (Oxford),
Chris Simpson (Liverpool JMU),
Ian Stevens (Birmingham),
Tom Theuns (Durham),
Phil Uttley (Southampton),
Peter Wilkinson (Manchester),
Graham Woan (Glasgow)

\clearpage
~
\thispagestyle{empty}
\clearpage
\pagenumbering{roman}

\tableofcontents
\clearpage
~
\clearpage

\pagenumbering{arabic}
\section{Executive Summary}

{\bf LOFAR}, the {\bf Lo}w-{\bf F}requency {\bf Ar}ray, is a
next-generation software-driven radio telescope operating between 30 and
240\,MHz, currently under construction in the Netherlands. This low
frequency radio band is one of the few largely unexplored regions of the
electromagnetic spectrum. The sensitivity and angular resolution offered
by LOFAR will be two to three orders of magnitude better than existing
telescopes, and as such it will open up this new window on the
Universe. LOFAR will impact on a broad range of astrophysics, from
cosmology to solar system studies: it will conduct the first studies of
the Epoch of Reionisation, carry out the deepest large--sky radio source
surveys ever, revolutionise the study of transient phenomena, make
measurements of ultra-high energy cosmic rays via radio emission from
air showers, and investigate the radio signatures of solar and
interplanetary activity. In addition, history indicates that exploring new
frequency windows has always led to unexpected discoveries.

There is growing European involvement in LOFAR, driven by the need to add
stations far from the main core in order to improve angular resolution.
LOFAR-UK is a project aimed at cementing UK participation in LOFAR via the
operation of four stations within the UK, as part of a European expansion
including Germany, France, Sweden and probably other European
countries. LOFAR-UK ground stations will allow the highest angular
resolution LOFAR observations, reaching sub-arcsecond scales at the
highest LOFAR frequencies, and as a result will also improve the
(confusion-limited) sensitivity limit of the telescope for deep
surveys. UK stations will also significantly enhance the instantaneous
(u,v)-plane coverage, essential for snapshots of transient phenomena.

LOFAR-UK will achieve involvement for UK astronomers in a world-leading
science facility operating in the immediate future.  It will allow the UK
to build up important scientific and technical expertise in `next
generation' radio astronomy in preparation for the Square Kilometre Array
(SKA). Noting the dramatic increase in the diversity of topics addressable
with the next generation of radio telescopes, LOFAR-UK will play an
important role in helping to broaden the UK community that has an interest
in radio astronomy: one of the key features of the LOFAR-UK consortium is
that it gathers together traditional `radio astronomy' groups with groups
with very limited experience in radio astronomy, but great interest in the
new science to be achievable with LOFAR and the SKA.

This White Paper outlines the strategic importance to the UK astronomy
community of gaining involvement in the LOFAR project, the scientific
interests of UK researchers in using the telescope, and the technical
challenges that will need to be overcome.

\clearpage
~
\clearpage

\section{Introduction}
\label{intro}

LOFAR is a next-generation software-driven radio telescope currently under
construction by ASTRON in the Netherlands. LOFAR will explore the
30-240\,MHz radio sky with two to three orders of magnitude more
sensitivity than previous surveys, and will have an enormous field of view
facilitating semi-continuous monitoring of more than half of the entire
sky at the lowest frequencies.

\begin{figure}[!b]
\centerline
{\psfig{file=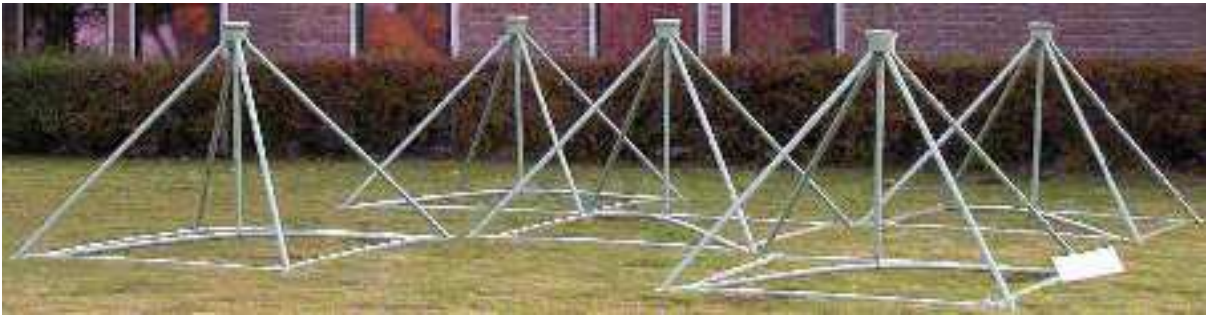,width=16cm}
}
\caption{\label{antenfig} \small Prototype low-band (30--80 MHz) LOFAR
 antennae of the type that have been deployed at the LOFAR Initial
 Test Station (ITS) and Core Station One (CS1). Eventually up to 5000
 such antennae will be distributed across the Netherlands, in 36-50
 stations each containing 96 antennae (plus 48 high-band tiles).}
\end{figure}

\begin{figure}[!t]
\centerline{\psfig{file=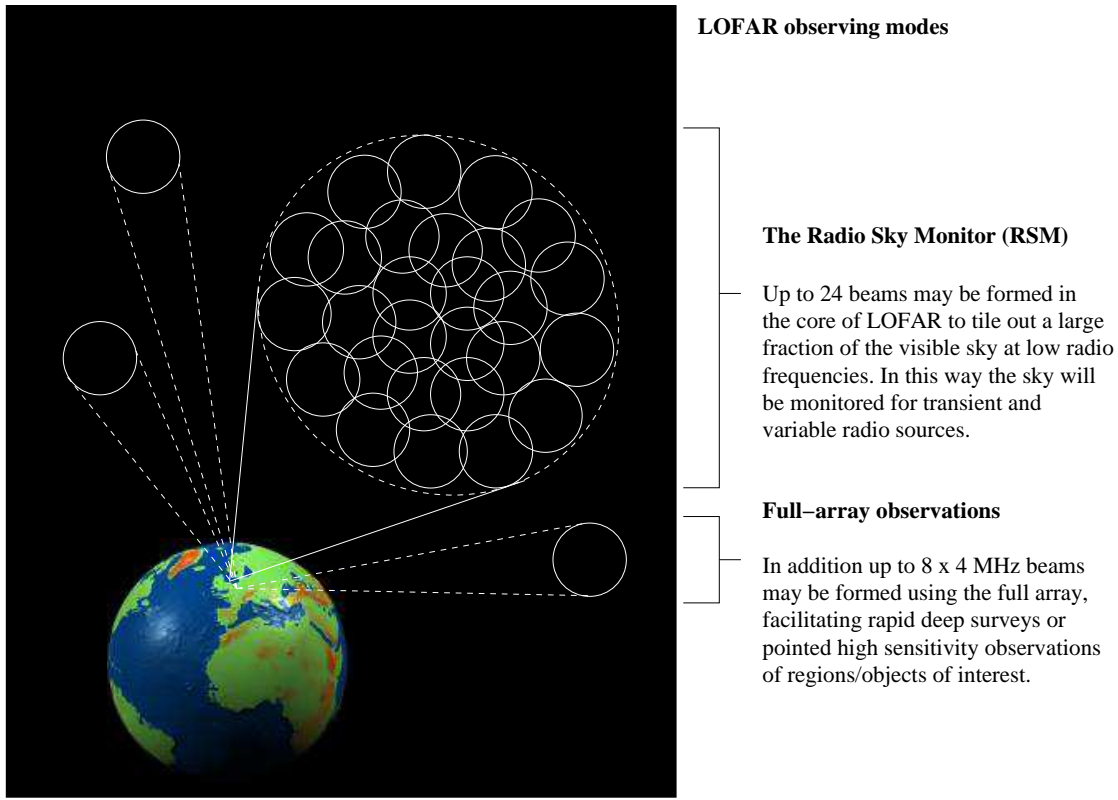,width=15cm}}
\caption{\label{obsmodesfig} \small The modes of operation of LOFAR. The
wide-field, multi-beam capability of the telescope will allow targeted
observations of specific targets, and systematic surveys, to be carried
out in parallel with deep monitoring of a large fraction of the visible
sky on a daily basis.  }
\end{figure}

The Dutch LOFAR will be located entirely within a region of $\sim 200$
km diameter in the north of the Netherlands. It will be composed of
between 36 and 50 stations (depending upon finances), approximately
half of which will be located in a 2\,km core close to Exloo. Each
station will contain 96 low-band (30-80\,MHz) and 48 high-band
(120-240\,MHz) antennae. These antennae will be extremely simple in
design and have no moving parts (see Figure~\ref{antenfig}). They
will, however, be sensitive to a large fraction of the sky, and by
{\em beamforming} at each station can be made to `look' in any
direction on the sky. The signal from each station is transported to
the LOFAR correlator (an IBM {\em BlueGene} supercomputer located at
The University of Groningen), where it is correlated and then passed
to a secondary computing cluster where images are formed and
distributed to the science teams. Only data transport and computing
limitations limit the number of beams available -- the current LOFAR
design allows for, e.g., 8 $\times$ 4 MHz beams, or a smaller number
with larger bandwidth (greater sensitivity) up to a maximum of 32
MHz. The LOFAR stations located in the core will be able to form more
beams in order to monitor a large fraction of the sky simultaneously.
Figure~\ref{obsmodesfig} illustrates the possible LOFAR observing
modes.

Progress with the construction and operation of LOFAR in The Netherlands
continues apace. The LOFAR Initial Test Station (ITS) operated
successfully with 16 prototype low-band antennae between 2004 and 2006. In
the fourth quarter of 2006 the LOFAR Core Station 1 (CS1) was completed at
Exloo in the Netherlands, close to the designated core of the full
array. CS1 comprises a central cluster of 48 low-band antennae and three
outliers each of 16 antennae, to a maximum baseline of 500m (in order to
better simulate the operation of a full LOFAR). The design for the
high-band antennae is at an advanced stage, and the {\em BlueGene}
supercomputer correlator is fully operational. Over the next two to three
years all of the Dutch stations will be constructed and connected and the
Dutch array will be complete. In the meantime data is continuously being
recorded and analysed as the array grows: the first official sky maps from
CS1 have been publically released, and LOFAR has also detected its first
pulsar and solar bursts (see {\it www.lofar.org} for more details).  In
November 2006, LOFAR had a Calibration Comprehensive Design Review (CDR),
to which the response was positive (in particular the CDR panel were happy
that issues associated with Radio Frequency Interference (RFI) would not
be a show-stopper).

LOFAR is therefore an innovative technology project in an advanced stage
of development, where the novel approach to data transport, data
processing and associated software development will lead to a step-change
in radio astronomy capabilities. It is certainly the most innovative and
ambitious radio astronomy project prior to construction of the Square
Kilometre Array (SKA), which is a long-term ($\sim 15$yr) project for the
world astronomy community and a cornerstone of the STFC Road Map. For
more details about LOFAR see {\it www.lofar.org}.

\subsection{International extensions of LOFAR}

The interferometric design of LOFAR means that additional stations can
relatively easily be added to the Dutch array, subject to the availability
of a fast internet connection to transport the data in real-time to the
{\it BlueGene} correlator in Groningen. There has been strong
international interest in extending the LOFAR array beyond the borders of
the Netherlands. The primary reason for this is that increasing the
longest baselines of the array leads to a corresponding increase in the
angular resolution of the observations; with international baselines, the
resolution at the highest LOFAR frequencies will be below an arcsecond.
Such angular resolution is important to accurately localise the detected
radio sources, and allow cross-matching of these with sources detected in
other wavebands. In addition, deep LOFAR observations quickly become
confusion-limited, and so an increase in angular resolution translates
directly to an increase in the sensitivity to which the array is able to
probe.

\begin{figure}[!t]
\begin{tabular}{cc}
\raisebox{3cm}{\parbox{9cm}{\Large{\bf See associated jpg file\\ LOFAR\_EUsites.jpg}}}
&
\psfig{file=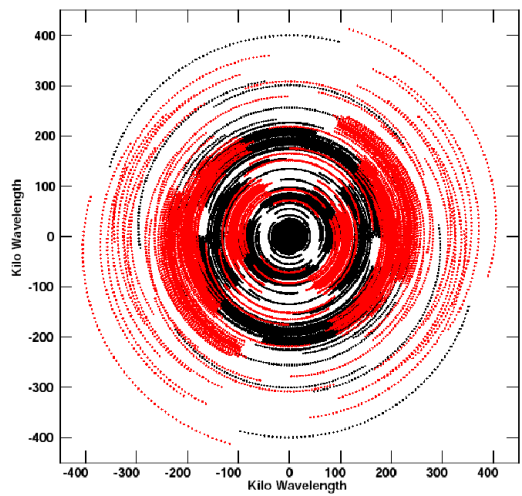,width=7cm}
\\
\end{tabular}
\caption{\label{eusites} \small {\it Left:} The locations of currently
funded international stations in Germany, France and Sweden, together with
the proposed 4 UK stations. {\it Right:} A simulation of the (u-v)
coverage (at 100\,MHz) provided by the currently funded Dutch and
international LOFAR stations, together with the proposed UK
stations. Baselines involving UK stations are shown in red.}
\end{figure}

Several European countries (including Germany, France, Sweden, Italy and
Poland, as well as the UK) are all investigating the possibility of
constructing LOFAR stations.  The most advanced of these is the German
Long Wavelength consortium (GLOW) which is proposing the construction of
six LOFAR stations within Germany (see {\it
www.mpifr-bonn.mpg.de/public/pr/white.paper.oct6.pdf}), with the first of
these at Effelsberg already being commissioned and funding for three more
confirmed. The French Long Wavelength consortium (FLOW) has also produced
a White Paper (see {\it
www.lesia.obspm.fr/plasma/LOFAR2006/FLOW\_Science\_Case.pdf}), proposing
the construction of one LOFAR station at Nan\c{c}ay. There is also
confirmed funding for a Swedish station at Onsala. The distribution of
these sites is shown in the left panel of Figure~\ref{eusites}. In
combination with these, UK stations would offer both the longest baselines
(highest angular resolutions) and also a significantly improvement in the
instantaneous (u,v) coverage for intermediate baselines, as shown in the
right panel of Figure~\ref{eusites} and discussed in more detail in
Section~\ref{technical}.

Since 2004, members of the UK community with an interest in LOFAR have
been meeting every few months to critically appraise the possibility of a
significant UK contribution to LOFAR. This has led to the establishment of
the {\it `LOFAR-UK Consortium'}.  In the space of two years the LOFAR-UK
consortium has established itself as one of the broadest UK astronomy
consortia. Over 100 members have voluntarily subscribed to the general
LOFAR-UK mailing list, which is comparable in size to the mailing list for
the Sloan Digital Sky Survey. Currently thirteen UK universities have
pledged funds to the project, totalling $\pounds$600k. In addition two --
University of Manchester and the RAL -- have pledged `in kind'
contributions (fibre connections and/or land) of equivalent or greater
value. An additional number of UK Universities are currently attempting to
secure funds to join the consortium.

The LOFAR-UK consortium was formalised in late 2006 via the signing of a
Memorandum of Understanding, leading to the formation of the LOFAR-UK
Management Council (MC). The MC has elected Rob Fender (Southampton) and
Steve Rawlings (Oxford) as Project Leader and Deputy Project Leader
respectively. Three project coordinators with different responsibilities
have also been appointed: Science Coordinator -- Philip Best (Edinburgh);
e-Science Coordinator -- Bob Nichol (Portsmouth); Technical Coordinator --
Rob Beswick (Manchester).  The University of Hertfordshire will act as
treasurers for the LOFAR-UK project.

The LOFAR-UK consortium has concluded through scientific simulations that
the optimal initial approach is the deployment of four LOFAR-UK stations,
subtending a range of angles and baseline lengths relative to the Dutch
LOFAR core. Leading candidates for these four sites are {\it Lord's
Bridge}, {\it Chilbolton}, {\it Jodrell Bank} and {\it Edinburgh}. The
Lord's Bridge and Jodrell Bank sites already have e-MERLIN fibre
connections; for Chilbolton and Edinburgh the cost of connecting to a dark
fibre network is under investigation, but in both cases a 10 Gbit/s
connection to the nearest regional JANET network is possible.  On a longer
timescale, there is an ambition to add additional UK stations, with
specific possibilities being central Wales and the north--east of England.

\subsection{This document}

The layout of this document is as follows. Section~\ref{context} details
LOFAR's context in the overall picture of UK radio astronomy. LOFAR will
build upon the UK's existing widespread expertise in radio astronomy and
interferometry, and will allow the UK to develop new scientific and
technical expertise in the many additional areas in which radio
astronomical facilities will contribute in the coming decades. As such,
LOFAR will provide the optimal stepping stone on the road to the Square
Kilometre Array which, through increased sensitivity and direct redshift
measurement, will totally revolutionise the field. The broad--base of the
LOFAR-UK consortium demonstrates how UK involvement in LOFAR will lead to
the development of a wide UK community with interest and experience in the
new science to be achievable with SKA.

Sections~\ref{reion} to~\ref{solar} detail the specific UK science
interests in using LOFAR.  LOFAR is a ground-breaking experiment which has
set the pace for developments in next generation radio astronomy, and will
have a profound impact on a broad range of astrophysics from cosmology to
solar system studies. Many of these science goals have already been well
described by the Dutch Science case for LOFAR (available at {\it
www.lofar.org/PDF/NL-CASE-1.0.pdf}), and so the emphasis in the current
document is upon:

\begin{itemize}
\item The ways in which the main LOFAR science goals can be enhanced
through UK involvement in LOFAR, through the provision of complementary
datasets and facilities available to UK researchers, and through the
expertise in the UK community.
\item New science capabilities that will arise from the improved
  angular resolution of LOFAR when UK (international) baselines are
  added. 
\item Additional science goals of interest to UK scientists.
\end{itemize}

The UK's science interests can be categorised under five broad
headings:

\begin{enumerate}
\item The Epoch of Reionisation
\item Low Frequency Surveys of the Radio Sky with LOFAR
\item Radio Transients (including Pulsars)
\item Ultra-High Energy Cosmic Rays
\item Solar and Heliospheric Physics
\end{enumerate}

These are detailed in Sections~\ref{reion} to~\ref{solar}
respectively. The first four of these correspond to the four Dutch Key
Projects for LOFAR. The fifth is an important additional research
interest, which is also shared by the German GLOW consortium, and has
been adopted as an International LOFAR Key Project.

Section~\ref{technical} details the technical issues surrounding the
LOFAR-UK project. The choice of sites within the UK for the construction
of LOFAR stations is discussed, as are the technical challenges relating
to connection of these sites to the main LOFAR array. In addition, since
Solar and Heliospheric Physics is not one of the Key Projects identified
by the Dutch LOFAR team, and hence has no pre-defined observing strategy,
the observational requirements for this aspect of LOFAR-UK science are
detailed. Finally, Section~\ref{consortsec} summarises the organisation
and management of the LOFAR-UK consortium, together with the current
funding status.
\clearpage

\section{LOFAR in the context of UK radio astronomy}
\label{context}

The UK has a strong history in radio astronomy, and is looking forward to
being an active and leading player in the field for the coming
decades. The low-frequency radio surveys conducted at Cambridge over the
past five decades have produced some of the most important and
well--studied radio source samples, providing a wealth of information on
the most powerful active galactic nuclei over the entire history of the
Universe. Over the same period, the Jodrell Bank observatory has been at
the cutting edge of the development and operation of long baseline and low
frequency radio interferometers. The development of the phase-stable
radio-linked interferometer led to the creation of MERLIN, the world's
largest permanently connected interferometer, which has a maximum baseline
length of 217km.

Current UK priorities for radio astronomy are the development of
e-MERLIN and the Design Study for the Square Kilometre Array
(SKADS). All of the leading institutions in these projects are members
of the LOFAR-UK consortium, and fully support the project. At low
radio frequencies, LOFAR is the most ambitious `SKA Science
Pathfinder'.  The digital/software basis of LOFAR was the inspiration
for the `all--digital' SKA concept being developed by UK groups as
part of SKADS for the low ($\lta 1$\,GHz) frequency band. The SKA is a
cornerstone of the STFC Road Map, and UK involvement in LOFAR would
enable UK astronomers to build up important scientific and technical
expertise.

\subsection{The University of Manchester and Jodrell Bank Observatory}

The University of Manchester's Jodrell Bank Observatory has been
developing and operating long-baseline and low frequency radio
interferometers for over 40 years, culminating in the creation of MERLIN,
the largest permanently connected interferometer in the world.  Initially
MERLIN's prime frequency was 408 MHz, not far above the LOFAR band: some
of its very first observations showed the existence of extended low
frequency emission around flat-spectrum quasars, leading to the
orientation-based unified scheme of flat- and steep-spectrum quasars
\citep{orr82}.  More recent 408 MHz observations, including a 220-km
baseline to Cambridge, were made in 1994 and highlights include the
detection of a large \HII\ region in M82 \citep{wil97}.

MERLIN has also operated at 151 MHz, during the 1985 solar minimum.  After a
period of `interference-busting', mostly involving locating sources of
sporadic broad-spectrum RFI, several observing programmes were completed,
notably on the bridges of radio galaxies \citep{lea89} and the extended
emission surrounding the jet of 3C273 \citep{dav85}.

MERLIN observations are now focussed on sensitive high-resolution imaging
at 5 and 1.5 GHz, with an angular resolution of 50 to 150 milli-arcseconds.
The e-MERLIN upgrade involves the installation of an optical fibre network
connecting the 5 remote MERLIN telescopes to Jodrell Bank at 30 Gb/s each,
using `dark fibre' and purpose-built transmission equipment. Together with
new receivers, IF equipment and a new correlator, this will provide
micro-Jy sensitivity.  First fringes with a prototype correlator are
expected in early 2008.  e-MERLIN will be the natural high-resolution
partner to the EVLA.  e-MERLIN will also provide a natural complement to
the extended LOFAR, providing sub-arcsecond imaging at 1.5 GHz for
comparison with LOFAR images below 300 MHz at similar resolution.
[Inclusion of LOFAR stations in the UK and in southern Germany will
provide baselines of up to 1200km, providing an excellent match to
e-MERLIN at 1.5 GHz].

Several of the key science goals for e-MERLIN have close parallels with
the science goals for a long-baseline LOFAR, including studies of distant
starburst galaxies, the co-evolution of galaxies and AGN, strong
gravitational lensing, and jets from X-ray binaries.  The large frequency
span of e-MERLIN and the extended LOFAR at comparable angular resolution
will benefit many of the programmes.  Of course, e-MERLIN does not have
the very wide field of view of LOFAR, but it is wide enough (10-30 arcmin
diameter) to play an important role in deep surveys, as well as follow--up
observations of objects detected and studied in other LOFAR projects.

Jodrell Bank Observatory and the University of Manchester have also played
a leading role in the development of data transmission for long-baseline
radio astronomy using e-VLBI, carrying out some of the first UK-NL
transmissions at $>$700 Mb/s, diagnosing bottlenecks in various e-VLBI
connections, and working on transmission techniques and protocols. They
are also working on direct digital connections into and out of the
e-MERLIN correlator, to allow multiple e-MERLIN telescopes to transfer
data to JIVE at 1 Gb/s each and to connect other European radio telescopes
to the e-MERLIN correlator at 4 Gb/s. They are also heavily involved in
the design and development work for the Square Kilometre Array as part of
the UK technical collaboration involving Cambridge, Manchester and Oxford.

\subsection{The University of Cambridge and the Mullard Radio Astronomy
  Observatory}

The Cavendish Astrophysics Group (formerly Radio Astronomy Group) played a
major role in the development of radio interferometry and the use of
deep, low-frequency, radio surveys for cosmology. A variety of
interferometers were developed to perform the 3C, 4C, 5C, 6C and 7C radio
surveys. The Mullard Radio Astronomy Observatory has been located at
Lord's Bridge near Cambridge for the past 50 years.  Surveys and other
experiments have operated there at frequencies from 38\,MHz through to
15\,GHz with the main Cambridge surveys being performed at 178\,MHz and
151\,MHz, in the middle of the LOFAR high--frequency band. These surveys
have made a considerable impact on a broad range of astrophysics. The 3C
and 4C surveys detected the most powerful active galactic nuclei (AGN)
over the history of the Universe, providing constraints on the evolution
of the black hole activity \citep[e.g.,][]{lon66} and acting as beacons to
the most massive galaxies at all redshifts \citep[e.g.,][]{lil84}. The 4C
survey also provided crucial evidence in support of Big Bang cosmologies.
Pulsars were discovered with the sensitive 81.5 MHz phased array.  The One
Mile and 5-km telescopes were pioneering instruments in the development of
true imaging radio interferometers.

Current experimental activities in the Cavendish Astrophysics Group
include radio, sub-millimetre and optical interferometry. At the Mullard
Radio Astronomy Observatory the principal instrument is a telescope for
blank-field Sunyaev Zel'dovich work called AMI (Arcminute MicroKelvin
Imager), which is nearing completion.  The observatory also hosts one of
the telescopes of the MERLIN interferometer.  The group also participates
in the construction and exploitation of other telescopes and experiments
sited around the world. They have provided common user receivers for the
JCMT (most recently the heterodyne array system HARP) and are building
prototype water vapour radiometers for ALMA. Experiments that they are
prime contributors to include the Very Small Array, a custom-built radio
telescope (operating between 26 and 36\,GHz) sited in Tenerife, for
imaging primordial anisotropies of the CMB, and most recently CLOVER, a
millimetre experiment for measuring CMB polarisation.  They are partners
with the University of New Mexico in the construction of a major new
facility for optical interferometry -- the Magdelena Ridge Optical
Interferometer -- which builds on earlier work with the Cambridge Optical
Aperture Synthesis Telescope (COAST).

The Cavendish Astrophysics Group are heavily involved in the design and
development work for the Square Kilometre Array as part of the UK
technical collaboration involving Cambridge, Manchester and Oxford: this
includes leading one of the main design studies within the European SKA
Design Study with responsibility for, among other areas, technical
simulations. They strongly support the push for UK involvement in LOFAR,
and have extensive technical and scientific expertise to offer to the
project.

\subsection{LOFAR-UK as a stepping-stone to the SKA}

UK astronomers have been central to the planning for the Square Kilometre
Array since its inception.  Peter Wilkinson (Manchester) published one of
the first papers outlining the basic concept of a large interferometer for
\HI\ surveys \citep{wil91}, and Wilkinson and Diamond (Manchester) have
been prominent, long-standing and influential members of the International
SKA Steering Committee.  Rawlings (Oxford) spent 2003-2005 as vice-Chair
and then Chair of the International Science Working Group: during this
period he co-edited the international SKA science case \citep{car04}

In 2004, Wilkinson was the UK coordinator for the SKA Design Study
(SKADS) which culminated in a successful bid for European Community
funding under the Framework 6 (FP6) programme. This proposal unifies
significant activity in several EC countries (France, Germany, Italy,
the Netherlands, Portugal, Spain, UK).  The UK SKADS consortium was
formed comprising three major University partners (Cambridge,
Manchester and Oxford) alongside three minor partners (Cardiff,
Glasgow and Leeds). This consortium successfully applied for PPARC
funding that has resulted in a comprehensive design study
incorporating science, data and network simulations, as well as a
hardware design project built around the concept of an all-digital
system. The UK SKADS programme will be funded from mid--2005 to
mid--2009, and has built up a team of more than 20 postdoctoral
researchers split across the participating UK institutions.

The UK SKADS team have led the development of the SKADS Benchmark Scenario
-- which couples the capabilities of phased-array technologies at low
($<1$ GHz) frequencies with those of small dishes at higher frequencies --
and recently costed this for consideration by the International SKA
Project Office and the wider international SKA effort. They have been
developing links with all of the international SKA pathfinder projects. In
the low-frequency (30--300\,MHz) band these include LOFAR, the Mileura
Wide-field Array (MWA) and the Long Wavelength Array (LWA). In the
mid-to-high frequency band (300\,MHz -- 20\,GHz), these include the Allen
Telescope Array (ATA), MWA and the Karoo Array Telescope (KAT).

In 2006, following comprehensive studies of several candidates sites, the
international SKA project short-listed two sites for further
consideration: Western Australia and South Africa. Some SKA pathfinder
projects are being developed on these sites. The only SKA pathfinder
project being developed in Europe is LOFAR. As the largest contributors to
the SKADS EC project are the Netherlands and the UK, it seems natural that
these countries collaborate in some way on the LOFAR project. The total
cost of UK involvement in LOFAR is tiny compared to the likely UK and
global investment in radio astronomy over the next two decades, but will
offer a unique opportunity to obtain hands-on experience both on a
technical and on a scientific level. It is also worth emphasising that the
southern hemisphere locations proposed for the SKA mean that its
construction will not render LOFAR redundant: it is envisaged that both
will operate in tandem, providing full--sky surveys and transient
monitoring.
\clearpage

\section{LOFAR-UK and the Epoch of Reionisation}
\label{reion}

Protons and electrons produced in the Big Bang combined together at a
redshift of $z\approx 1100$ to form neutral hydrogen. UV and X-ray
radiation from the first stars and black holes (AGN) then ionised the
hydrogen again\footnote{Stars or AGN are by far the most likely reionising
sources, but the possibility of something more exotic, such as decaying
particles, cannot be excluded. In such a case, LOFAR's reionisation
measurements could have fundamental implications for physics.}, and later
radiation from galaxies and quasars kept the bulk of the Universe highly
ionised up to now. However, little is known about when and especially how
the reionisation happened. LOFAR will probe the reionisation epoch by
searching for the redshifted 21-cm signal that arises from neutral
hydrogen in the intergalactic medium (IGM), provided that the spin
temperature of the \HI\ is de-coupled from the Cosmic Microwave Background
temperature.  This signal disappears as the IGM gets ionised, and
measurements of the distribution of neutral and ionised regions around
reionisation will provide a wealth of information about the first stars
and galaxies.

Study of the Epoch of Reionisation is one of the Dutch Key Projects for
LOFAR, and has been one of the key drivers for the design of the
telescope, in particular the concentration of a large fraction of the
stations within a central core. The addition of longer baselines to LOFAR
is not essential for the detection of the reionisation signal {\it per
se}, but will play an important role in the identification, localisation
and spectral characterisation of the contaminating foreground signals,
which are the largest impediment to the detection of the redshifted 21-cm
signal. In addition to providing long-baseline stations, the UK will bring
to the LOFAR consortium extensive expertise in cosmological simulations of
large-scale structure formation, galaxy formation, and radiative
transfer. Providing good models for how galaxies form will be essential to
maximise the information that can be obtained from the LOFAR data, and
realise the full potential of having this exciting new window on the
Universe.

\subsection{Introduction: The Epoch of Reionisation}

According to the present view of the early Universe, hydrogen atoms
were first formed about a million years after the Big Bang, when the
primordial matter cooled to a temperature of about 3000K. The Universe
then became dark and it cooled further due to the general Hubble
expansion. These ``Dark Ages'' came to an end many hundreds of millions
of years later, when the first stars and black holes started producing
light. Ionising radiation from these, and subsequently forming objects,
then began to warm and ionise the Universe, until it again became
highly ionised. The time at which 50\% of the Universe was ionised is
termed {\bf the Epoch of Reionisation (EoR)}. Although it is important
to determine the redshift of the EoR, the main scientific interest is
in {\em how} the Universe went from neutral to ionised, and the
properties of the sources that caused the transition.

Current observational constraints on the EoR are not very strong. They are
based on measurements of cosmic microwave background (CMB) temperature
anisotropies and polarisation, Gunn-Peterson absorption in quasar spectra,
and the temperature evolution of the IGM.

\begin{enumerate}
\item After reionisation, free electrons Thomson-scatter CMB photons,
smoothing out the temperature anisotropies on small scales. When
measurements of temperature anisotropies are combined with data on the
large-scale structure of galaxies, the EoR is constrained to be below
$z\sim 30$. Thomson scattering also induces polarisation in the CMB, and
(under the simplified assumption of a sudden reionisation history) the
Wilkinson Microwave Anisotropy Probe (WMAP) team used the 3-year dataset
to constrain the EoR to be around $z\sim 12$
\citep[Figure~\ref{fig:reion};][]{spe07}, although there is some
degeneracy with other cosmological parameters. Note that this is
significantly lower than the redshift $z\sim 20$ estimated from the WMAP
1st-year polarisation data.

\begin{figure}[!t] 
\begin{center}
\psfig{file=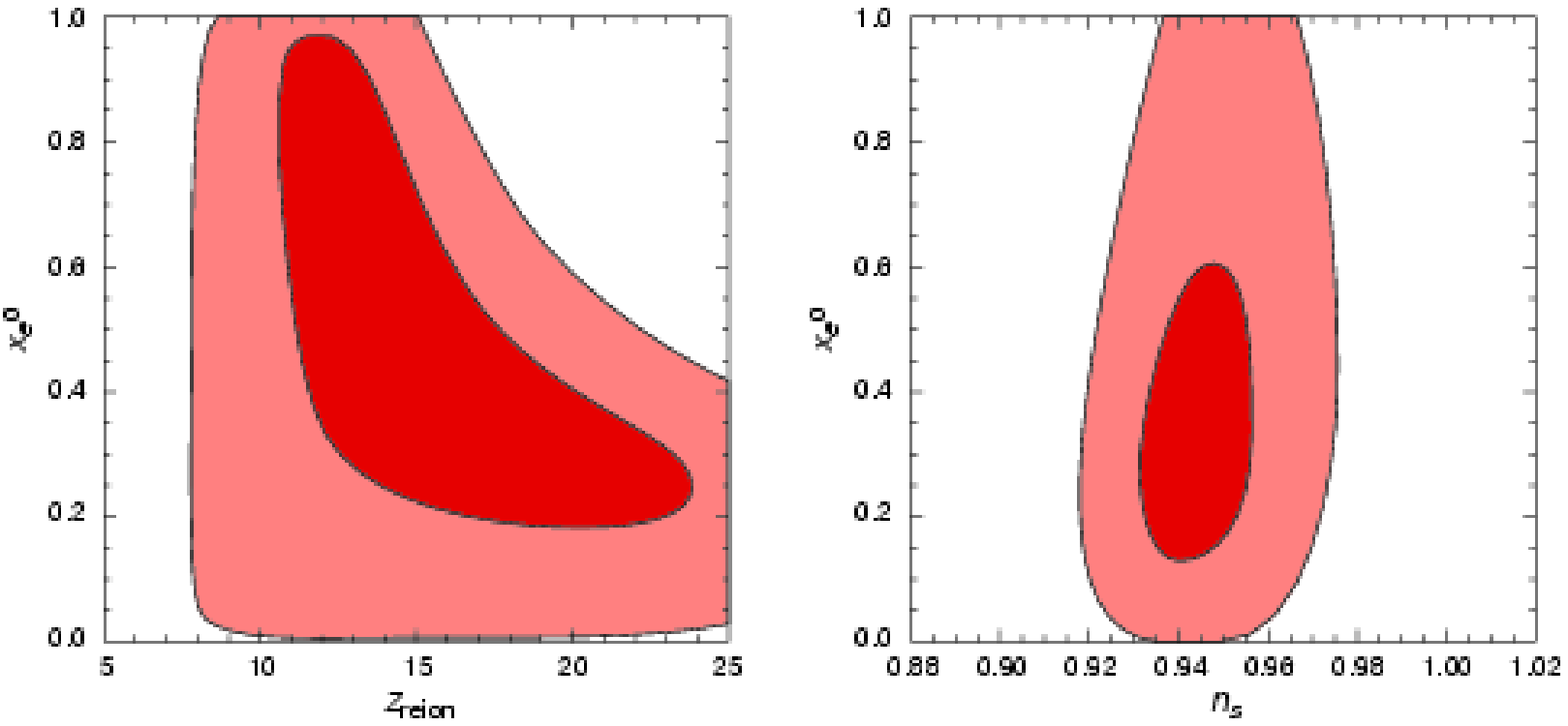,width=65mm,clip=}
\end{center}
\caption{\label{fig:reion} \small WMAP year 3 joint constraints on the
reionisation redshift, $z_{\rm reion}$, and ionised fraction,
$x_e^0$. These assume that the IGM became partly ionised at $z_{\rm
reion}$ to an ionisation fraction of $x_e^0$, and then became fully
ionised at $z=7$ \citep[from][]{spe07}.}
\end{figure}

\item The absence of a Gunn-Peterson trough in the spectra of $z\le 6$
QSOs shows that the IGM was highly ionised by $z \approx 6$
\citep[e.g.,][]{bec01}.  The mean flux decrement increases rapidly before
$z\approx 6$ \citep{bec01,fan02}, and indicates that the Universe is
becoming more neutral. This is also indicated by the sizes of ionised
regions around high-$z$ QSOs \citep{mes07}, although systematics
complicate this interpretation \citep{bol07}. However, such measurements
do not directly determine the redshift at which the ionised fraction was
50\%.

\item As the Universe gets photo-ionised, it also gets heated. After
reionisation, Compton cooling again cools the IGM and this can be used to
constrain the EoR \citep[Figure~\ref{fig:theuns};][]{the02} to $z\sim 10$,
although it is dependent on assumptions about the heating rate after
reionisation and the reionisation of elements heavier than hydrogen.

\begin{figure}[!t]
\begin{center}
\psfig{file=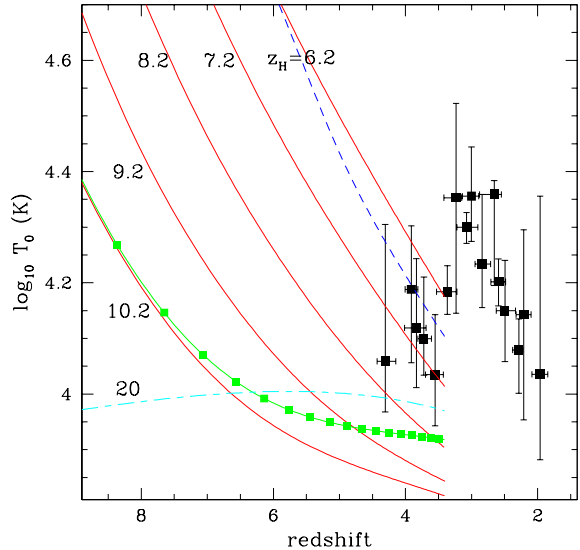,width=9cm}
\vspace*{-0.5cm}
\end{center}
\caption{ \small The evolution of the temperature, $T$, of the IGM. The
red lines show the model predictions (not including radiative transfer)
for the temperature evolution for different (labelled) \HI\ reionisation
redshifts, while the black points show measurements of $T$ from quasar
absorption lines. The rise in the measured temperature at $z\sim 3$ is
thought to be caused by reionisation of \HeII, and so the constraint on the
reionisation redshift of \HI\ comes from comparing models and data at
$z\sim 4$ \citep{the02}.  Adiabatic expansion cooling causes the Universe
to be colder than observed at $z=4$ if the EoR is too early, while
reionisation at later times fit the data better. These models assume a
very high post-reionisation temperature, to show that, even for such
(unrealistically) high initial temperatures, reionisation should not have
happened much above $z=10$ to avoid the IGM being to cool at $z=4$.}
\label{fig:theuns}
\end{figure}

\item The level of UV background light post-EoR also constrains the EoR
and the spectra of the sources responsible. Current models
\citep[e.g.,][]{gne00,ben06} tend to over-produce the post-EoR UV
background, resulting in an IGM too transparent to Ly$\alpha$ scattering,
as compared with the observational data. Attempts to combine the measured
evolution of the mean scattered Ly$\alpha$ flux with numerical simulations
of the IGM suggest that there is no rapid evolution of the reionising
source emissivity beyond $z > 5$, and that the EoR is at $z \lta 11$
\citep{mei05}.
\end{enumerate}

In summary, there are good indications that reionisation started somewhere
around $z \lta 12$, and was mostly completed by at least $z \gta 6$.
However, none of the existing constraints are very tight, and current
models struggle to reproduce all of the data. Improved observational
constraints on the epoch and mechanism of reionisation are urgently
required.

In the redshift range $6<z<11$, the 21-cm hyperfine line of neutral
hydrogen will fall within the LOFAR high frequency band, and will be
observable provided that the spin (excitation) temperature of the
transition has decoupled from the CMB temperature \citep[e.g.][]{mad97}.
This line will disappear as the Universe gets ionised, and hence LOFAR
offers an opportunity to probe in great detail the epoch of reionisation
and the nature of the sources of ionising photons in the early Universe. A
firm detection of the EoR by LOFAR would be of great importance, whilst
even a null detection would significantly improve our current knowledge of
both the epoch and mechanism of reionisation, providing important
constraints on the early history of radiation sources. The PAST and MWA
({\it web.haystack.mit.edu}) experiments in China and Australia,
respectively, have similar goals.

Four of the most important questions that need addressing about this
major event in the history of the Universe are:

\begin{itemize}
\item When did reionisation occur?
\item How rapidly did the IGM change from being mostly neutral to
  mostly ionised?  
\item What were the sources of ionisation, and how did they affect the
    global progression of the reionisation?
\item How did reionisation affect subsequent galaxy formation?
\end{itemize}

LOFAR aims to provide the data to answer these questions, but it will be
necessary to make realistic models of reionisation in order to take full
advantage of all of the information that LOFAR observations will provide.

Section~\ref{obssig} discusses in some detail the signal expected from
reionisation. Section~\ref{foregrnd} considers the issue of foreground
emission sources, and the techniques needed to remove these in order to
detect the reionisation signal. Section~\ref{galform} discusses the
requirements on modelling to interpret this signal, and
Section~\ref{reionuk} details the contribution that the UK community can
offer in this respect.

\subsection{Observational 21cm signatures of Reionisation}
\label{obssig}

In the simplest idealised model of reionisation, the IGM everywhere
changes abruptly from being neutral to ionised at the reionisation
redshift, $z_r$. If the spin temperature ($T_s$) of the 21-cm hyperfine
transition has already decoupled from, and been raised above, the CMB
temperature ($T_{CMB}$) before reionisation, then this will manifest
itself in 21-cm radiation as a global, all-sky spectral signal: 21-cm
radiation will be detected in emission at wavelengths $\lambda\ge 21\times
(1+z_r){\rm cm}$, but not below that \citep{sha99}. If $T_s \gg T_{CMB}$,
then this step in brightness temperature will depend only on the cosmic
baryon density and the reionisation redshift: for $z_r \approx 10$ it will
have an amplitude of 10-20mK and occur at a frequency around 130\,MHz. In
principle, LOFAR will be easily able to detect such a signal after only a
few hundred hours of observing. Measuring the frequency at which the step
in emission occurs will then directly determine $z_r$, while the width of
the step in frequency will yield the duration of the reionisation
epoch. Although the reality of extracting such a signal from the
foreground contamination will be non-trivial (see Section~\ref{foregrnd}),
determining these parameters is one of the key science goals of the
Netherlands LOFAR consortium.

Complications to this simple picture arise because (1) the \HI\ spin
temperature needs to decouple from the CMB temperature for any 21-cm
signal to be seen (in either emission or absorption), and (2) the
reionisation is expected to occur inhomogeneously, with \HII\ regions
being ionised around individual sources and growing until they
eventually overlap. Even within the neutral part of the IGM, the gas
density and kinetic and spin temperatures will be inhomogeneous. 

The \HI\ spin temperature, $T_s$, can be coupled to the gas kinetic
temperature, $T_k$, by collisions between atoms \citep[and with ions
and electrons;][]{sco90}, and by scattering of Lyman-$\alpha$
photons. In the absence of these processes, $T_s$ will relax to the
CMB temperature, and the \HI\ gas will be invisible in 21-cm radiation
(either in emission or absorption). Atomic collisions are likely to be
important only in fairly overdense regions \citep[e.g.,][]{kuh06}, but
non-ionising UV-radiation from the first galaxies can couple the spin
and kinetic temperatures even in low-density gas through a process
known as \lq Lyman-$\alpha$ pumping\rq\, or the \lq
Wouthuysen-Field\rq~ effect. This is expected to couple $T_s$ to $T_k$
throughout most of the IGM, even when only a small fraction of the IGM
has been reionised \citep{mad97,hir06}. Several recent papers discuss
these processes in detail, but reach differing conclusions:
\citet{kuh06} and \citet{ili02} argue that a strong emission signal
will be produced by collisional coupling in mini-haloes, whereas
\citet{oh03} suggest that Lyman-$\alpha$ pumping dominates the
coupling under almost all circumstances. There is clearly much scope
for further theoretical modelling to put the basics of the 21-cm
signal on a firmer footing.

\begin{figure}[!h] 
\begin{center}
\psfig{file=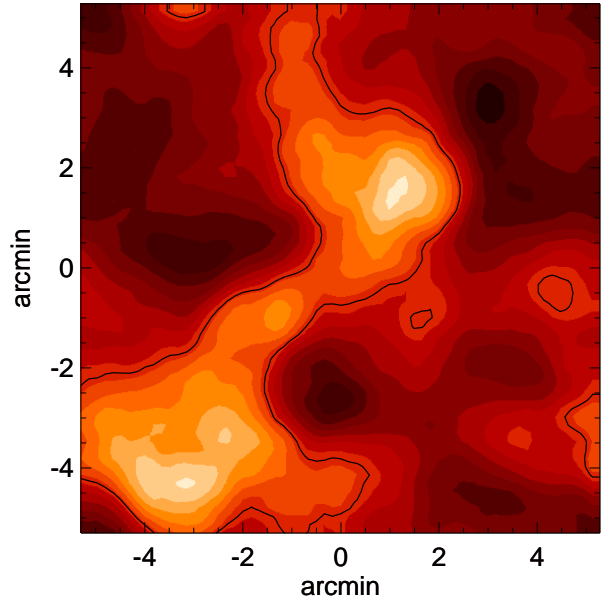,width=8cm,clip=}
\end{center}
\caption{\label{eorfig} \small Simulated radio map of redshifted 21--cm
emission against the CMB at $z=8.5$, on the scale of $20h^{-1}$ (comoving)
Mpc \citep[from][]{toz00}. The point spread function of the synthesised
beam is assumed to be a spherical top--hat with a width of 2 arcmin. The
frequency window is 1 MHz around a central frequency of $150$ MHz.  The
colour intensity goes from $1$ to $6\,$ $\mu$Jy per beam.  For clarity, the
contour levels outline regions with signal greater than $4\,\mu$Jy per
beam.}
\end{figure}

\begin{figure}[!h]
\begin{center}
\psfig{file=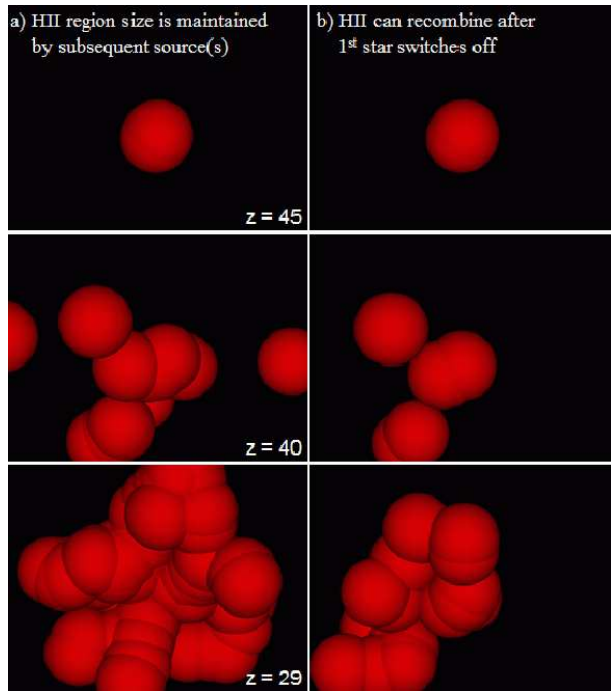,width=80mm}
\end{center}
\caption{\small The growth of overlapping \HII\ regions around early
  ionising sources \citep[from][]{ree05}.}
\label{fig:reed}
\end{figure}

\begin{figure}[!h]
\begin{center}
\psfig{file=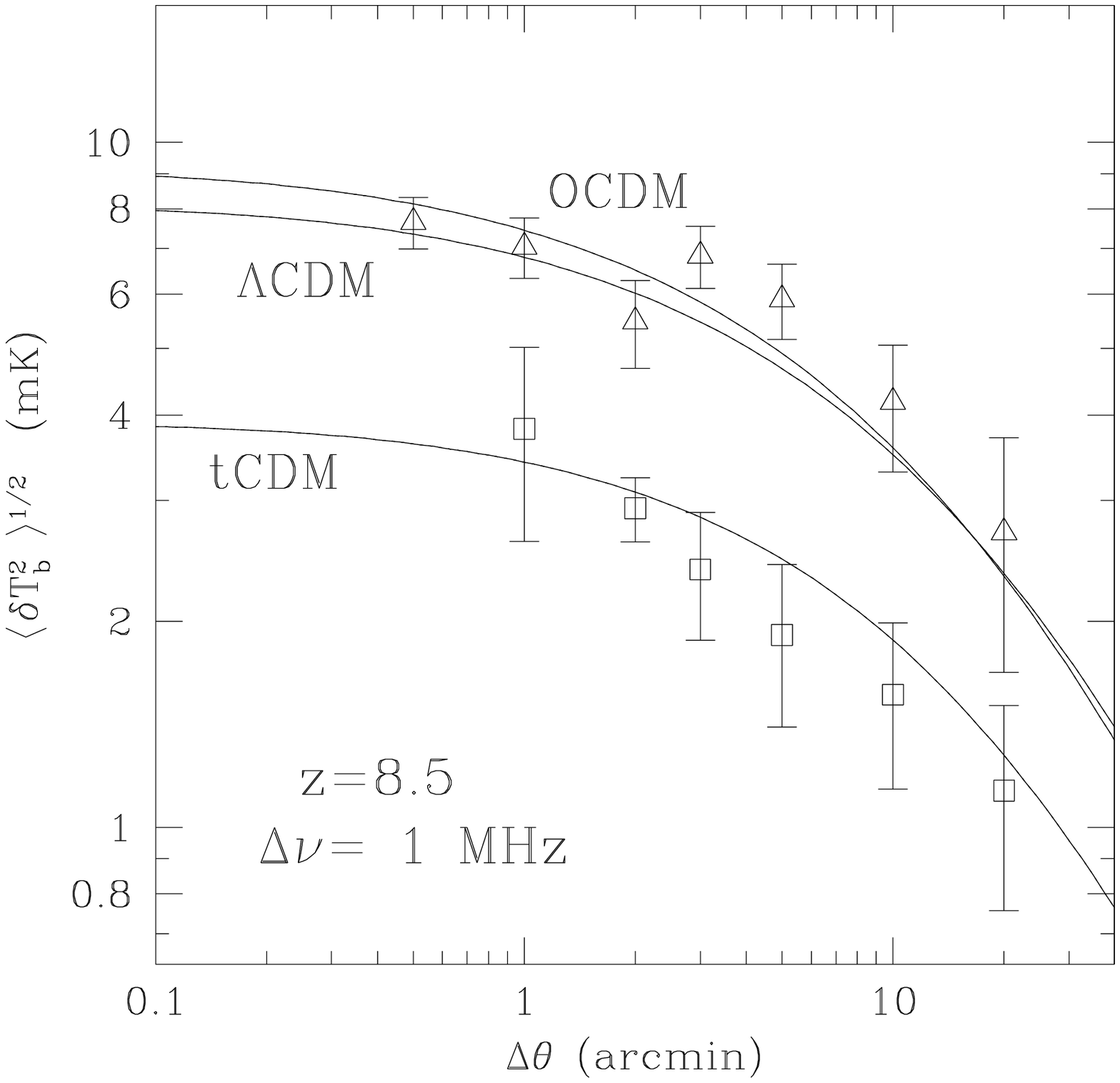,width=85mm}
\end{center}
\caption{\small The predicted rms brightness temperature fluctuations as a
function of angular scale \citep[from][]{toz00}. The smooth curves show
predictions based on linear theory, while the symbols show estimates based
on N-body simulations.}
\label{fig:tozzi}
\end{figure}

Once collisions or Lyman-$\alpha$ scattering have coupled $T_s$ to $T_k$,
the gas will appear in either emission or absorption in the 21-cm line,
depending on whether $T_k > T_{CMB}$ or $T_k < T_{CMB}$. Gas in halos and
filaments can be shock-heated, but for gas close to the average IGM
density, the only effective heating mechanism prior to reionisation
appears to be heating by X-ray photons produced by shock-heated gas in
galaxy halos, by star-forming regions, by QSOs, or by mini-QSOs
\citep{mad97,che07}. (Note that the energy required to have $T_k>T_{\rm
CMB}$ is only 0.004 eV per particle at $z \approx 10$.) Thus, halos and
filaments are expected to appear in emission in 21-cm prior to
reionisation, but the lower density IGM should appear in absorption at
early times, and in emission later (after it has been heated by X-rays;
cf. Figure~\ref{eorfig}).  Prior to full reionisation, the IGM will
therefore be a mixture of ionised regions surrounding sources, embedded in
partially ionised and neutral regions (Figure~\ref{fig:reed}). This will
lead to brightness fluctuations with structures up to a degree in
size. The simulations of \citet{toz00} -- see also \citet{cia03} and
\citet{fur04} -- suggest that the fluctuations in the brightness
temperature should decrease with increasing angular scale
(Figure~\ref{fig:tozzi}).

Although the 21-cm signal from \HI\ around the reionisation epoch is
expected to have structure down to very small angular scales (0.1 arcmin
or less), the low surface brightness of the signal means that it may only
be possible to detect it over receiver noise if it is smoothed over scales
of a few arcminutes on the sky, which in turn implies that only the
central core of the LOFAR array will be effective for measuring angular
structure in the signal from the reionisation epoch. In order to provide a
sufficiently large collecting area to calibrate the faint signal from the
EoR, approximately half of the Dutch LOFAR stations will be located within
the ($\sim 2$\,km) core. The design of LOFAR is thus optimised to detect
these fluctuations on a scale of a few arcmins at $z=8.5$.

\subsection{Removal of foreground signals and detection strategies}
\label{foregrnd}

The biggest hindrance to the detection of an EoR signal is the
contaminating foreground signals, which collectively exceed the EoR signal
by 5 orders of magnitude. These comprise unresolved radio sources (the
majority of which will be starburst galaxies at these low flux densities),
free-free and synchrotron emission from the Galaxy, and free-free emission
from electrons in the intergalactic medium
\citep[e.g.][]{dim02,oh03,gne04,mor04,bow07}. Galactic and extragalactic
radio recombination lines (RRLs) are also sources of contamination
\citep{oh03}, but more readily manageable: the frequencies of RRLs from
the Galaxy are well-known and avoidable, while the remaining extragalactic
RRLs are removable as they occur over narrow frequency ranges
\citep{gne04}.

The most straightforward detection of the EoR is a global whole-sky signal
\citep{sha99}. A strategy for removing the foreground contamination would
be to perform measurements over wide differences in frequency, to obtain a
fiducial measurement before (or after) reionisation is complete. The most
straightforward implementations of this approach are unlikely to succeed
because of the magnitude of the contamination \citep{gne04}. There may,
however, still be scope for developing this approach using more subtle
statistical methods.

Two alternative techniques are to chop the signal over angle or over
frequency \citep{mad97}. These methods may be used to image individual
growing HII regions or to measure the fluctuations in the signal in a
statistical sense \citep{toz00}. Extracting an EoR signal by differencing
measurements separated by angle, however, is unlikely to adequately cancel
the effects of the contaminants \citep{gne04}. The most promising approach
is therefore to difference the signal by frequency, for individual patches
of the sky: this takes advantage of the smooth spectra of the
contaminating sources, compared with the relatively large frequency
variations of the 21cm signal arising from the spatial fluctuations of the
IGM separated in redshift \citep{mad97,gne04}. \citet{oh03} point out that
the changing beamwidth with frequency makes the elimination of the slowly
varying (with frequency) Galactic foreground not so straightforward, but
\citet{gne04} argue that angular correlations will facilitate the removal
of the contamination, and conclude that the strategy should be successful
at extracting an EoR signal for moderate angular resolution observations
($10-20$ arcminutes). They also argue that if Galactic synchrotron and
thermal emission are structured more strongly than contaminating
extragalactic point sources on small angular scales, they could be
difficult to remove. Currently the angular structure of these Galactic
contaminants is unknown.

In order to isolate and spectrally characterise these discrete sources, so
that they may be removed from the EoR map, long baselines (tens to
hundreds of kilometres) are crucial; these will be provided by the
international extension to LOFAR. In addition, careful thought must be put
into planning the observing strategy. Most of the emphasis has been on
detecting the statistical fluctuations
\citep{toz00,mor04,zal04,bha05,bow07}, but LOFAR is well-suited to imaging
the growing and overlapping HII regions as well
\citep{mad97,toz00,zar05}. Techniques for imaging the HII regions deserve
further development.

\subsection{Reionisation and galaxy formation}
\label{galform}

Probing the EoR using the redshifted 21-cm line will open a new window on
the very early Universe, currently out of reach of any other
observatory. While measuring the redshift and duration of reionisation
will provide important global constraints on the sources responsible for
reionising the Universe, obtaining detailed constraints on the physical
nature of these sources (e.g., their luminosities and halo masses), and
connecting them to the process of structure formation, will require
analysing 3D maps of the spatial and frequency structure of the 21-cm
emission. Both the construction and analysis of such maps are very
challenging tasks.  Several recent papers discuss strategies for
determining the 21-cm power-spectrum from the data, for example,
\citet{zal04}, \citet{mor05} and \citet{fur06}.

Interpreting the 21-cm data requires modelling of the sources of
ionisation, and investigating how rival models can be
distinguished. Necessary ingredients for the simulations are large-scale
structure formation, star formation and feedback, formation and accretion
onto black holes, emission and heating by X-rays, and radiative transfer
of the Lyman-$\alpha$ and ionising radiation to compute the 21-cm signal.

\begin{figure*}
\begin{tabular}{cc}
\psfig{file=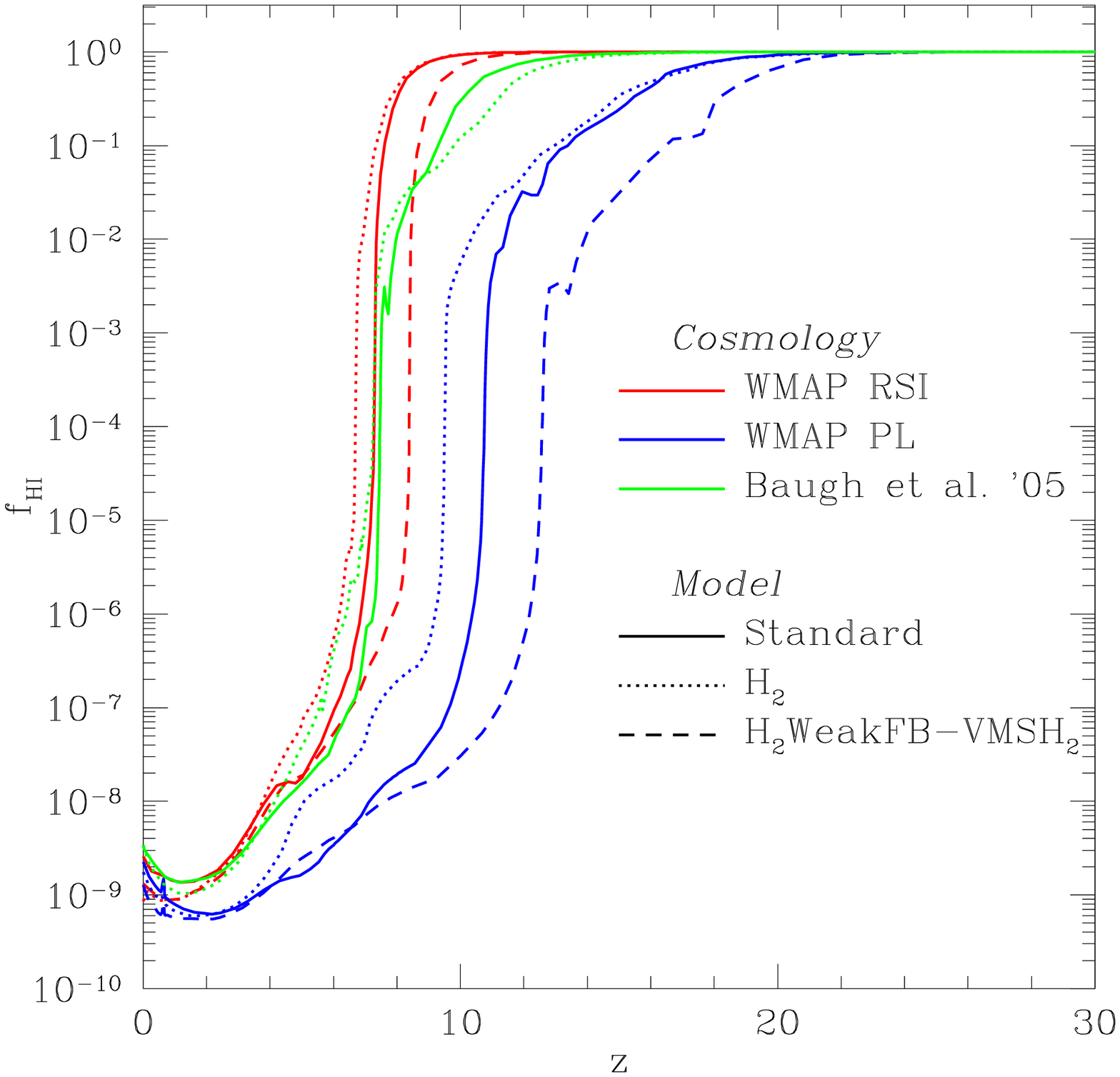,width=80mm} & \psfig{file=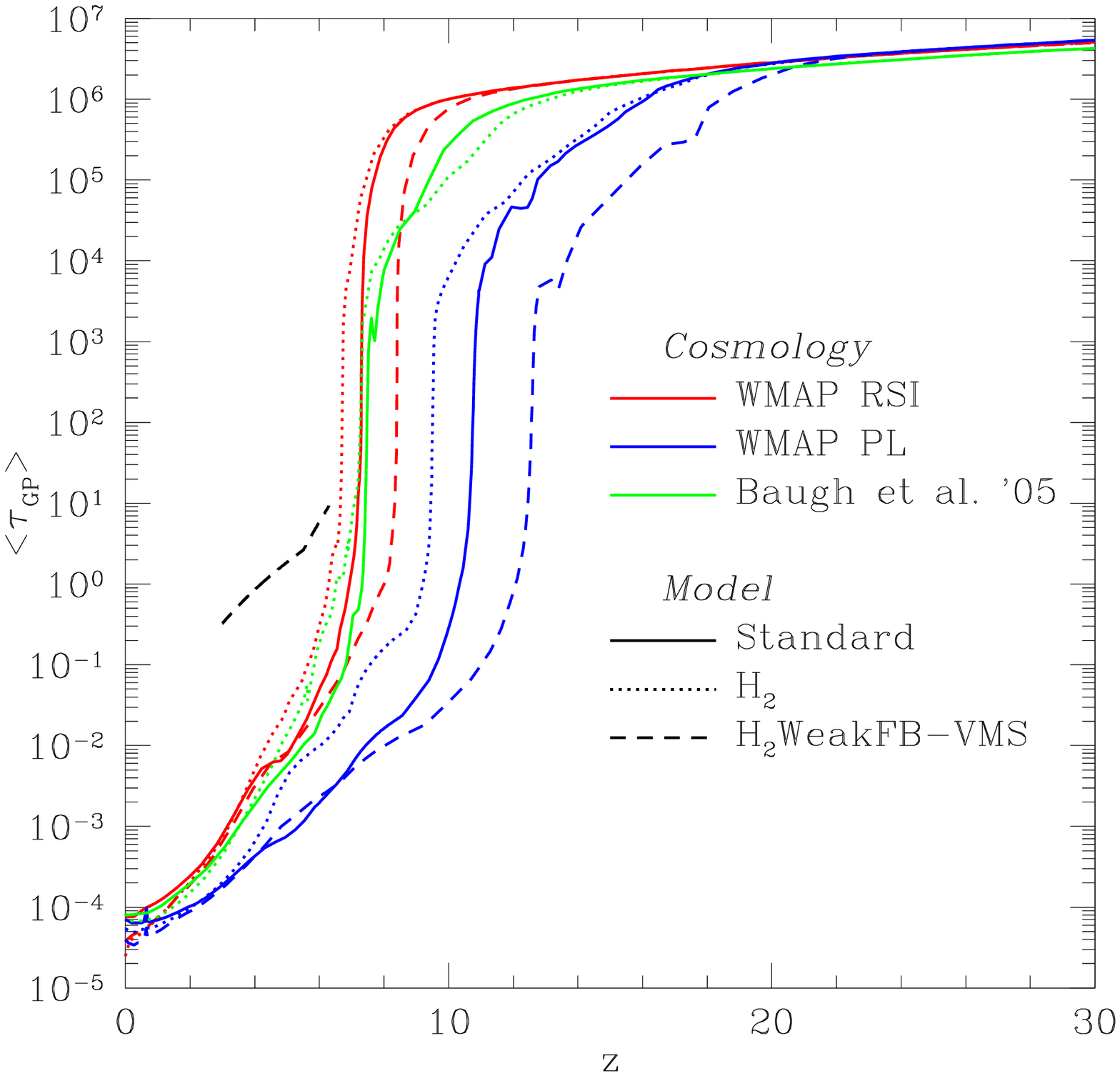,width=80mm}
\end{tabular}
\caption{\small The mean mass-weighted \HI\ fraction (left-hand panel) and
Gunn-Peterson optical depth (right-hand panel) as functions of redshift
for various models in three cosmologies \citep[see key in figure for
colour and line type coding; from][]{ben06}. The dashed black line in the
right-hand panel indicates the fit to the mean optical depth reported by
\citet{fan06}.}
\label{fig:tauGP}
\end{figure*}

The 21-cm data should be able to constrain many properties of
high-redshift galaxy formation that are currently poorly-known, such as
the ubiquity and initial mass function (IMF) of zero-metallicity
population~III stars, the presence (or not) of a substantial population of
very early X-ray sources, and the escape fraction of ionising photons from
small galaxies. An example calculation is illustrated in
Figure~\ref{fig:tauGP}, which shows predictions for reionisation from a
physically-motivated galaxy formation model. The different curves show how
the results depend on different cosmological parameters and on different
assumptions about gas cooling and about the IMF of the first stars.

LOFAR\ can potentially constrain many of the properties of the high-$z$
galaxies responsible for reionisation. However, it is very important to
understand the limits of the models. This can only be done if a large
enough range of models is investigated, by a set of relatively independent
groups. Current model predictions from different groups vary relatively
widely, even for basic observables such as the EoR, and much more so when
realistic physics such as radiative transfer is included. In order to
obtain constraints on structure formation from the LOFAR power-spectrum it
will be essential to compare the data with a suite of models. An analagous
situation arose with the 2dF Galaxy Redshift Survey: although little
modelling was required to {\em measure} the 2dF galaxy-power-spectrum,
mock galaxy catalogues were essential to allow cosmological constraints to
be obtained from the measurements.

Finally, it is worth noting that with integration times of several
beam-years, it may in principle be possible for LOFAR to measure the
21cm emission power spectrum with sufficient accuracy to detect Baryon
Acoustic Oscillations \citep{mao07}. Such a detection would prove the
cosmological nature of the reionisation signal, by matching the
observed BAO angular scale with the reionisation brightness
temperature step. The BAO signal could also be used to constrain
cosmological models, particularly those that predict an early onset
for dark energy.

\subsection{Reionisation and the UK}
\label{reionuk}

The cosmology community in the UK is well-placed to make a substantial
contribution to interpreting and exploiting LOFAR\ data on the epoch of
reionisation. For extracting the full cosmological information from the
21-cm maps of the reionisation epoch, it will be essential to have
high-resolution numerical simulations which include all of the relevant
physics of galaxy formation, radiative transfer and gas heating and
cooling. Although they will not be required to determine the exact time
the EoR took place, simulations and analytical models will be required to
allow the full scientific impact of all the data provided by
LOFAR. Moreover several key physical processes that govern the EoR are
still insufficiently understood, with competing groups reaching different
conclusions, demonstrating the need for further thorough theoretical
investigations.

Several UK universities have a proven track-record in performing such
simulations and theoretical modelling. For example, the Virgo consortium,
which is a collaboration between several groups in the UK with strong
links to the Netherlands (Leiden Observatory) and Germany (MPA in
Garching) is a world-leader in developing simulation codes, running them
on very large computers, and comparing the results in detail to data
extending over the whole electromagnetic spectrum.
\clearpage

\section{Deep Extragalactic Surveys with LOFAR-UK}
\label{surveys}

``Deep Extragalactic Surveys with LOFAR'' is the key project which has
aroused the most widespread interest amongst UK researchers.  A broad
range of scientific topics can be addressed by deep and sensitive surveys
of the low frequency radio sky. Many of these have already been well
described by the Dutch science case for LOFAR surveys, including:

\begin{itemize}
\item Detecting the most distant radio galaxies in the Universe, and using
  these to study the most massive galaxies at early epochs, and early
  cluster formation.
\item Studying intracluster magnetic fields, using diffuse radio emission
  in galaxy clusters.
\item Probing galaxy evolution by studying radio--selected star
  forming galaxies across a wide range of cosmic epoch.
\item Investigating the large-scale structure of the Universe through
  radio source clustering.
\item Constraining the physics of radio sources, and their evolution.
\item Delineating the spatial distribution of the interstellar medium
  (ISM) in nearby galaxies.
\item Studying galactic sources, such as supernova remnants, \HII\
  regions, exo-planets and pulsars.
\item Exploring new parameter space for serendipitous discovery.
\end{itemize}

The current document does not attempt to repeat these discussions, but
rather focusses upon the ways in which these can be enhanced through UK
involvement in LOFAR, due to (i) the improved angular resolution of the
array when UK (international) baselines are added, providing new science
opportunities; and (ii) the addition of complementary datasets, observing
capabilities, expertise and manpower that the UK community will
provide. In addition, there are a large number of additional science goals
of interest to the UK community which were not discussed in the Dutch
science case.

UK researchers have already formed part of the LOFAR Surveys Working
Group, helping to design the format of the surveys that LOFAR will carry
out. We envisage a very productive and collaborative working relationship
between LOFAR-UK and the Dutch Surveys team.

\subsection{Complementary observations of LOFAR deep survey regions}

In the current LOFAR Sky Survey plan, it is proposed to survey the entire
$2\pi$ steradians of the northern sky at frequencies of 30, 60, 120 and
200\,MHz (plus 15\,MHz if the sensitivity of the instrument is still
reasonable at that frequency). In addition, there will be a deeper survey
at 120 and 200\,MHz survey, expected to be over approximately 250 square
degrees (hereafter referred to as `LOFAR-deep'), reaching the confusion
limit of $\approx$6\,$\mu$Jy rms at 200\,MHz. An even deeper survey at
200\,MHz over a few square degrees, pushing into the confusion limit, is
also under consideration.  Given that most of the faint and distant radio
population appear to have steep radio spectral indices \citep[$\alpha \sim
1$, where $S_{\nu} \propto \nu^{-\alpha}$; e.g.,][]{mar06}, the 200\,MHz
`LOFAR-deep' survey will have a sensitivity equivalent to $1 \sigma \sim 1
\mu$Jy at 1.4 GHz.  Such depths have not yet been approached, requiring,
in principle, a year or so of exposure with telescopes like the VLA or the
GMRT, and then only over $\lta 1$\,deg$^{2}$.  The survey speed of LOFAR
will outstrip even the EVLA by at least a factor of $\sim 50$, and will
not be bettered until the SKA is operational.

The vast improvement that LOFAR will offer over previous radio surveys
will undoubtedly result in a large number of significant advances in our
understanding of the Universe. In modern astronomy, however, it is often
the combination of datasets from different international facilities which
results in the most significant breakthroughs. We discuss here a number of
different complementary surveys of the LOFAR survey regions which the UK
is currently involved in preparing or undertaking; combining these with
the LOFAR survey data will greatly increase the scientific potential of
the LOFAR surveys.

\subsubsection{Optical and near-IR surveys of the LOFAR survey regions}

\begin{figure}[!b]
\begin{tabular}{cc}
\psfig{file=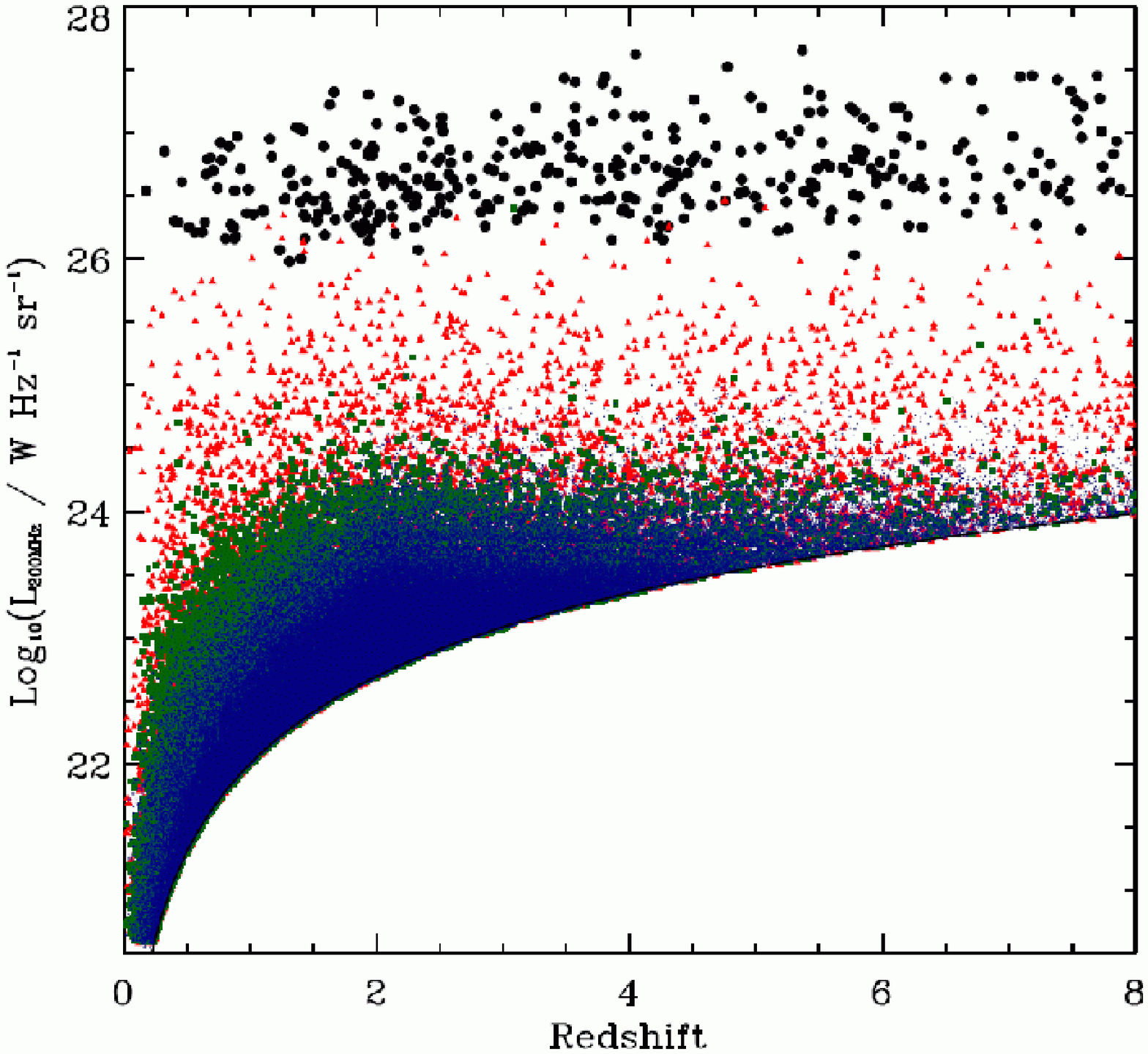,width=7.4cm,clip=}
&
\psfig{file=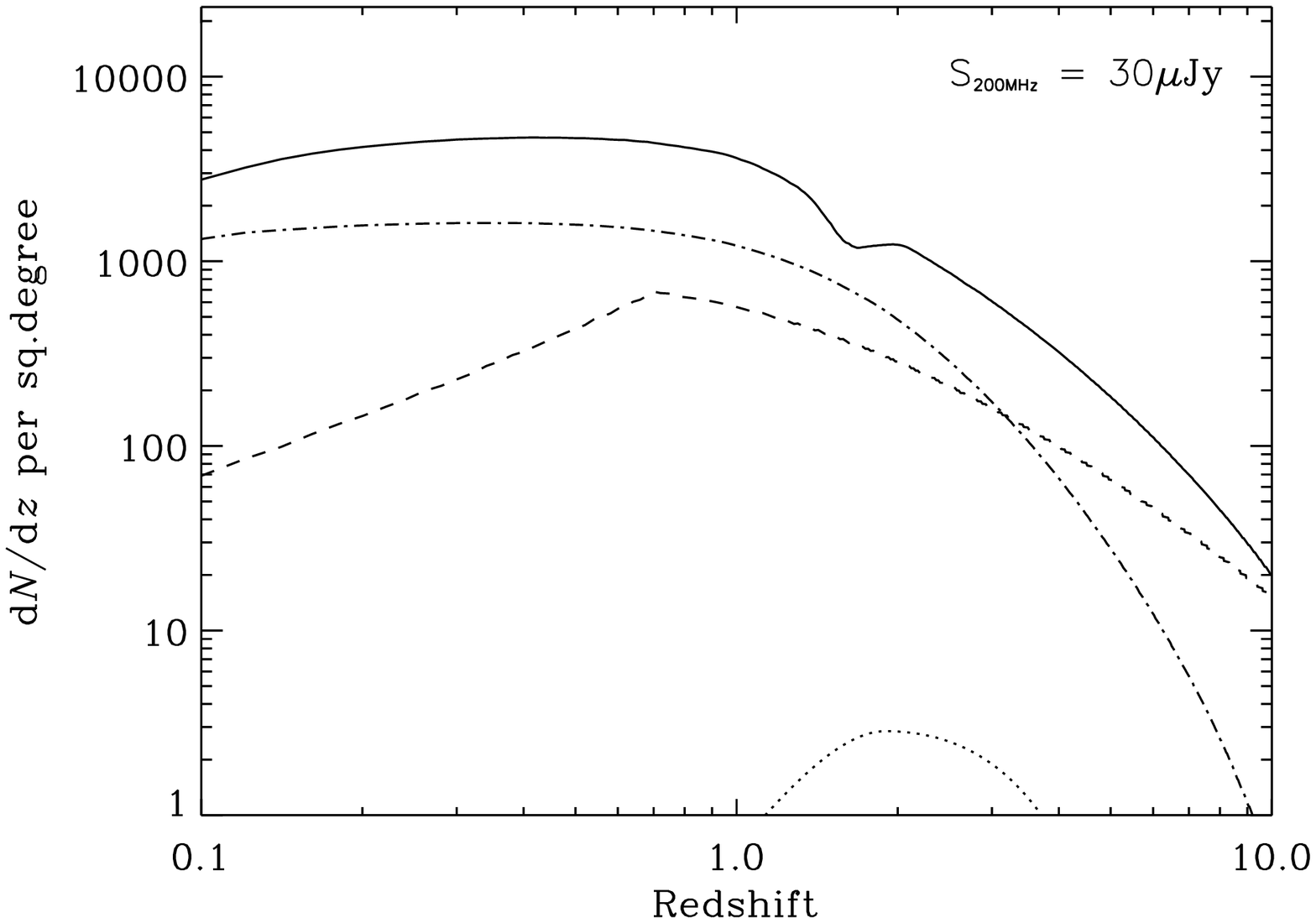,width=8.6cm,clip=}
\end{tabular}
\caption{\label{fig:pz} \small {\it Left:} A simulated $L_{\rm 200MHz}-z$
(radio luminosity versus redshift) plane for 1 deg$^{2}$ of the
`LOFAR-deep' survey, to an approximate flux density limit of $S_{\rm
200MHz} = 30\mu$Jy. The model splits the LOFAR population into four
sub-populations \citep{jar04}: starbursts (blue); radio-quiet quasars
(RQQs; green), FR\,I radio galaxies (red); and FR\,II radio galaxies
(black; in this case calculated for $100 ~\rm deg^{2}$).  This survey will
provide volume-limited samples of AGN (RQQ, FR\,I and FR\,II) out to very
high redshift, and also detect most of the luminosity density in
starbursts out to $z \sim 2$, and the most extreme starbursts to much
higher redshift.  {\it Right:} The number of sources per square degree in
the deep LOFAR survey, split into the various sub-populations. The solid
line represents the star-forming galaxies, the dot-dashed line represents
the radio-quiet quasars, the dashed line shows the FR\,I radio galaxies
and the dotted line denotes the FR\,II radio galaxies. One can see that
such a survey is dominated by the star-forming galaxies. [NB. The slight
kink in the star-forming galaxies curve is an artefact of the
two-population modelling of the evolving star-forming population.]}
\end{figure}

At the 200\,MHz confusion limit, the LOFAR-deep survey will have the
sensitivity to probe volume-limited samples of AGN, as well as large
populations of star-forming galaxies to high redshift (e.g.,
Figure~\ref{fig:pz}; see Section~\ref{sfsec} for details). The addition of
UK and international LOFAR stations will also enable LOFAR-deep to provide
sub-arcsecond angular resolution. It will detect tens of millions of
objects in huge volumes of the high-redshift Universe but, crucially, only
in 2D projection. We have learnt from studies of the local Universe that
the shift from 2D to 3D tends to transform hints on key issues in galaxy
formation and cosmology \citep[e.g.,][]{efs90} to firm measurements,
e.g.,\ how local galaxies trace the dark matter and hence $\Omega_{\rm M}$
\citep[e.g.,][]{per01}, and believable constraints on exotica like the
mass of neutrinos \citep{elg02} and dark energy \citep{per07}. A critical
complement to the LOFAR surveys is thus the addition of deep optical
and/or near-infrared observations.

\begin{figure}[!t]
\begin{tabular}{cc}
\psfig{file=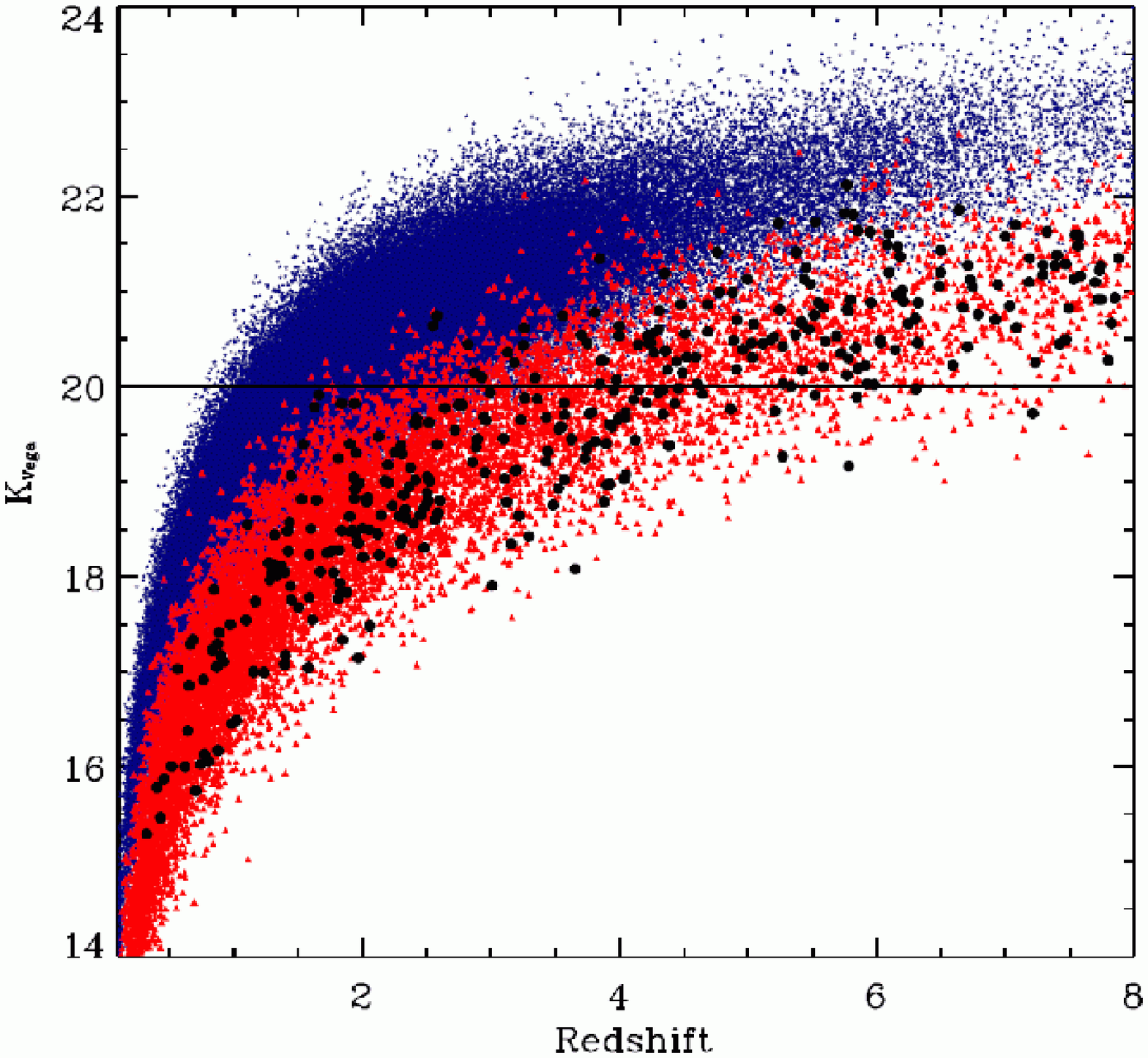,width=7.9cm,clip=}
&
\psfig{file=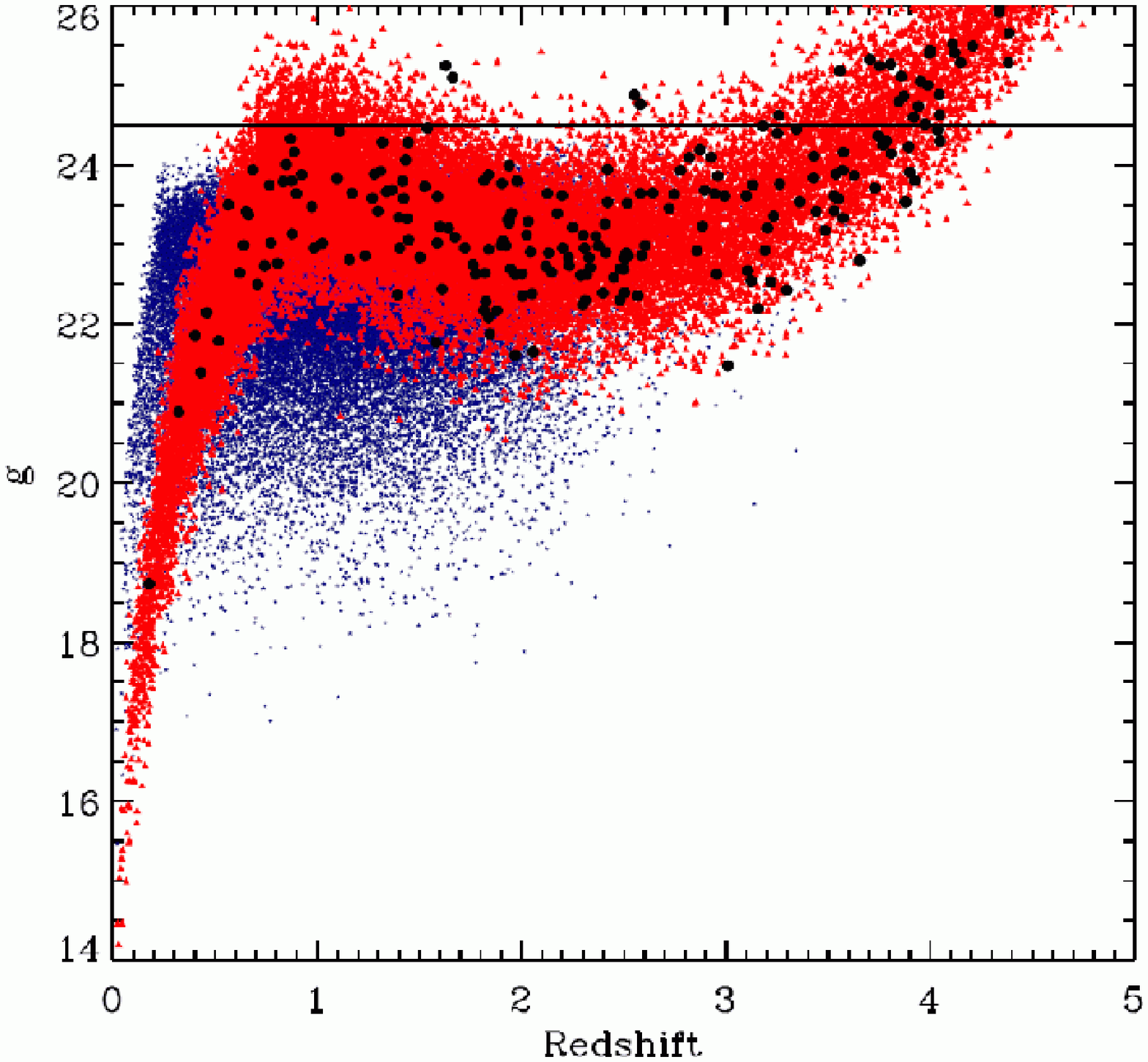,width=7.9cm,clip=}
\end{tabular}
\caption{\label{fig:kz} \small {\it Left:} A simulated $K-z$ (near-IR
Hubble) diagram for 1 deg$^{2}$ (100 deg$^{2}$ for FR\,IIs) of the
`LOFAR-deep' survey.  The LOFAR sources are separated into three
sub-populations \citep[see][]{jar04}: starbursts (blue); FR\,I radio
galaxies (red); and FR\,II radio galaxies (black). LOFAR will also detect
radio--quiet quasars (RQQs), which for clarity are not plotted [Type-2
RQQs will follow the locus of the FR\,Is; Type-1 RQQs will be brighter and
be detected as sources with large point-source contributions].  The
horizontal line shows the practical limit for WFCAM imaging over a
$\sim$250 square degree area; to this limit essentially all FR\,Is (and
RQQs) would have detections to $z \sim 2.5$.  Any FR\,IIs (a fact
determinable from radio data alone) with no K--band counterpart to $K
\approx 20$ will have large redshift. {\it Right:} A simulated $g-z$
diagram for these same sources. The horizontal line shows the approximate
limit of both the Pan-STARRS 3$\pi$ survey observations.  In observations
to these limits essentially all FR\,Is (and RQQs) would be detected at
both $K$ and $g$ (and $i$), yielding, via a slight variant of the $BZK$
method \citep{dad04}, accurate photometric redshifts below $z \approx 1.2$
and a clear separation between `red and dead' FR\,Is and blue starbursts
in the $1.2 < z < 3$ regime. The hard upper edge to the blue points is an
artefact of the simplistic modelling of the relationship between radio and
(rest-frame) UV flux.}
\end{figure}

The Wide Field Camera (WFCAM) on the United Kingdom's InfraRed Telescope
(UKIRT) is an extremely powerful instrument for wide--field near-infrared
studies. With four 2048x2048 Rockwell devices, it offers an exposed solid
angle of 0.21 square degrees in a single shot, leading to easily the
fastest survey rate of any IR instrument in the northern hemisphere.
Commissioned in 2004, WFCAM is being used to carry out a number of
different sky surveys, under the collective name of the UKIRT Infrared
Deep Sky Survey (UKIDSS). These include two Galactic surveys (the Galactic
Plane Survey, and the Galactic Clusters Survey), and three extragalactic
surveys of differing areas and depth: the Large Area Survey (LAS; 4000
deg$^2$ to $J=20.0$, $H=18.8$, $K=18.4$), the Deep Extragalactic Survey
(DXS; 35 deg$^2$ to $J=22.5$, $K=21.0$) and the Ultra Deep Survey (UDS;
0.77 deg$^2$ to $J=25$, $H=24$, $K=23$). These surveys will be incredibly
powerful when combined with the LOFAR data, particularly the northern DXS
fields which offer a natural starting point for the 250 square degree deep
LOFAR surveys, and the UDS which would be an obvious target for an
extremely deep LOFAR pointing (as would the UltraVISTA COSMOS field,
although, like the UDS, this is equatorially located). In addition to
these surveys, general observer time with WFCAM is available for smaller
programmes, whilst beyond 2009 there will be a new opportunity to propose
for additional large surveys. An ambition of the LOFAR-UK team is to use
WFCAM to survey the full 250 square degrees of the LOFAR-deep survey
regions down to $K=20$. This is sufficiently deep to detect essentially
all of the LOFAR AGN population at $z \lta 2.5$ (see
Figure~\ref{fig:kz}a).

In the optical bands, data of great value to LOFAR will be taken by the
Pan-STARRS project (Panoramic Survey Telescope \& Rapid Response System;
{\it pan-starrs.ifa.hawaii.edu/public/}). Pan-STARRS is a multi-colour
optical survey to be carried out between 2008 and 2011 on a dedicated 1.8m
telescope in Hawaii. The Pan-STARRS project will survey the entire 3$\pi$
steradians of sky visible from Hawaii in the $g$, $r$, $i$, $z$ and $y$
bands, to a 5$\sigma$ depth of $g \approx 24.6$ (1--2 magnitudes deeper
than the Sloan Digital Sky Survey, which does not reach sufficient depth
to get redshifts, either spectroscopically or photometrically, for the
majority of the faint radio sources). The Pan-STARRS data will
provide optical identifications and colour information for the majority of
AGN detected by the whole suite of LOFAR surveys (e.g., see
Figure~\ref{fig:kz}b), and photometric redshifts for all those at $z \lta
1$. In addition, 10 fields of approximately 7 square degrees each will be
visited many times to produce a Medium Deep Survey, with limiting depths
of $g \approx 27.3$, $r \approx 26.9$, $i \approx 27.9$, $z \approx 26.3$,
$y \approx 24.8$. These 10 fields have been chosen to have the best
possible complementary data, particularly in the near--IR (e.g.,
UKIDSS-DXS and VISTA:VIDEO regions), and it is likely that many of these
regions will be selected for the LOFAR deep survey fields.  The Medium
Deep Survey data will identify essentially all of the AGN and starburst
galaxies detected by LOFAR, and provide accurate multi-band photometric
redshifts for these out to at least $z \sim 2$.

The Pan-STARRS survey is a private survey, being carried out in
collaboration between the University of Hawaii, Harvard University,
Johns Hopkins University, Las Cumbres Observatory, the Max Planck
Institutes in Heidelberg and Munich, and a UK University consortium
consisting of Durham, Edinburgh and Queen's Belfast. Edinburgh and
Durham are both members of the LOFAR-UK consortium, and so researchers
at these universities will have access to the Pan-STARRS data. In
addition, it has been agreed that the Pan-STARRS data will be made
publically available approximately 1 year after the end of the survey
(i.e., in about 2012), which is a timescale comparable to that on
which LOFAR will be producing significant quantities of sky survey
data.

Deep optical and near-IR data are essential for full optimisation of the
LOFAR-deep survey because:

\begin{itemize}
\item We need to locate objects in 3D, requiring some sort of photometric
redshifts. For LOFAR AGN, the well-known tightness of the $K-z$ relation
(Figure~\ref{fig:kz}a) -- i.e.\ the fact that radio jets emerge from only
the most massive galaxies \citep[e.g.,][]{bes98,mcl04} -- means that a $K$
data point alone is sufficient to provide a photometric redshift; for
LOFAR starbursts, $K$ plus optical will largely suffice.

\item With $\sim$arcsecond beam sizes and complicated radio source
structures \citep[e.g.,][]{sim06}, reliable `follow-up' spectroscopy of
LOFAR sources with fibre instruments like FMOS would be inefficient
without optical or near-IR identifications.

\item A fundamental property of any astronomical object is its mass.  The
$K$-band traces old stellar populations so that $K$-band selection broadly
maps onto a stellar-mass selection criterion. On the scales of galaxies,
this is a good proxy for halo (dark-matter) mass selection.

\item Optical and near-IR data allow a characterisation of the
environments of the radio sources \citep[e.g.,][]{sta05,vbr06,str06}.
Near--IR data reaching $K \sim 20$ would allow sub-$L_*$ luminosities to
be probed at $z < 2$, and thus offer a sensitivity which reaches well
below those of the richest clusters detected by X-ray and Sunyaev
Zel'dovich (SZ) surveys.
\end{itemize}

\subsubsection{Westerbork and GMRT surveys of `LOFAR-deep'}

LOFAR-UK and LOFAR-NL have begun jointly to survey 18 square degrees of
the LOFAR deep field region using the Westerbork telescope at 1.4\,GHz, to
a roughly uniform rms sensitivity of $11 \mu$Jy. This will provide $\gta 1
\sigma$ detections of the majority of $S_{\rm 200 MHz} \gta 30 \mu$Jy
source in the deep LOFAR surveys. The LOFAR-UK consortium has agreed to
manage the effort of calibrating and reducing these data, on the
understanding that reduced data products will be made available to the
whole LOFAR-NL + LOFAR-UK survey team prior to any science exploitation.

Although this WSRT survey covers only a subset of the region that will be
surveyed by the LOFAR deep survey, the sky area is large enough to sample
the full range of environments (from richest clusters to voids) over a
wide range of cosmic epochs. The survey covers the northern Spitzer
Wide-area InfraRed Extragalactic (SWIRE) survey regions ($\approx
9$deg$^{2}$ in the Lockman Hole, $\approx 9$deg$^{2}$ in Elais-N1), which
are the only large sky patches in the northern sky with comprehensive
multi-waveband datasets, and will therefore be likely targets of the first
wide and deep LOFAR surveys. As well as the mid-IR SWIRE data (at 3.6,
4.5, 5.8, 8.0, 24.0, 70 \& 160$\mu$m), deep GMRT radio data ($\sim$60$
\mu$Jy rms sensitivity at 610 MHz) have been taken by members of the
LOFAR-UK consortium, whilst other datasets are currently being built up,
or planned, and will become available for exploitation whilst the WSRT
survey is underway (e.g., the UKIDSS-DXS survey in the near-IR, discussed
above).

The WSRT survey will take place over the period 2007-2009, and has a
number of different goals:

\begin{itemize}
\item In combination with the existing 610\,MHz GMRT data, the WSRT
data will allow the radio spectra of the brightest $\sim 10^{5}$ radio
sources in this sky region to be determined, prior to the onset of the
LOFAR surveys. These will provide an accurate sky model for
calibration of the LOFAR array, greatly aiding early observations.

\item In combination with the LOFAR data, the WSRT data will determine
the radio spectra of the objects in the deep LOFAR surveys (both
spectral indices and curvature), allowing separation of flat- and
steep-spectrum objects in the deep LOFAR surveys as soon as they
become available. This is an essential first step in the
characterisation of the LOFAR population and will allow, with full
photometric redshift information, the first robust calculation of the
radio luminosity function \citep[e.g.,][]{dun90} and hence the cosmic
heating effect of radio sources \citep[e.g.,][]{raw03}. This will also
allow the isolation of rare but interesting sub-populations of objects
(e.g., most distant AGN, cluster haloes etc) from the LOFAR data.

\item The WSRT observations will provide a dataset deep enough to allow
`stacked' analyses of sub-populations of distant AGN and starbursts,
selected in non-radio wavebands, before any LOFAR data have been
taken. This will allow the generation of a full simulated LOFAR sky,
deviations from which in the real data will pinpoint exotic objects
quickly.

\item The WSRT survey will provide an early target list for follow-up
observations, for example, for spectroscopic surveys with FMOS.
\end{itemize}

Whilst the Westerbork survey will accomplish these important goals, it
still only covers a small proportion of the LOFAR deep survey
area. Therefore, in collaboration with Dutch researchers, the LOFAR-UK
consortium has submitted (and had the first 120 hours of time allocated
for) a GMRT proposal (PI: Green) to survey the entire 250 deg$^2$ at
610\,MHz to an rms sensitivity of 70$\mu$Jy. This will combine and extend
the various distinct GMRT surveys that UK and Dutch researchers have so
far carried out (totalling about 30 deg$^2$ in different well-studied
regions). As with the surveys discussed above, the LOFAR-UK Universities
will make fully reduced data available to the entire LOFAR consortium.

This GMRT survey will provide radio spectral index information for all of
the brighter sources within the LOFAR deep survey region.  Although it is
impossible with GMRT to reach the same effective sensitivity level as the
200\,MHz LOFAR data over such a large sky area, the GMRT survey does reach
the sensitivity level required for optimal comparability with the whole
suite of LOFAR surveys. LOFAR surveys at 30, 60, 120 and 200\,MHz will be
available over the entire 250 square degrees, and the least sensitive of
these will be the 60\,MHz survey.  This has a (1$\sigma$) rms sensitivity
of $\approx 0.33$ mJy beam$^{-1}$ which, for a typical non-thermal
spectral index of 0.7, corresponds to 70$\mu$Jy beam$^{-1}$ at
610\,MHz. Thus, for typical radio sources detected at each of 30, 60, 120
and 200\,MHz, the GMRT data will add a valuable flux density measurement
at 610\,MHz.

\subsection{Starforming galaxies in the deepest radio surveys}
\label{sfsec}

For many years, radio surveys have been known for their ability to find
the rarest most powerful active galactic nuclei (AGN) out to very high
redshifts, and lower power examples in the nearby Universe. However, the
next generation of radio surveys conducted with the new radio telescopes
will break into new parameter space for extragalactic surveys. Due to the
massive increase in sensitivity of LOFAR over previous telescopes, the
most numerous extragalactic sources will no longer be the AGN, but
starburst galaxies. A 200\,MHz LOFAR-deep survey to a 30$\mu$Jy flux
density limit will detect galaxies with star formation rates of $10
M_{\odot}$ per year out to redshift $z \approx 2$, whilst objects with
more extreme star formation rates of $100 M_{\odot}$ will be detectable
out to $z \sim 6$.

Using the radio-luminosity functions derived from previous
low-frequency surveys for the more powerful radio source populations,
such as the Fanaroff \& Riley Class I and II radio galaxies
\citep[FR\,Is and FR\,IIs;][]{fan74}, along with a prescription for
the radio luminosity of radio-quiet quasars derived from the X-ray
luminosity function of \citet{ued03}, we are able to estimate the
contribution of the AGN to any LOFAR survey. This follows the work
described in \citet{jar04} for the Square Kilometre Array
(SKA). Furthermore, we are also able to estimate the contribution to
the total source population from star-forming galaxies using the
luminosity function from \citet{yun01} along with an assumed evolution
which ensures that the source counts at both mid- and far-infrared
wavelengths are not exceeded \citep[see][]{bla99}. Assuming that a
deep LOFAR survey could reach an rms sensitivity of
$\approx$6\,$\mu$Jy at 200\,MHz, this would imply a total source
density of around 20,000 per square degree. As shown in
Figure~\ref{fig:pz}b, the vast majority of these sources (about 80 per
cent) will be star-forming galaxies.

\begin{figure}[!t]
\begin{tabular}{cc}
\raisebox{0.4cm}{\psfig{file=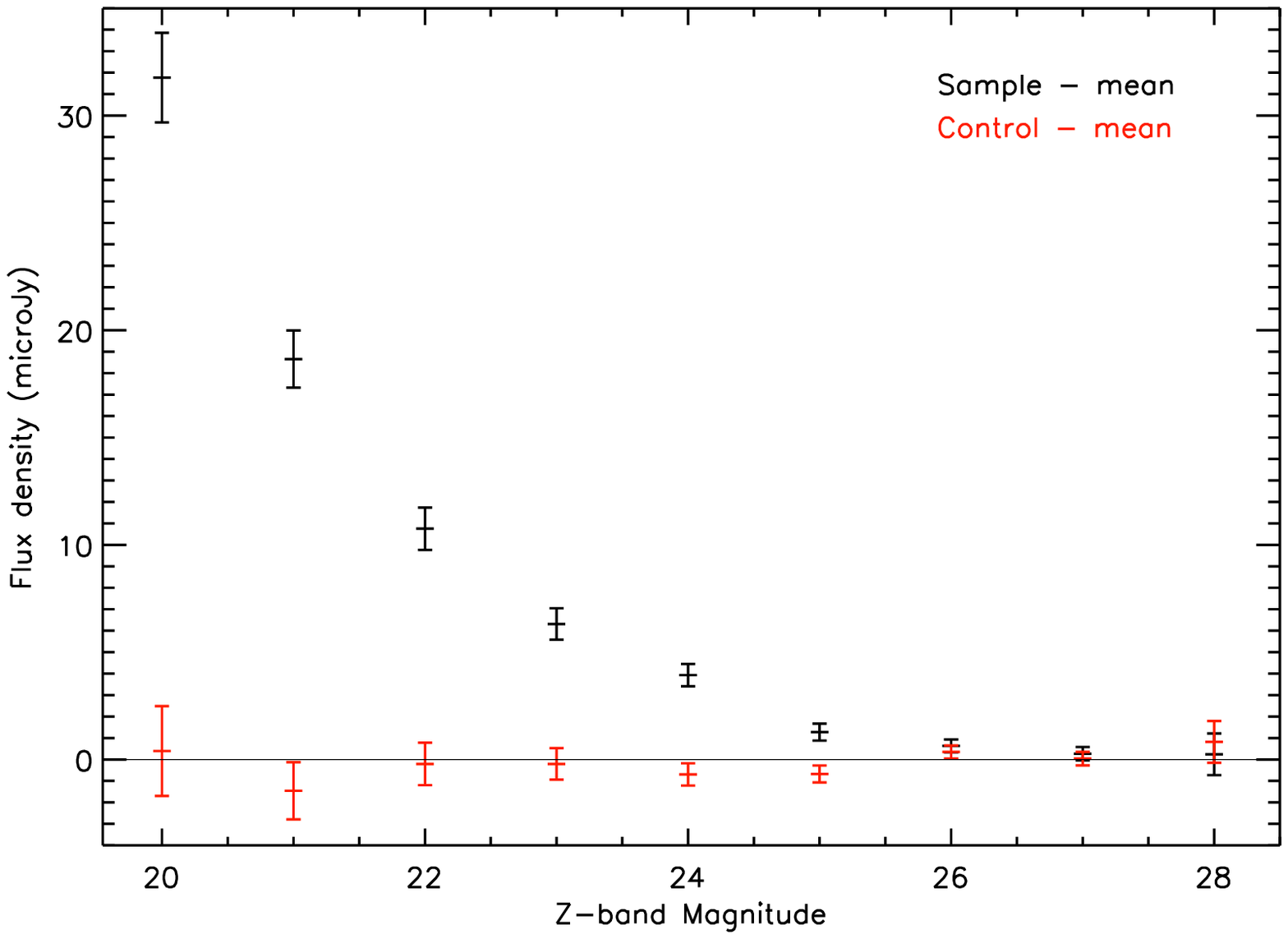,width=8.8cm,clip=}}
&
\psfig{file=hdfstack_23.ps,width=6.2cm,clip=}
\end{tabular}
\caption{\label{hdfstackfig} \small {\it Left:} Statistical detection of
the $\mu$Jy radio emission from $>$8000 optically selected galaxies within
an 8.5$\times$8.5 square arcminute field centred on the HDF-N (excluding
all radio emission brighter than 20$\mu$Jy \,beam$^{-1}$). Radio emission is
statistically detected from galaxies brighter than magnitude 25. The red
points represent a control sample of randomly selected positions away from
any catalogued source. {\it Right:} The average radio emission from 957
galaxies in the HDF-N with magnitudes between 22.5 -- 23.5. The contours
are $\sqrt2$ times 340\,nJy\,beam$^{-1}$.}
\end{figure}

An alternative method of estimating the source density and population mix
expected for deep LOFAR surveys is to consider the results of existing
deep radio surveys.  \citet{mux05} analysed deep 1.4\,GHz MERLIN and VLA
observations of an 8.5$\times$8.5\,arcmin~area centred upon the Hubble
Deep Field North, and detected 92 radio sources to a detection threshold
of 40\,$\mu$Jy. Around 70\% of radio sources fainter than about 100$\mu$Jy
are associated with star-forming or composite AGN-starburst galaxies.  In
deep {\it HST} ACS {\it z}-band images of this same area, $\sim$13,000
galaxies are detected. Whilst the vast majority of these galaxies are not
individually detected at radio wavelengths, it is possible to
statistically detect these faint optical systems in the deep radio
data. The left panel of Figure~\ref{hdfstackfig} shows the binned radio
flux of $\sim$8000 of these sources (excluding radio sources $>$20$\mu$Jy
and all potentially confused sources) versus their optical
brightness. Optical sources as faint as 25th magnitude are statistically
detected at the level of a few microJansky; this is illustrated by a
`stacked' average image of the radio emission from $\sim$1000, 23rd
magnitude galaxies, shown in the right panel of Figure~\ref{hdfstackfig}.
These results imply that approximately half of the $\sim$2700 galaxies
brighter than 24th magnitude will have radio flux densities of greater
than 4\,$\mu$Jy at 1.4\,GHz within this 8.5$\times$8.5 arcmin
field. Assuming that the radio emission arising from this faint radio
source population is related to star-formation with a radio spectral index
$\sim$0.7, the LOFAR-deep survey with a detection limit of 30$\mu$Jy at
$\sim$200\,MHz will detect a couple of tens of thousands of `normal'
star-forming galaxies per square degree.

Obviously, with these source densities, confusion becomes a major issue.
This is particularly true with the current Netherlands-only LOFAR, where
the spatial resolution is $\sim 3-4$~arcsec, limiting the depth to which
any deep survey could reach. However, with UK and other international long
baselines, LOFAR would be able to probe to much deeper fluxes; a deep
LOFAR survey with international baselines, pushing into the confusion
limit, could conceivably probe star formation rates approaching a solar
mass per year at $z \sim 2$. This is a depth that will probably not be
feasible with other telescopes (radio or otherwise) over such large areas
of sky until the SKA is operational.

It is not only the flux density limit to which star-forming galaxies can
be detected that is improved by the addition of international baselines,
but also the angular resolution with which they are observed. Increasingly
it is being recognised that, to understand astrophysical phenomena, a
pan-chromatic view of the Universe is required. LOFAR will uniquely
contribute to this view by providing high-sensitivity observations of the
little-explored low frequency Universe. The radio flux density is closely
related to the star formation rate \citep[e.g.,][]{con92}, and will
provide one of the most reliable ways of measuring the star formation
rates of high redshift galaxies; LOFAR will thus provide an extremely
powerful additional piece of information in the multi-wavelength studies
of galaxy formation and evolution being undertaken using both current and
next-generation astronomical instruments (e.g., e-MERLIN, EVLA, eVLBI,
ALMA, NGST, VLTI, Spitzer, Herschel, Chandra, Swift, etc). In order to
achieve this, however, LOFAR observations must have comparable angular
resolutions to these other instruments (i.e., ideally $<1$arcsec) and also
must be able to be astrometrically aligned with sub-arcsecond
precision. Increasingly, as higher sensitivity observations are made at
all wavelengths and fields become more crowded with sources, it is
essential that observations are able to both separate and correctly
identify sources at different wavelengths.

\subsubsection{The star-formation history of the Universe}

As discussed above, a 200\,MHz LOFAR-deep survey to a 30$\mu$Jy flux
density limit will detect galaxies with star formation rates of $10
M_{\odot}$ per year out to redshift $z \approx 2$, which is the epoch 
at which the cosmic star formation history is believed to have peaked. 
Combining such radio data with deep optical and near-IR data (e.g. from 
UKIDSS DXS) will enable considerably more detailed study of star formation
at this key epoch.

In order to fully understand the basic features of galaxy formation
and evolution, one must understand the volume-averaged star formation
rate as a function of epoch, its distribution function within the
galaxy population, and its variation with environment. Surveys of the
star-formation rate as a function of epoch suggest that the
star-formation rate density rises as $\sim$\,(1+z)$^{4}$ out to at
least z\,$\sim$\,1 \citep[e.g.][]{lil96} then flattens around $z \sim
2$, although different star-formation indicators still give widely
different measures of the integrated star-formation rate density
\citep[see][]{sma02}. The problem is that currently used star
formation indicators either need large corrections for dust (e.g. the
UV flux), large extrapolations for faint sources below the sensitivity
limit (e.g. sub-mm sources), or are based on small-field surveys with
large uncertainties due to the effects of large-scale structure and
cosmic variance. Use of the radio flux density as a star formation
rate indicator in wide-field LOFAR surveys would overcome all of these
issues.

It is not only the global average star formation rate which is
important for our understanding of galaxy formation and evolution, but
more crucially the nature and distribution of the star forming
galaxies at high redshifts. Many recent results point towards the star
formation in the most massive galaxies occurring earlier than that in
lower mass galaxies -- so-called `down-sizing', or anti-hierarchical
growth \citep[e.g.][]{cow96}. Massive galaxies must therefore form
stars rapidly at an early epoch, and then have their star formation
truncated, for example by feedback from AGN
\citep[e.g.][]{bow06,cro06,bes06b}. Determining the characteristic
mass at which star formation begins to be truncated, as a function of
cosmic epoch, would determine the physical processes involved in the
downsizing activity and place tight constraints upon galaxy formation
models. At low redshift, the star formation rates of galaxies are also
greatly influenced by the environment in which they reside, with star
formation strongly suppressed in dense environments. To what extent is
it the build-up of galaxies into groups and clusters since $z \sim 1$
that drives the decline in the cosmic star formation rate?

These issues can only be addressed by examining how the relationships
between star-formation rate and both galaxy mass and environment
evolve with redshift. Deep LOFAR radio data would provide
star-formation rates for large samples of galaxies over the redshift
range $z = 0$--3, across the peak star formation epoch, for which deep
optical and near-IR observations (e.g. from UKIDSS, VISTA surveys,
Pan-STARRS) are providing photometric redshift and stellar mass
estimates, and from these also estimates of environment.  Such data
would provide a uniquely powerful tool for measuring the growing
influence of environment on star-forming galaxies, and determining how
the characteristic stellar mass of star forming galaxies changes with
both environment and epoch. Such observations would yield key tests of
current galaxy formation models \citep[e.g.][]{ben00b,bau05,bow06}.

An extremely deep LOFAR pointing in the UKIDSS UDS or UltraVISTA COSMOS
field would enable such studies to be extended back to the earliest cosmic
times, $z \gta 6$. The UDS will have near-IR data from WFCAM to $J=25$,
$H=24$, $K=23$, over 0.77 sq.\ deg., together with extremely deep Subaru
SuprimeCam data at optical wavelengths, and longer wavelength data from
Spitzer. These will permit accurate photometric redshifts and stellar mass
estimates out to $z > 4$ \citep[cf.][]{cir07}.  UltraVISTA will have even
deeper IR data from VISTA, to $Y=26.7$, $J=26.6$, $H=26.1$, $K=25.6$, over
0.75 sq.\ deg., together with existing deep multiwavelength data, further
extending the redshift range of study.  Although both of these fields are
at equatorial locations, the exquisite complementary data provides a
strong argument for carrying out exceptionally deep LOFAR 200MHz
observations, probing into the confusion limit. Not only will this allow
detection of galaxies with more typical star formation rates out to higher
redshift, but also through radio stacking analyses it will be possible to
determine mean star formation rates as a function of galaxy properties
(redshift, mass, environment, colour, etc) out to $z \sim 6$.

\subsubsection{Comparison with sub-millimetre  and mid-to-far infrared surveys}

The SCUBA-2 instrument for the JCMT is due to be commissioned in 2008,
and will be the first large-format ``CCD-like" camera for
sub-millimetre astronomy. It will allow large areas of sky to be
mapped simultaneously at 850 and 450$\mu$m, at speeds up to a thousand
times faster than the current SCUBA camera. A number of legacy surveys
will be carried out using SCUBA-2, of which the most relevant for
LOFAR studies of distant starbursts is the SCUBA-2 Cosmology Survey.

In the first two-year plan, the SCUBA-2 Cosmology Survey aims to map 20
square degrees of sky at 850$\mu$m down to the confusion limit at an rms
level of 0.7\,mJy/beam. This survey will be carried out in fields with
existing multi--wavelength data (XMM-LSS, Lockman Hole, Chandra Deep Field
South, ELAIS N1, COSMOS, and Bootes). The best weather conditions will be
used to map a smaller area to a much deeper depth at 450$\mu$m (where the
confusion limit is deeper due to the higher angular resolution): this will
reach 0.5\,mJy/beam at 450$\mu$m over 0.6 deg$^2$ (GOODS-N, GOODS-S, UDS,
and COSMOS).  Ultimately it is proposed to extend these surveys to 70 and
2 square degrees at 850 and 450$\mu$m respectively.

Deep LOFAR observations of the SCUBA-2 Cosmology Survey regions will allow
the host galaxies of essentially all of the sub-mm sources to be
identified.  Previous radio surveys to a limit of around 7$\mu$Jy/beam rms
at 1.4\,GHz have succeeded in identifying over two-thirds of the sub-mm
sources \citep[e.g.,][]{ivi07}. GMRT observations at 610\,MHz have
demonstrated that the radio spectra of sub-mm sources show no evidence of
turning over at lower radio frequencies \citep{iba07}, and so 200\,MHz
observations to an rms level of 6$\mu$Jy/beam will effectively reach a
factor of $\sim 5$ deeper than current 1.4\,GHz limits.

\begin{figure}[!b]
\begin{tabular}{cc}
\raisebox{3cm}{\parbox{9cm}{\Large{\bf See associated jpg file\\ 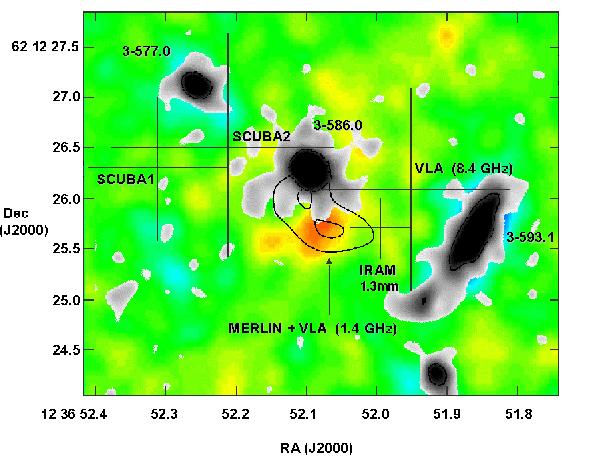}}}
&
\psfig{file=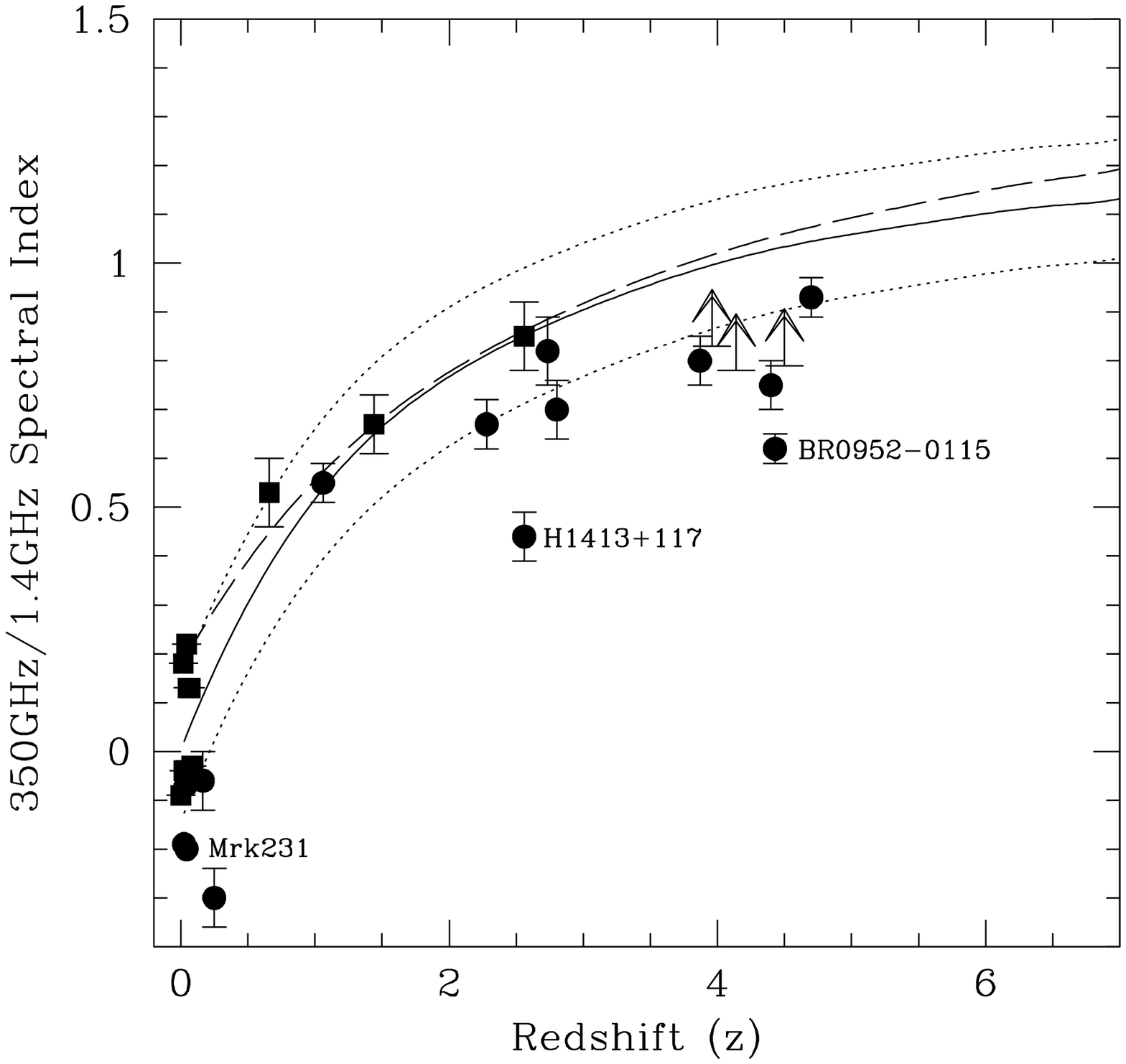,width=7cm,clip=}
\end{tabular}
\caption{\label{hdf850} \small {\it Left:} Positional information on
HDF850.1, overlaid on a $K-I$ colour image of the field
\citep[from][]{dun04}. The SCUBA1 and SCUBA2 crosses represent position
estimates from two different reductions of the SCUBA data. The positions
from VLA 8.4\,GHz and IRAM PdB 1.3\,mm detections are also shown. In all
cases the size of the cross indicates the 1$\sigma$ error in the relevant
position. The contours show a high resolution MERLIN+VLA 1.4\,GHz radio
map. Only with this sub-arcsec resolution was the host galaxy of HDF850.1
able to be associated with an optically faint extremely red galaxy. This
demonstrates the need for international baselines to increase the angular
resolution of LOFAR. {\it Right:} The modelled and observed radio to
sub-mm flux ratio as a function of redshift \citep[from][]{car00}. This
flux ratio allows a reliable redshift estimate out to $z \sim 3$.}
\end{figure}

Radio identifications of the sub-mm sources are of great importance
because they provide accurate astrometric positions of the sub-mm sources:
at 850$\mu$m the beam size of the JCMT is 14 arcsec, and so without
follow-up identifications it is impossible to reliably identify the host
galaxy of most sub-mm sources. The international baselines of LOFAR are
extremely beneficial in this respect, and in some case necessary, in order
to provide sufficiently high astrometric accuracy. For example, only by
using very accurate absolute positions from radio observations was it
possible to identify the brightest sub-mm source in the Hubble Deep Field
North \citep[HDF850.1;][]{mux05,dun04} with an optically faint, and
extremely red, $z \sim 4$ galaxy situated within $\sim$1 arcsec of several
brighter, nearby elliptical galaxies (Figure~\ref{hdf850}, left panel). In
addition, radio detections of the sub-mm sources provide an immediate
redshift estimate, since the radio to sub-mm flux ratio is a sensitive
indicator of redshift \citep[cf. Figure~\ref{hdf850}, right panel;
e.g.,][]{car99}.  This is because of the strong positive k-correction at
sub-mm wavelengths.

The radio data will be equally valuable in identifying sources
detected by the Spitzer, Herschel and Akari satellites. Spitzer is
currently operational, and there is strong UK involvement in the
Spitzer SWIRE Legacy Program which has imaged over 50 square degrees
in seven infrared bands between 3.6 and 160$\mu$m. Akari (previously
Astro-F), a Japanese satellite with UK involvement, was launched in
2006 and has begun mapping the entire sky in 6 infrared bands from 9
to 180$\mu$m, with especially deep observations in the North Ecliptic
Polar (NEP) region. The Herschel satellite is due to be launched in
2008, and will carry out deep infrared surveys between 60 and
500$\mu$m. In each of these cases, the resolution at far-IR and sub-mm
wavelengths is such that high angular resolution follow-up will be
essential to identify the host galaxies: LOFAR will provide this.

\subsubsection{Gigamasers: pinpointing luminous starbursts at very high
  redshift} 

Powerful OH masing is relatively common amongst IR-luminous galaxies: the
most recent survey \citep{dar02} found that at least a third of
ultraluminous IR galaxies (ULIRGs, $L_{\rm FIR} \ge 10^{12} L_{\odot}$)
support megamasers or gigamasers ($L_{\rm OH} \ge 10 L_{\odot}$ or $L_{\rm
OH} \ge 10^{3} L_{\odot}$, respectively). If starbursts are responsible
for a significant fraction of the luminosity of ultraluminous and
hyperluminous IR galaxies, as currently thought \citep[e.g.,][]{far02a},
then the associated turbulence may enable low-gain unsaturated masing
\citep{bur90}. The earliest OH maser-line observations appeared to
demonstrate a quadratic relationship between OH and far-IR (FIR)
luminosities, $L_{\rm OH}$ and $L_{\rm FIR}$. This is believed to be due
to the abundant flux of FIR photons pumping an OH population inversion in
the star-forming molecular gas \citep{baa85,baa89}. Since FIR and radio
luminosities ($L_{\rm rad}$) are well correlated \citep[e.g.,][]{hel86},
emission stimulated by the background radio continuum would then yield
$L_{\rm OH} \propto L_{\rm FIR} L_{\rm rad} \propto L_{\rm FIR}^2$.

\citet{tow01} argued that this quadratic dependence would yield OH masers
detectable amongst the high-redshift sub-mm galaxy population, with peak
luminosity densities up to two orders of magnitude greater than those seen
in current samples. This would allow the determination of an accurate and
relatively unbiased redshift distribution for sub-mm galaxies, as well as
constraining the mass of their black holes, determining their geometric
distances, and even probing the evolution of fundamental constants
\citep[e.g.,][]{bar05,lo05,kan05,cap06}. Although there remains some
dispute about the precise dependence of $L_{\rm OH}$ on $L_{\rm FIR}$ ---
correcting large maser samples for Malmquist bias \citep{kan96} favours a
weaker dependence \citep[$L_{\rm OH} \propto L_{\rm FIR}^{1.2\pm
0.1}$;][]{dar02}, but using a smaller, complete sample continues to
suggest a quadratic relationship \citep[$L_{\rm OH} \propto L_{\rm
FIR}^{2.3 \pm 0.6}$;][]{klo04} --- it is reasonable to expect OH maser
emission from a high proportion of distant starbursts, with $L_{\rm OH}$
in the range 10$^{4-5} L_{\odot}$ for $L_{\rm FIR} \sim 10^{13} L_{\odot}$
and mJy-level peak flux densities.

Maser searches have clear advantages over other methods of detecting
distant starbursts: i) the bandwidth requirement for blind detection of OH
megamasers is low, $<$1\,GHz ($\nu_{\rm obs}$ = 165--835\,MHz for $z$ =
1--10), with $z \sim 7$ accessible to LOFAR at 200\,MHz; ii) the
instantaneous survey area is limited only by the primary beam ($>$1
deg$^2$ for an OH line search with LOFAR at 200\,MHz); iii) interferometry
permits some rejection of radio-frequency interference; iv) the position
of an emission line can be pinpointed accurately, on the sky and in
redshift space; finally, v) the dual-line 1665/1667\,MHz OH spectral
signature can act as a check on the line identification and the reality of
detections.

\citet{ivi06} tested this technique using a well-studied, FIR-luminous
galaxy, with a well-determined redshift. LOFAR, with its large
instantaneous bandwidth, will enable the first {\em blind} search for
high-redshift gigamaser emission. It will capable of probing starburst
galaxies at $z \sim 7$, the progenitors of the luminous quasars found by
SDSS at $z \sim 6$, and will be able to do so whilst conducting its deep
sky surveys. The improved angular resolution that international baselines
will provide will be beneficial in determining accurate positions, to
allow detailed follow-up observations of any high-redshift gigamasers
discovered.

\subsubsection[Extending the radio--infrared correlation]{Extending the
  radio--infrared correlation to lower radio frequencies and luminosities}

\begin{figure}[!b]
\begin{center}
\psfig{file=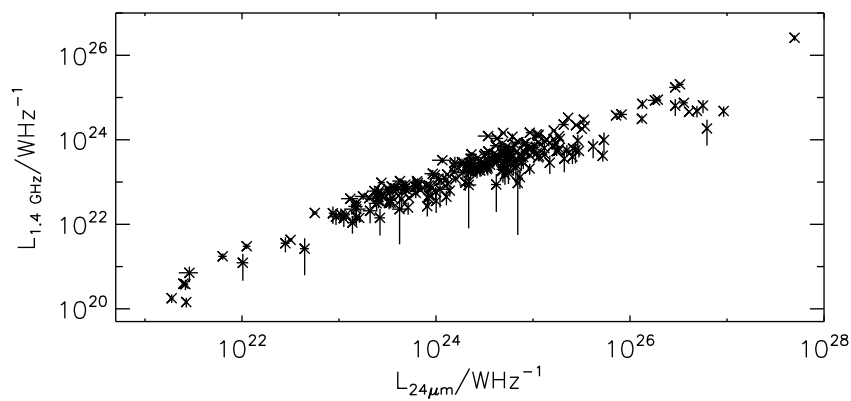,width=12cm,clip=}
\end{center}
\caption{\label{radirfig} \small The 20\,cm radio versus 24\,$\mu$m IR
luminosity correlation for 24\,$\mu$m selected Spitzer sources within the
Hubble Deep Field North.  Luminosities have been {\it k}-corrected in the
IR assuming a SED slope equivalent to Arp220 and in the radio assuming a
(1$+z)^{0.7}$ boosting \citep[from][]{bes06a}.}
\end{figure}

Over 25 years ago it became clear that the radio and global infrared
emission from galaxies were tightly correlated. In the 1980s, data from the
IRAS satellite demonstrated that this correlation extended over many
orders of magnitude, from nearby dust-rich dwarfs to ultra-luminous
infrared galaxies (ULIRGs).  More recently, comparisons of observations
from NASA's Spitzer IR satellite and radio observations have extended this
correlation both to the mid-IR and over a still wider range of
luminosities \citep[see Figure~\ref{radirfig};][]{app04,bes06a}. This
correlation can be extended to even fainter luminosities (L$_{25\,\mu{\rm
m}}\approx10^{20}$\,W\,Hz$^{-1}$) by considering discrete regions within
individual nearby star-forming galaxies \citep{mur06}. The correlation
between the radio and infrared emission arises because both are related to
the star formation processes; the infrared emission is produced from dust
heated by photons from young stars, while the radio emission arises from
synchrotron radiation produced by the acceleration of charged particles
from supernova explosions.

While recent deep field observations at radio (using MERLIN, VLA \& the
WSRT) and IR (primarily using Spitzer) wavelengths have made significant
advances in understanding the relationship between decimetre radio
emission and mid-IR emission in star-forming galaxies \citep[both nearby
galaxies and out to z$\sim$4; e.g.,][]{gar02}, the relationship between
the infrared emission and that at low radio frequencies is still
unexplored. LOFAR will explore the lower frequency (30-240\,MHz) radio
emission from millions of star-forming galaxies and, in conjunction with
the current generation of IR satellites such as Spitzer, will extend the
radio-IR correlation to lower radio frequencies and luminosities, allowing
the investigation of any evolution of this relationship with redshift,
luminosity and radio frequency.

Although surveys using LOFAR with 100\,km baselines would be able to
detect a large number of sources and contribute to this work, the lack of
resolution will result in the confusion limit being quickly reached and
hence significantly reduce both the number and luminosity range of sources
that can be observed. The addition of international baselines would be of
great benefit.

\subsection{Radio--loud AGN and their influence on galaxies and the
large--scale environment}  

It is now crystal clear that even the most basic features of the local
galaxy population, such as its luminosity function, cannot be understood
by studying stellar populations in isolation. It has become apparent that
AGN play a key role in galaxy evolution, with AGN outflows being
responsible for controlling or terminating the star formation of their
host galaxies. At least two modes of feedback are needed: a slow accretion
mode \citep{cro06,bow06}, the observational manifestation of which is the
\citet{fan74} Class I (FR\,I) radio galaxy population, and a fast
accretion mode associated with quasars \citep[e.g.,][]{sil98}. To
understand galaxies we need to identify and quantify the physical
processes which control the AGN feedback process, including the AGN
duty-cycle, as well as to determine the influence of the AGN on the
large-scale properties of clusters and protoclusters.  As one must study
all this as a detailed function of both epoch and environment, adequate
sampling of the highest redshifts and rarest environments demands the use
of much larger sky areas than are available in existing `deep fields'.

As well as delivering a transformational increase in radio survey speed,
the low radio frequency of LOFAR is also liable to be crucial. Although
there are many complications in the physics of radio sources
\citep[e.g.,][]{blu00}, it remains crudely true that the observed, as
opposed to intrinsic, duty cycle of radio-jet and radio-starburst activity
increases with decreasing frequency: this is because of the longer
synchrotron lifetimes of the lower-energy relativistic particles probed at
lower frequencies. LOFAR frequencies may well be low enough to probe all,
or at least most, of both modes of AGN feedback in the observable
Universe.

\subsubsection{The nature and evolution of radio--AGN feedback on galaxy scales}

\begin{figure*}[!bt]
\begin{tabular}{cc}
\psfig{file=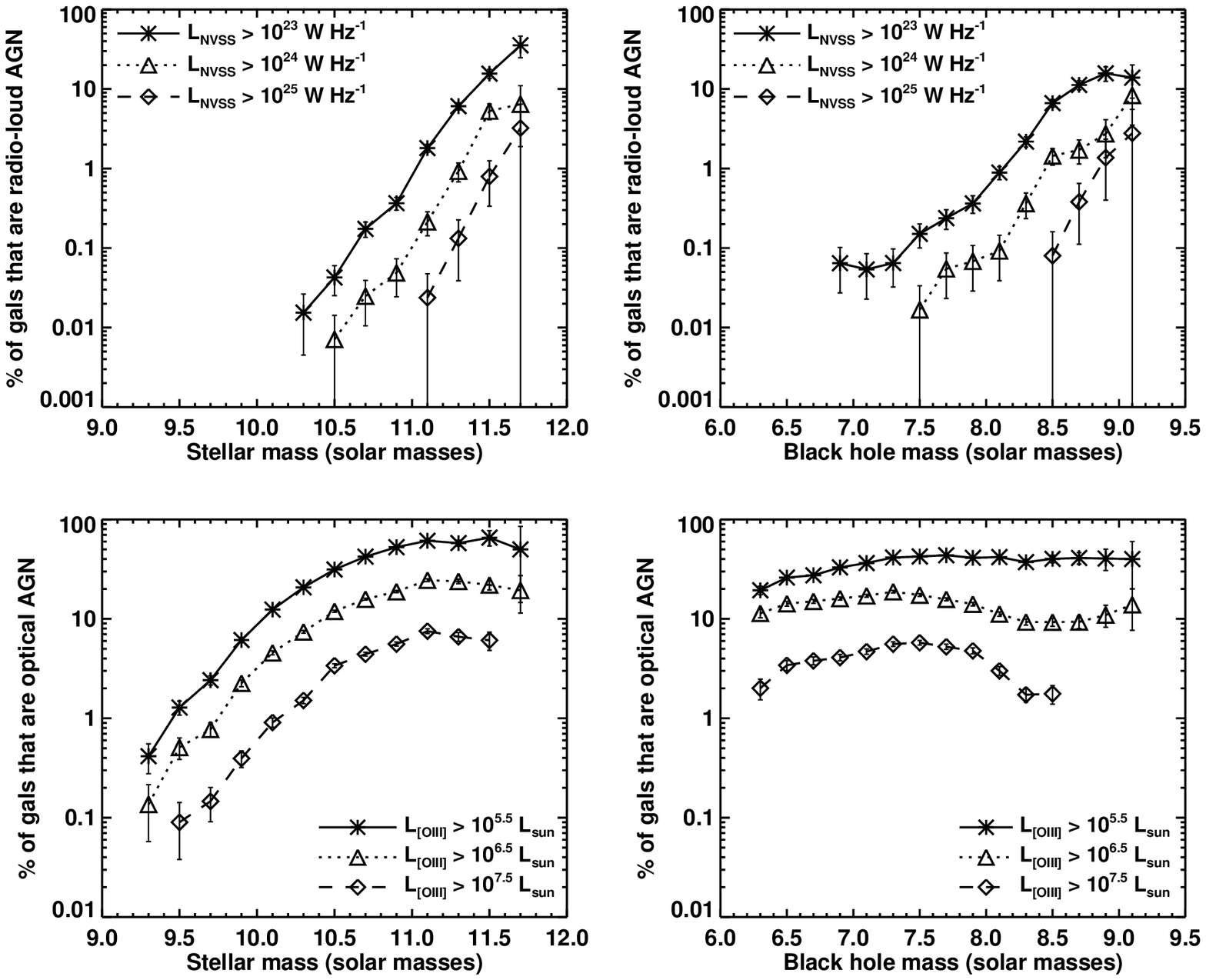,angle=0,width=7.8cm,clip=}
&
\raisebox{0.2cm}{\psfig{file=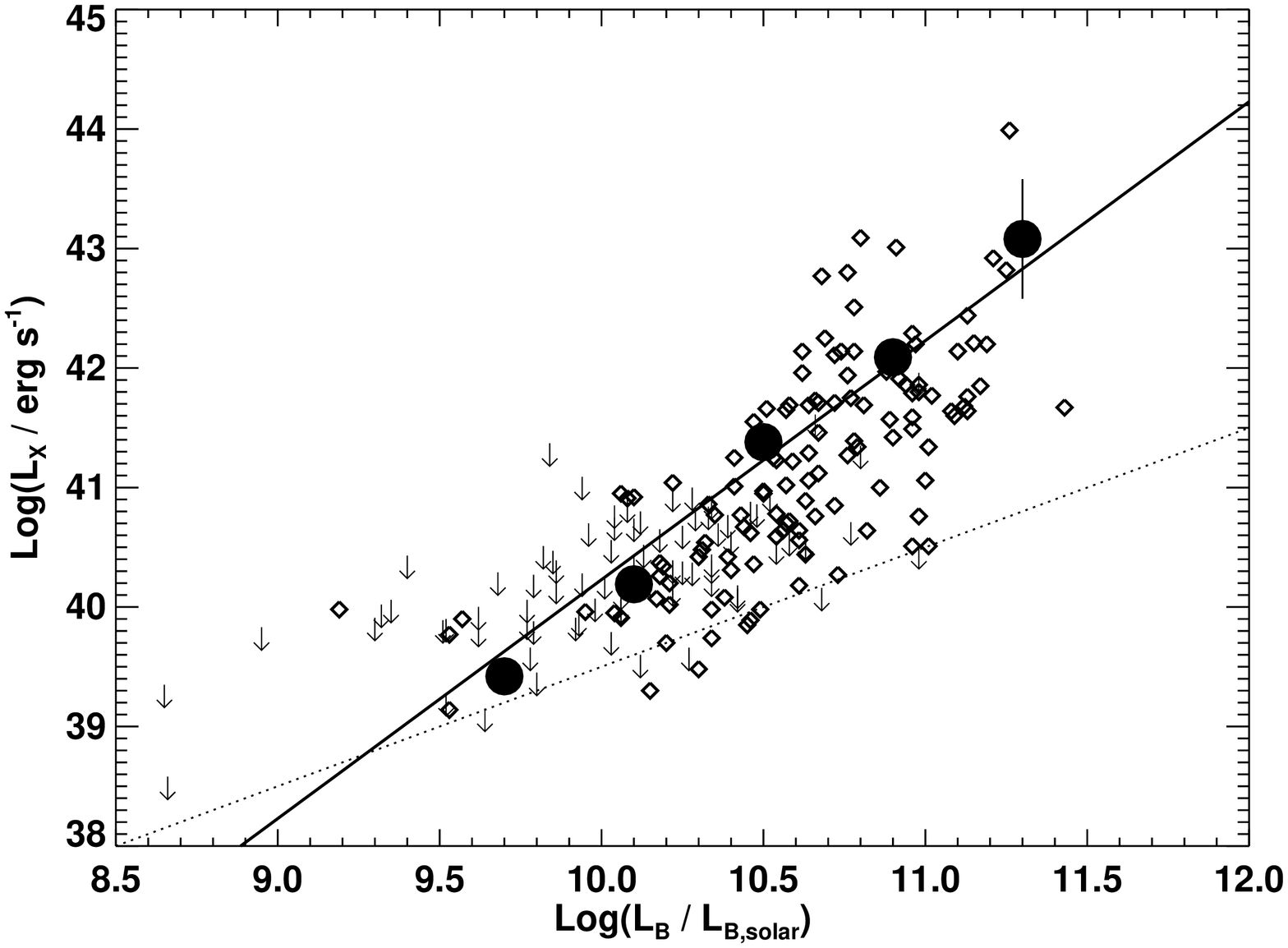,angle=90,width=8.2cm,clip=}}
\end{tabular}
\caption{\label{sdss_rad1} \small {\it Left:} the fraction of galaxies
which are radio--loud AGN, as a function of stellar mass, for different
cuts in radio luminosity. The radio-loud AGN fraction is a remarkably
strong function of stellar or black hole mass and has the same slope at
all luminosity limits \citep[from][]{bes05b}. {\it Right:} The radio-AGN
heating versus radiative cooling balance in elliptical galaxies
\citep[from][]{bes06b}. The data points show bolometric X-ray luminosity
($L_X$; the rate at which elliptical galaxies radiate energy away from
their hot gas haloes), versus optical luminosity ($L_B$) for elliptical
galaxies from the sample of \citet{osu01}. The large filled circles show
the mean values of $L_X$ for galaxies in 5 bins of $L_B$. The solid line
shows the prediction for the amount of heating produced by recurrent
radio-loud AGN activity: this balances radiative cooling losses remarkably
well, across the full range of optical luminosities (masses).}
\end{figure*}

Substantial progress in understanding the process of radio--AGN feedback
in the nearby Universe has recently been made by \citet{bes05b}, who
combined the Sloan Digital Sky Survey (SDSS) with the NVSS and FIRST radio
surveys to determine the statistical relationships between radio activity
and galaxy\,/\,black hole mass.  They found that the fraction of galaxies
that host radio--loud AGN (with $L_{\rm 1.4GHz} > 10^{23}$W\,Hz$^{-1}$) is
a strong function of stellar mass, rising from nearly zero below a stellar
mass of $10^{10} M_{\odot}$ to more than 30\% at stellar masses of $5
\times 10^{11} M_{\odot}$ (see Figure~\ref{sdss_rad1}a). \citet{bes05b}
also showed that this steep mass dependence of the radio-loud fraction
mirrors the mass dependence of the expected accretion rates on to the
central black hole from gas in the hot haloes surrounding massive
elliptical galaxies. They argued that low luminosity radio--loud AGN are
fuelled by cooling of this hot gas. \citet{bes06b} took this argument one
step further by using estimates of the mechanical energy output associated
with radio--loud AGN activity to calculate the time--averaged energy
output associated with recurrent radio source activity in a galaxy of
given mass. They showed that, for massive elliptical galaxies, the
radio--source heating roughly balances the radiative energy losses from
the hot gas surrounding the galaxy (Figure~\ref{sdss_rad1}b), and that the
recurrent radio--loud AGN activity may therefore offer a self-regulating
feedback mechanism capable of controlling the rate of growth of galaxies.

To complete this picture, it is now essential to trace the evolution of
these relations: how does the fraction of galaxies hosting a powerful
radio source evolve with redshift, as a function of stellar or black hole
mass?; how does the balance between radiative cooling and AGN heating vary
with redshift, mass and environment? \citet{sad07} have begun to address
these questions by demonstrating that the radio luminosity function of
luminous red galaxies (LRGs; i.e., the most massive galaxies) evolves out
to $z \sim 0.5$ as $(1+z)^2$, but further study is hampered by the lack of
depth in both optical and radio datasets. The LOFAR deep survey is
sufficiently deep that it will detect essentially all radio--loud AGN in
the Universe out to very high redshifts. Combining these data with deep
optical and/or near-infrared data, particularly those from the
Pan-STARRS Medium Deep Survey, will provide both photometric redshifts and
stellar mass estimates. It will be therefore be possible to make
definitive measurements of how the mass--radio relations, the radio source
duty cycle, and the heating rate vary as a function of epoch across most
of the age of the Universe.

Important questions that will be addressed include: 

\begin{itemize} 
\item Do the high redshift samples show the same steep mass dependence
of the radio-loud AGN fraction as the $z\sim 0$ SDSS sample (implying
that the same physical processes are at work)?

\item Does the location of the relations evolve with redshift?  This
would imply either an evolution in the efficiency of production of
radio emission, or an evolution in the fuelling rate. This is the key
input for parameterising AGN feedback in galaxy formation models.

\item Does the balance between AGN heating and radiative cooling
losses change?
\end{itemize}

The combination of sensitivity and high angular resolution offered by
LOFAR (with international baselines) will also allow a more detailed
investigation of the mechanisms by which AGN energy may be transferred
from radio-loud AGN to their galaxy (and cluster) environments.  Recent
X-ray observations with Chandra and XMM-Newton have revealed that several
mechanisms are in operation, including heating by strong and weak shocks
and by subsonic expansion \citep[e.g.,][]{kra03,fab03,cro03,for05}.  To
date, the only clear examples of heating by strong shocks are in massive
elliptical galaxies, where small (few kpc-scale), over-pressured radio
sources can inject amounts of energy comparable to the thermal energy of
the galaxy ISM, permanently increasing its entropy \citep{kra03,cro07}.

The FIRST radio survey (currently the best available) is only able to
resolve these low-luminosity galaxy-scale radio sources (like Centaurus A)
sufficiently that they are identifiable as double-lobed, and allow a
minimum pressure to be calculated, out to redshifts $z \sim 0.04$. As
such, only a small number of these sources are known, and their importance
for galaxy feedback is unclear. However, given their lifetimes of $\sim
10^{6}$ years, they may constitute an important population. The detection
of similar radio-lobe sources in a few active spiral galaxies
\citep[e.g.,][]{gal06} increases the importance of understanding this type
of low-luminosity radio outburst.  LOFAR, with international baselines,
will enable these radio sources to be detected and resolved to a redshift
of $z \sim 0.5$, over large sky areas, enabling population statistics to
be compiled for the first time. Studies of the environmental dependence of
this type of radio outburst will be crucial to understand whether they are
triggered by galaxy mergers (as hinted at by the examples known to date),
and to constrain their impact in different host galaxy populations.

\subsubsection{Feedback from FR\,IIs?}

The local Universe SDSS surveys do not contain many sources more powerful
than $L_{\rm 1.4GHz} \sim 10^{25}$W\,Hz$^{-1}$, due to the rarity of these
sources and the small volume studied. Above this luminosity, most radio
sources are FR\,IIs. The origin of the dichotomy between FR\,I and FR\,II
sources is still a matter of debate, with both intrinsic properties (black
hole spin or accretion flow) and the extrinsic environment (jet disruption
through interactions) argued to play a role \citep[see discussion
in][]{sne01}. The strong correlation seen between radio and emission-line
AGN activity in powerful FR\,II sources indicates that the fuelling (and
hence triggering) method may be very different from that postulated for
the FR\,I sources in the SDSS study. In such a case, it is reasonable to
assume that the mass dependence of the radio--loud fraction would break
down. However, this remains untested, and the role that FR\,II sources may
play in AGN feedback is also unknown.

In addition to determining the fraction of low luminosity radio sources as
a function of galaxy mass at moderate redshifts, the LOFAR surveys will
provide significant samples of radio sources more powerful than the FR\,I
-- FR\,II divide (cf. Figure~\ref{fig:pz}a). The radio--loud fraction
versus mass relation for the FR\,II sources can therefore be determined,
and compared to that of the FR\,I sources at the same redshift, in order
both to quantify the role that the FR\,II sources play in AGN feedback,
and also to greatly enhance our understanding of the differences between
the two radio classes.

\subsubsection{AGN feedback and the larger--scale environment}

Gas in the central regions of clusters of galaxies often has a radiative
cooling timescale very much shorter than the Hubble time. In the absence
of a heating source, a cooling flow would be expected to develop, whereby
the temperature in the central regions of the cluster drops and gas flows
inwards at rates of up to $\sim 1000 M_{\odot}$\,yr$^{-1}$ \citep[see][for
a review]{fab94}. However, recent XMM-Newton and Chandra observations of
cooling flow clusters have shown that the temperature of cluster cores
does not fall below $\sim 30$\% of that at large radii, and that the
amount of cooling gas is only about 10\% of that predicted for a classical
cooling flow \citep[e.g.][]{pet01,dav01,tam01,kaa01}. This implies that
some heating source must balance the radiative cooling losses, preventing
the gas from cooling further.

\begin{figure*}[t]
\centerline{
\psfig{file=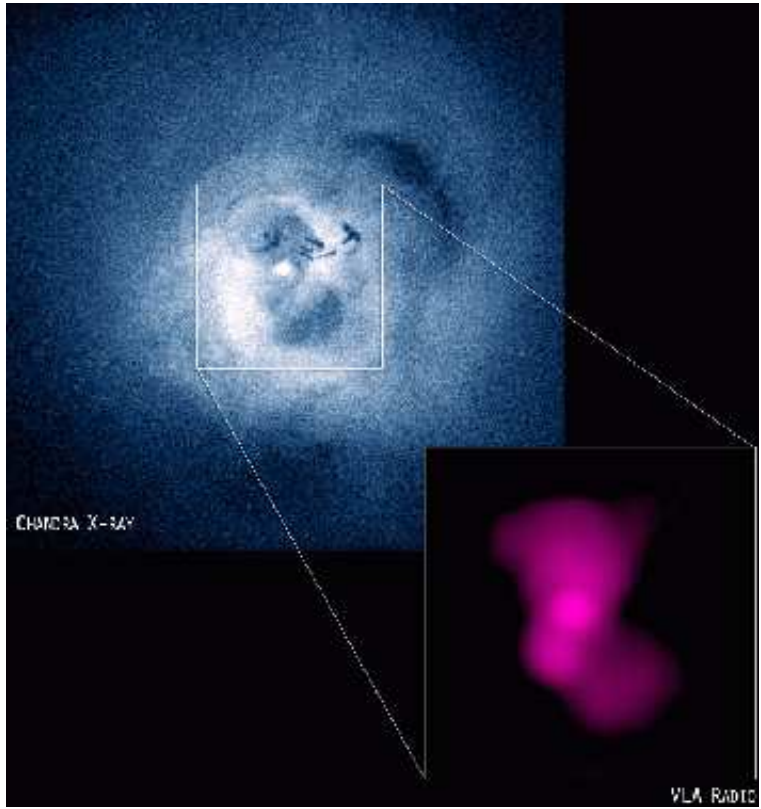,angle=0,width=10cm,clip=}
}
\caption{\label{bubfig} \small A Chandra X-ray image of the Perseus
cluster (credit: A.\ Fabian) together with the 328\,MHz VLA image of the
same region (credit: G.\ Taylor). Evacuated cavities are clearly visible
in the central regions of the X-ray image of the cluster, co-incident with
the location of the radio emission. An older, radio-quiet, cavity is also
visible towards the north-west of the cluster.}
\end{figure*}

Heating by radio sources associated with the brightest cluster galaxies
(BCGs) has gained popularity in recent years, as X--ray observations have
revealed bubbles and cavities in the hot intracluster medium of some
clusters, coincident with the lobes of the radio sources
\citep[e.g.][cf. Figure~\ref{bubfig}]{boh93,car94b}.  These are regions
where relativistic radio plasma has displaced the intracluster gas,
creating a low--density bubble of material in approximate pressure balance
with the surrounding medium, which then rises buoyantly and expands. For
some clusters the (pV) energy contained within the evacuated bubbles has
been shown to be sufficient to balance the cooling losses, at least for a
short period of time \citep[a few $\times 10^7$\,yr; e.g.\
][]{fab03,bir04,dun05}.

In some clusters, such as Perseus and Hydra-A, extended cavity systems are
seen, with the inner cavities filled with radio plasma but the outer
cavities being radio-quiet (cf. Figure~\ref{bubfig}). These larger radius
cavities may be associated with previous radio outbursts; their older ages
mean that the relativistic electrons will have lower Lorentz factors, and
so emit at lower frequencies: characterisation of these radio-quiet
cavities using LOFAR observations will help to determine the duty cycle of
the radio-AGN activity of the BCGs. LOFAR will also have the sensitivity
to probe the diffuse cluster sources out to much higher redshifts, which
is important to understand the evolution of clusters and cooling flows.

There are only a few high-redshift precursors of the Abell cluster
population (the largest collapsed dark-matter haloes) per square degree
\citep[e.g.,][]{vbr06}, but the 250 square degrees of the proposed
LOFAR-deep survey is sufficient to obtain samples of $\sim 25$ clusters in
each of ten small ($\Delta z = 0.1$) redshift bins over the range $0.5
\leq z \leq 1.5$. These proto-clusters can be identified using the
near--IR and optical surveys of this region (Pan-STARRS, for example, will
select clusters out to $z \approx 1$ in the 3$\pi$ survey and to $z
\approx 1.5$ in the Medium Deep Survey regions); the LOFAR deep survey
will then enable a definitive measurement of the how the mass--radio
relation, the radio source duty cycle and the heating rate vary as a
function of environment and epoch. Studies using clusters selected from
the SDSS data suggest that, locally, the radio--AGN feedback is stronger
in richer environments \citep{bes07}: how does this environment--activity
relation evolve out to earlier cosmic epochs? These investigations are
important to understand the extent to which radio AGN feedback can be
responsible for the substantially lower rates of star formation activity
in dense environments \citep[cf.][]{raw04b}.

We note that it is essential that the individual patches of the LOFAR deep
survey are large enough to ensure that we can investigate superclustering
-- the clustering of these clusters \citep[cf.][]{swi07}. Even on
half-arcminute scales the surface density of old passively evolving
galaxies at high redshift varies by several tens of percent
\citep[e.g.,][]{yam05}, whilst significant overdensities in quasi-linear
structures on scales up to $\sim 100$ Mpc (corresponding to angular sizes
of $\sim 2 \times 2 ~ \rm deg^{2}$ at $z \sim 1$) have previously been
detected by radio surveys \citep{bra03}.  This suggests that patch sizes
of $\gta 5$ by 5 degrees will be necessary and sufficient, both to detect
the boundaries of individual superclusters and to measure the statistics
needed to confront models with data.  This is an important measurement
because the space density of superclusters is one of the sharpest
discrepancies between theory and observation in the local Universe
\citep{ein06}.

\subsubsection{Radio galaxy structure and electron energy distributions}

To understand the large-scale energy input from radio-loud active
galaxies into the IGM and into the hot baryonic atmospheres of groups
and clusters of galaxies \citep[e.g.,][]{cro06,bow06,roy04,rus04}, it
is vital to understand the details of the radio-loud AGN `feedback'
process: how much energy is transported by an observed AGN jet, in
what form (or forms) is it stored in the observed jets and lobes, and
how is it transferred to the external medium?

The observed radio-through-X-ray synchrotron and X-ray inverse-Compton
emission from jets and lobes indicates that the lobes contain, at a
minimum, electrons, positrons and magnetic field. The e$^+$e$^-$
population is somewhat energetically dominant \citep[e.g.,][]{cro05} and
has a roughly power-law energy distribution with a steep index: thus the
low-energy electrons completely dominate the relativistic particle
component of the lobe energy density and pressure. The relativistic
particles are thought to be accelerated at relatively compact sites
(predominantly in the jets for low-power radio galaxies; predominantly in
compact hot spots at the ends of the jets for high-power objects). The
size scales of these regions are of the order of a few kpc, compared to
hundreds of kpc for the large-scale lobes.

By providing arcsec-scale (kpc-scale for typical radio galaxy distances)
resolution at low frequencies, LOFAR + UK long baselines will address two
critically important issues. Firstly, it will allow the observation, via
synchrotron emission, of the population of low-energy leptons in the lobes
(with $\gamma \approx 1000$, where $\gamma$ is the electron Lorentz
factor) which are currently only observable by their X-ray inverse-Compton
emission \citep[e.g.,][]{har05}.  Direct comparisons between the
structures of synchrotron and inverse-Compton emission for electrons at
these energies will be possible and this will determine the distribution
of energy density in the other component of the relativistic plasma, the
magnetic field. Secondly, it will probe much lower-energy electrons in the
high magnetic field regions of particle acceleration, the jets and
hotspots. Existing observations hint at the possibility of a low-energy
cutoff in particle acceleration at energies around $\gamma = 500$ --
$1000$ \citep{car91,har01}, but the difficulty of making these
observations at high resolution means that the evidence comes from only a
few bright (and atypical) sources.  LOFAR will probe electron energies
down to $\gamma \approx 100$ in the high-field acceleration regions of
many radio sources. Measurement of the low-energy electron spectrum in the
acceleration regions will not only reveal the energy spectrum that is
injected into the lobes, which is required if the energetically dominant
lepton population is to be understood, but also constrain the still poorly
understood particle acceleration mechanism(s); this will give clues about
how the bulk kinetic energy of jets is translated into the internal energy
of the observed relativistic plasma. The high resolution provided by the
long international baselines is vital to both problems, to allow the
compact particle acceleration regions to be separated from the lobes and
to allow point-to-point mapping of the lobe electron spectra and magnetic
field strength. UK involvement in LOFAR will therefore enable the
resolution of a number of long-standing problems in radio-galaxy physics,
that are crucial to understanding galaxy feedback.

\subsubsection{High redshift quasars and the link to star formation}

It can be seen from Figure~\ref{fig:pz} that the LOFAR sensitivity will be
adequate to detect all the quasar activity, even from so-called
`radio-quiet' quasars (RQQs), out to very high redshift. Many of these
radio--quiet quasars are type-2 quasars, where the central AGN is obscured
at optical wavelengths. Indeed, this population is already being
discovered in small numbers by the cross-matching of deep radio
(VLA\,/\,GMRT) and Spitzer catalogues \citep{mar05}. Matching of these
objects to X-ray datasets has revealed that the majority are Compton thick
and either present only in the hardest X-ray bands of XMM, or absent
altogether \citep{sim06,mar07}. This strongly implies that most
high-redshift quasar activity has yet to be detected directly. 

A near-IR survey to $K \approx 20$ will identify the host galaxies of
essentially all of these radio-quiet quasars out to $z \gta 2$ (see
Figure~\ref{fig:kz}), whilst the addition of the optical datasets will
provide optical detections (and hence photometric redshifts for the type-2
objects) out to a similar redshift. This will allow accurate measurements
of how the quasar activity evolves with redshift, which is vital for
studies of black hole growth, quasar feedback, and the link with star
formation, as well as more esoteric things like the gravitational wave
background.

A prediction of all quasar feedback theories \citep[e.g.,][]{sil98} is a
strong link between quasar and starburst activity, and there is evidence
that this occurs in individual objects. Coupling the LOFAR-deep survey
with deep K--band data (plus probably some high-resolution radio follow-up
with e-MERLIN) it will be possible to study directly, and in precise
detail, how quasar activity occurs alongside star formation as a function
of both epoch and environment. The radio emission of starburst galaxies
measures the time-averaged star-formation rate, independent of dust, and
LOFAR's sensitivity and frequency range will provide a unique resource for
mapping this out across the huge areas needed to obtain representative
samples of environments at each epoch.

\subsubsection{Evolution of the radio luminosity function and
  high-redshift radio galaxies}

In combination with optical and near--IR data to provide photometric
redshifts out to $z \gta 2$, the LOFAR surveys will enable the evolution
of the radio luminosity function (RLF) to be measured down to radio
luminosities an order of magnitude fainter than previous studies
\citep[cf.][]{cle04}, and with a sample size two orders of magnitude
larger. This will therefore provide a definitive measurement of how
strongly the evolution of the RLF varies with radio luminosity: in the
optical and X-ray wavebands, large differences in evolution are being
found between the epoch of peak activity of high luminosity and low
luminosity sources \citep[e.g.,][]{has05}, and so determining whether the
same holds for radio--loud AGN will address the question of what makes an
AGN radio-loud.  Comparing the cosmic evolution of the FR\,I and FR\,II
radio sources at the same radio luminosity will also shed light on the
origin of the different morphological classes of radio source \citep[an
intrinsic property of the central engine, or an environmental effect;
cf.][]{sne01,rig07}.

Another critical question is the evolution of the radio luminosity
function beyond $z \approx 2.5$, where there has been much debate in the
literature as to the presence or absence of a high redshift cut-off
\citep[e.g.,][]{jar01}. This is important to determine because, since
strong radio activity is only produced by the most massive black holes ($M
\gta 10^9 M_{\odot}$; e.g., Dunlop et~al 2002), the cosmic evolution of
powerful radio sources offers the cleanest way to constrain the evolution
of the top end of the black-hole mass function.  To a 30$\mu$Jy flux
density limit at 200\,MHz, there are expected to be $\sim 100$ radio--loud
AGN per square degree with $z \gta 4$ (cf. Figure~\ref{fig:pz}).
Extremely deep optical and near-IR data will be available over a few tens
of square degrees of the LOFAR deep survey (e.g., in the Pan-STARRS Medium
Deep Field regions), producing a sample of several thousand $z > 4$ radio
source candidates, identified through photometric redshifts. At these
redshifts Ly$\alpha$ is easily observable within the optical band, and so
measuring the redshifts of candidate high-redshift sources, and thus
determining the space density of radio sources at $z \sim 4$ (which would
unambiguously answer the question of the high redshift cut-off), is an
eminently achievable goal.

Since high-redshift radio galaxies are among the brightest known galaxies
in the early Universe \citep{bre02a}, follow-up studies of these $z\sim4$
radio galaxies will constrain the evolution of such massive galaxies.
Recently, it has been discovered that powerful radio galaxies are often
surrounded by significant galaxy overdensities, whose structures have
sizes of a few Mpc and velocity dispersions of a few hundred km s$^{-1}$
\citep[e.g.,][]{ven02}; the inferred masses of these structures are
several $\times 10^{14}$ M$_\odot$, consistent with them being the
precursors of rich clusters (protoclusters). Deep optical and near-IR data
are a vital starting point for studying the environments of these $z>4$
radio galaxies, which will constrain the formation of clusters at the
earliest epochs.

\subsubsection{Powerful radio galaxies within the Epoch of Reionisation}

Particularly interesting would be the discovery of powerful radio galaxies
at $z \gta 7$, since this would allow studies of the interstellar medium
through redshifted 21 cm absorption studies at the end of the reionisation
epoch \citep[cf.][]{car02}. Figure~\ref{fig:pz}a shows predictions for the
number of FR\,II radio sources expected in the LOFAR-deep survey and, as
this is predicated on the most reliable estimates we have for the decline
in the space density of radio sources at high redshift, this is highly
encouraging. Although the uncertainties in this rate of decline are
extremely large \citep[e.g.,][]{jar01}, the LOFAR-deep area of 250 square
degrees is large enough to virtually guarantee objects at $z \gta 7$, even
if the decline is, say, ten-times more severe than the current
extrapolation.

These $z \gta 7$ objects will be blank in the deep optical imaging and in
near-IR observations to $K \approx 20$, but such datasets are deep enough
that other blank--field radio sources will be rare. Since there exist
radio properties that are predictive of high radio power or high redshift
\citep[FR\,II, steep-spectral index etc; e.g.,][and references
therein]{deb00,cru06}, the candidates for $z \gta 7$ will be able to be
promptly followed up spectroscopically on 8-m telescopes.  There is
therefore an excellent chance that radio-loud objects will be found in the
EoR, and a chance that one will be found with a high enough radio flux
density to allow \HI\ absorption measurements with LOFAR \citep{car02}.

\subsection{Cosmology with the LOFAR sky surveys}

\subsubsection{Spectroscopic follow-up of LOFAR-deep, and Dark Energy}

The detection of the signatures of Baryon Acoustic Oscillations (BAO) at
low redshift \citep[e.g.,][]{per07} has led to the realisation that they
can be used to determine the dark energy component of the Universe, if
better statistics can be achieved in the large cosmic volumes available at
high redshift. State-of-the-art BAO surveys (e.g., with AA$\Omega$ on the
Anglo-Australian telescope) are aiming to provide information on the dark
energy parameter, $w(z)$, at redshifts $z \sim 0.7$. To complement this, a
key future goal is to measure the BAO length-scale to an accuracy of $\sim
2$\% at redshift $z \gta 1$; to achieve this, redshifts for a well-defined
sample of at least several hundred thousand galaxies will need to be
measured. 

The UK has a broad interest and strong scientific leadership in the study
of BAO. It is also involved in the development of the required
instrumentation to pursue these studies at $z \gta 1$, namely the next
generation of multi-object near-infrared fibre spectrographs, such as FMOS
\citep{dal06}. The FMOS spectrometer, currently under commissioning at
Subaru, has been built by a consortium of UK, Australian and Japanese
groups, and provides a 30$^\prime$ field sampled with 400 fibres by a pair
of NIR OH-Suppression spectrographs covering $0.9-1.8$\,$\mu$m.  Utilising
emission lines like H$\alpha$, deep FMOS observations will be sensitive to
even moderately ($\sim 1 M_{\odot} ~ \rm yr^{-1}$) star-forming galaxies
at $z \gta 1$. If the down-sizing scenario proves to be predictive, this
means that FMOS will spectroscopically identify all but the most massive
systems via their emission lines, whilst the most massive systems should
be bright enough in continuum to allow the measurement of absorption-line
redshifts. The UK--Japan agreement for FMOS means that the UK will get
access to FMOS for 30\% of the nights that FMOS is operational on Subaru.

The LOFAR-deep survey will provide an ideal dataset from which to select
targets for BAO surveys with FMOS, for a number of reasons. As
demonstrated in Figure~\ref{fig:pz}b, LOFAR-deep delivers above the
critical density d$n / $d$z$ of objects ($\sim 2000 ~ \rm deg^{-2}$) that
is needed for power-spectrum measurements at $z \approx 1.5$ to be
cosmic-variance (rather than shot-noise) limited. In addition, the objects
identified are those that, through their radio emission, are known to be
either starbursts or AGN, and therefore likely to have narrow H$\alpha$
emission; this means that redshifts can be obtained in relatively short
spectroscopic observations. Furthermore, the few hundred square degree
coverage proposed for the LOFAR-deep survey is approximately the size
needed to produce the required $\sim 2$\% accuracy in the BAO length-scale
measurement. Finally, the LOFAR population peaks in precisely the redshift
regions to be targeted by FMOS. The LOFAR-deep survey can thus
be an extremely valuable resource for future BAO surveys, such as the
proposed joint UK\,/\,Japan\,/\,Australian FMOS survey {\it FastSound}.

One critical constraint on the LOFAR-deep survey for BAO work, however, is
that because the angular scale of BAOs is relatively large (around 2.5
degrees at $z \sim 1$) contiguous regions of at least 5 by 5 square
degrees are required, and even larger regions are preferable. In order to
avoid aliasing issues, it is better if the deep fields are nearly
symmetrical, and for optimal follow-up the fields should be widely spread
in right ascension.

Optical and near-IR surveys of the LOFAR-deep regions play two crucial
roles with regard to BAO studies. Firstly, they will provide accurate
positions for each LOFAR source, allowing precise positioning of
spectroscopic fibres. Secondly, they can be used to select $z > 1$ blue
emission-line targets, whilst ensuring that a narrow range of mass
(constant $K$ at a given $z$) is targeted. This is important as mixing
populations of different, unknown, bias could easily distort the BAO
features in the power spectrum.

\subsubsection{Strong gravitational lensing}
\label{lenssec}

Strong-gravitational lenses are systems in which the gravitational field
of a foreground galaxy multiply images a background source. They are
important because studies of lens systems probe mass profiles of galaxies,
including the dark matter, and are beginning to give us major insights
into galaxy structure and evolution. For example, on the scale of a few
kiloparsecs, lensing studies, in some cases together with stellar
dynamical information, have shown that the total (baryonic\,+\,dark) mass
profiles of $z\sim0.5$ galaxies are approximately isothermal
\citep[e.g.,][]{coh01,wuc04,koo06}. It is even reaching the stage where
the light and dark matter profiles may be separable using combined HST
imaging and lensing studies \citep[e.g.,][]{war05}.

About 120 lens systems are currently known. The largest number ($\sim$40)
have been discovered in the Sloan Lens ACS (SLACS) survey \citep{bol06}
based on identification of multiple-redshift systems in the SDSS. A
further $\sim$40 have been discovered in radio surveys, the largest number
(22) in the Cosmic Lens All-Sky Survey (CLASS) of flat-spectrum radio
sources \citep{mye03,bro03a}. SLACS survey lenses are well suited for mass
profiles, since the background sources are extended; CLASS lenses have
been used for mass determinations but also, via variability studies, for
extraction of the Hubble constant \citep[e.g.,][]{big99,fas00,wuc04}.

Although they make up one-third of known lens systems (a fraction which
will decrease in the short term as surveys such as SLACS and CFHTLS
increase) radio lenses are very important in two major areas.

First, CDM simulations of galaxy formation are just beginning to have the
resolution to probe sub-galactic scales and make predictions about the
internal structure of galaxies. For example, they over-predict the number
of Galactic satellites \citep{moo99,kly99} and an immense amount of work
has been devoted to understanding this. Recent work \citep[e.g.,][]{moo06}
has addressed the important role of baryons in the centres of galaxies,
and secure theoretical predictions may be available in the next
decade. Gravitational lenses are vitally important for providing
observational tests of these theories, as they are the only way of
detecting lumpy dark-matter substructure in $z\sim0.5$ galaxies. This is
possible because lumps of dark matter close to the line of sight to a
lensed image will perturb its position and, especially, its magnification
in such a way that the lens can no longer be modelled by a simple, smooth
model \citep{mao98,dal02,met02}.  Radio lenses are important here because
the fluxes of the components are much less affected by micro-lensing by
stars in the lens galaxy, because radio sources are bigger than optical
quasars. The big problem is the small number (8) of four-image radio
lenses. A larger sample is urgently needed for progress in this area.

Second, images forming close to the centre of the lens galaxy allow probes
of its gravitational potential within the central 100\,pc
\citep{wal93,rus01,win03}. This is only possible in radio lenses, as the
optical picture is contaminated by light from the lensing galaxy. The
lensed image becomes stronger as the central potential becomes less
singular, and this therefore offers a probe of the steepness of the
central stellar cusp together with the mass of the central black
hole. Furthermore, central black holes may produce additional images
which, if detectable, measure black hole masses more directly.

In terms of practicalities, about 1 in 700 radio sources at $z \ge 1$ are
lensed. The major requirement for a radio survey is therefore a large
number of sources, which LOFAR will provide: in 10$^8$ sources one expects
100,000 lenses. Direct identification of the lenses requires high angular
resolution, however, since the typical separation of lensed images is only
about 0.5$^{\prime\prime}$: the higher the resolution, the more lenses
will be identified.  Surveys at lower (2-3$^{\prime\prime}$) resolution
are still useful, and in conjunction with optical data can be used to
vastly increase the efficiency of targeted searches with the EVLA/e-MERLIN
\citep{jac07}, but surveys with long-baseline LOFAR with $\sim 1''$
resolution could be expected to find many lens systems directly. How many
could be found immediately depends critically on the stability of the PSF,
but it is likely that between 50 and 100 systems would be found
straightforwardly \citep{jac02,wuc06}. Note that direct surveys with the
EVLA, limited to 1 in 700 efficiency, cannot be done without unreasonably
large amounts of observing time. Pre-filtering using a large-area LOFAR
survey with long baselines could increase the efficiency by an estimated
factor of 10, making a further EVLA survey practical.

\subsubsection{Weak lensing with LOFAR}

Weak gravitational lensing represents an important tool for probing the 
dark matter distribution in the Universe. Background galaxy images are 
distorted (sheared, or flexed) by the gravitational potential of the 
intervening dark matter distribution, causing an overall alignment in the 
appearance of galaxy images at all wavelengths. This allows detailed 
measurements of the projected gravitational potential throughout a vast 
(typically $z=0\rightarrow 1$) region of the cosmos. Furthermore, one can 
examine the dark matter profile of mass concentrations in the Universe, 
such as galaxies and clusters \citep[e.g.,][]{hoe04}.

Weak lensing provides a surprising variety of cosmological information on
a wide range of scales. It affords measurements of the overall density of
dark matter \citep[e.g.,][]{bro03b} and dark energy \citep{hea06}; equally,
it has recently been shown that higher-order (flexion) weak lensing
statistics can efficiently probe small-scale structure on the scale of
tens of kpc \citep{bac06}. In addition, correlations between lensing-based
dark matter maps and light maps will allow detailed understanding of
galaxy bias \citep{tay04}.

LOFAR is potentially well-placed to explore the dark matter distribution
using weak lensing, as it is able to observe a substantial density of
background sources ($n\simeq 30$ per square arcmin at 200\,MHz, $n\simeq
10$ per square arcmin at 120\,MHz) on which lensing can be measured.
However, galaxy shape information needs to be accurately assessed in this
technique, and this will not be possible with standard LOFAR. A
Netherlands-only LOFAR has a beam size of 3--4\,arcsec at 200MHz ($\sim 6$
arcsec at 120MHz), which is much larger than the typical size of the radio
sources that will dominate the deep LOFAR surveys: \citet{mux05} showed
that the extremely faint radio sources in the Hubble Deep Field and
Flanking Fields (1.4\,GHz flux densities down to $40\mu$Jy) have angular
sizes in the range 0.2 to 3 arcsec.

For this reason, there is a strong motivation from weak lensing studies
for a UK contribution expanding the baseline of LOFAR to $\sim1000$km.
This would lead to a beam size of $\simeq0.5$arcsec at 200MHz, providing
sharper galaxy shape information than is available with any ground-based
optical lensing survey. Even at 120MHz, the beam size would be sub-arcsec,
comparable to the effective PSF for most optical surveys
\citep[c.f. CFHTLS,][]{hoe06}.

In addition, the large survey areas projected for LOFAR will make its
lensing results competitive with the next generation of optical lensing
surveys (e.g., Pan-STARRS). Two LOFAR surveys in particular are of
interest. Firstly, 120 and 200\,MHz surveys covering $2\pi$ steradians,
with reasonable source density and beam size described above, will allow a
full-sky map of the projected dark matter distribution, with an accuracy
equivalent to $\simeq10^{13}M_{\odot}$ on 1Mpc scales at z=0.2.  The shear
power spectrum, simply related to the dark matter power spectrum, will
additionally be measured with S/N$>$5 on {\it each mode} of the power
spectrum from l=60 to 440; averaging over modes allows measurement of the
power to, e.g., one percent accuracy for 10--100Mpc scales at $z=1$.

Secondly, the deeper 200MHz survey covering 250 square degrees is also of
interest: while this will not provide such accurate statistics for the
overall dark matter distribution, the higher number density of sources
makes this survey suitable for 3-D dark matter maps \citep{bac03} and dark
matter evolution studies \citep{bac05}. The size of the survey makes
photometric (e.g., Pan-STARRS deep surveys, WFCAM on UKIRT) and
spectroscopic redshift follow-up plausible for this purpose.

\subsection{LOFAR studies of local galaxies}

\subsubsection{Low frequency observations of nearby starburst galaxies}

Starburst galaxies are galaxies which have a rapid and efficient
star-formation rate which cannot be maintained for the galaxy's
lifetime. The centres of these sources are generally heavily obscured at
optical wavelengths by large reservoirs of dust and gas which fuel the
ongoing star-formation. Observations at radio wavelengths are one of the
few ways in which the on-going physical processes can be studied:
supernovae, supernova remnants and compact H{\sc ii} regions can all be
observed at radio wavelengths. Nearby starburst galaxies provide an ideal
laboratory within which large samples of these sources can be
investigated.

The radio emission from starburst galaxies comprises a mixture of thermal
and non-thermal radiation.  At decimetric and centimetric wavelengths,
the radio spectra of star-forming galaxies is dominated by non-thermal
synchrotron radiation from accelerating charged particles associated with
supernova explosions and supernova remnants. At higher frequencies,
free-free radio emission and emission from dust become dominant. At the
lowest radio frequencies ($\lta$1\,GHz) this radio emission is absorbed by
intervening ionised foreground material, via free-free processes. When
directly compared with higher frequency observations, observations of the
radio emission in the LOFAR band can therefore be used to provide a vital
high resolution probe of the ionised material in the centres of galaxies.

\begin{figure}[t!]
\begin{center}
\begin{tabular}{cc}
\psfig{file=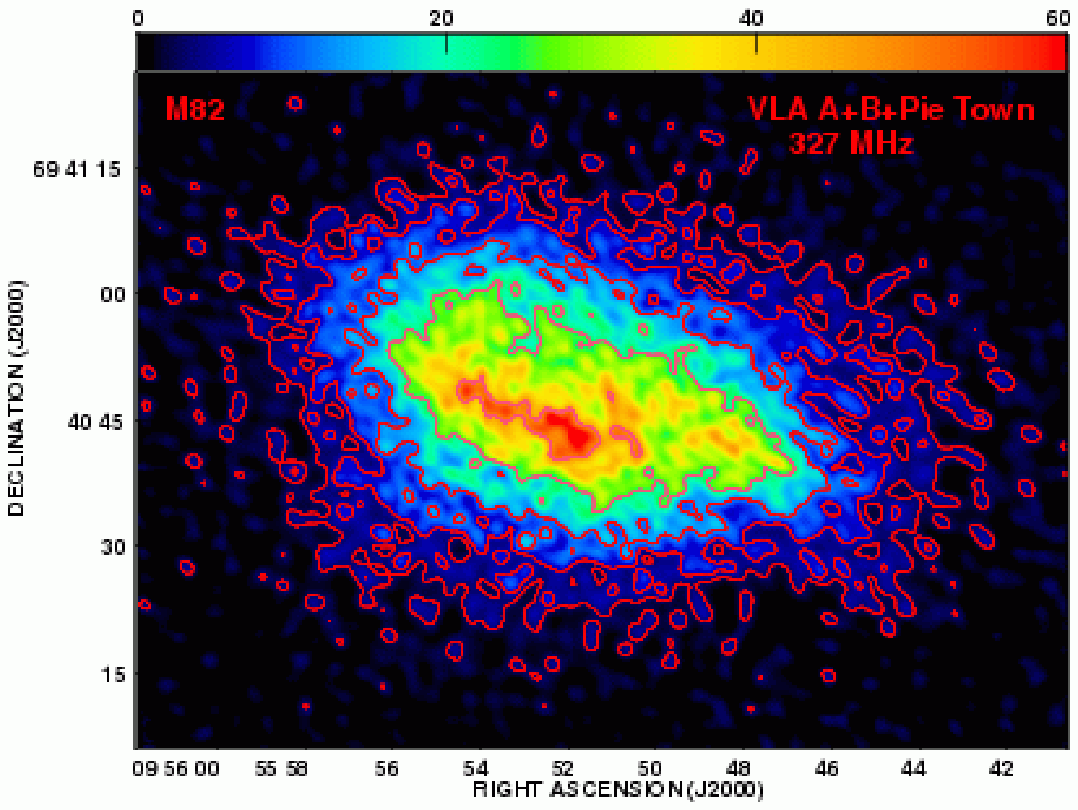,angle=270,width=7.5cm,clip=}
&
\psfig{file=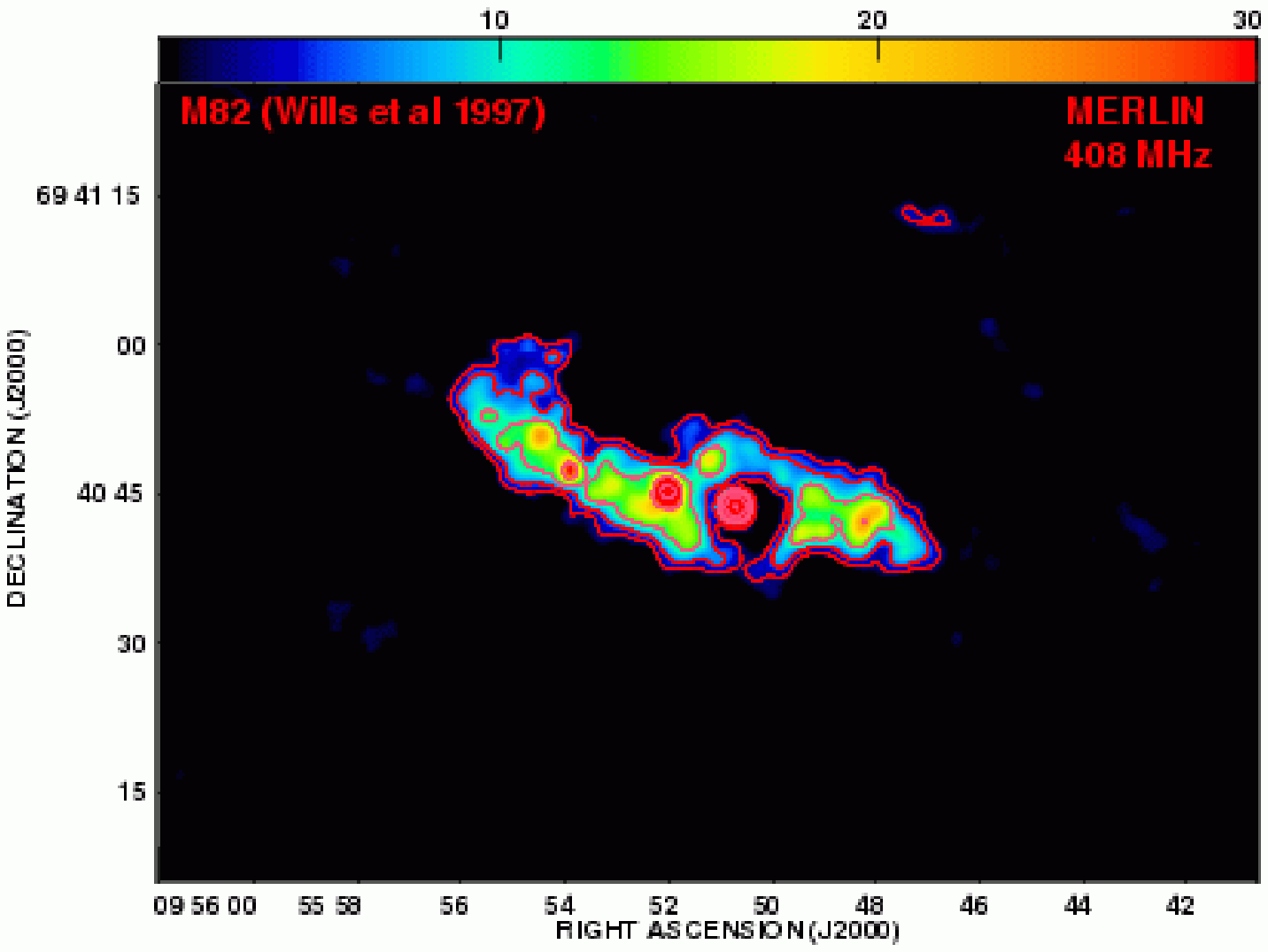,angle=270,width=7.5cm,clip=}
\end{tabular}
\psfig{file=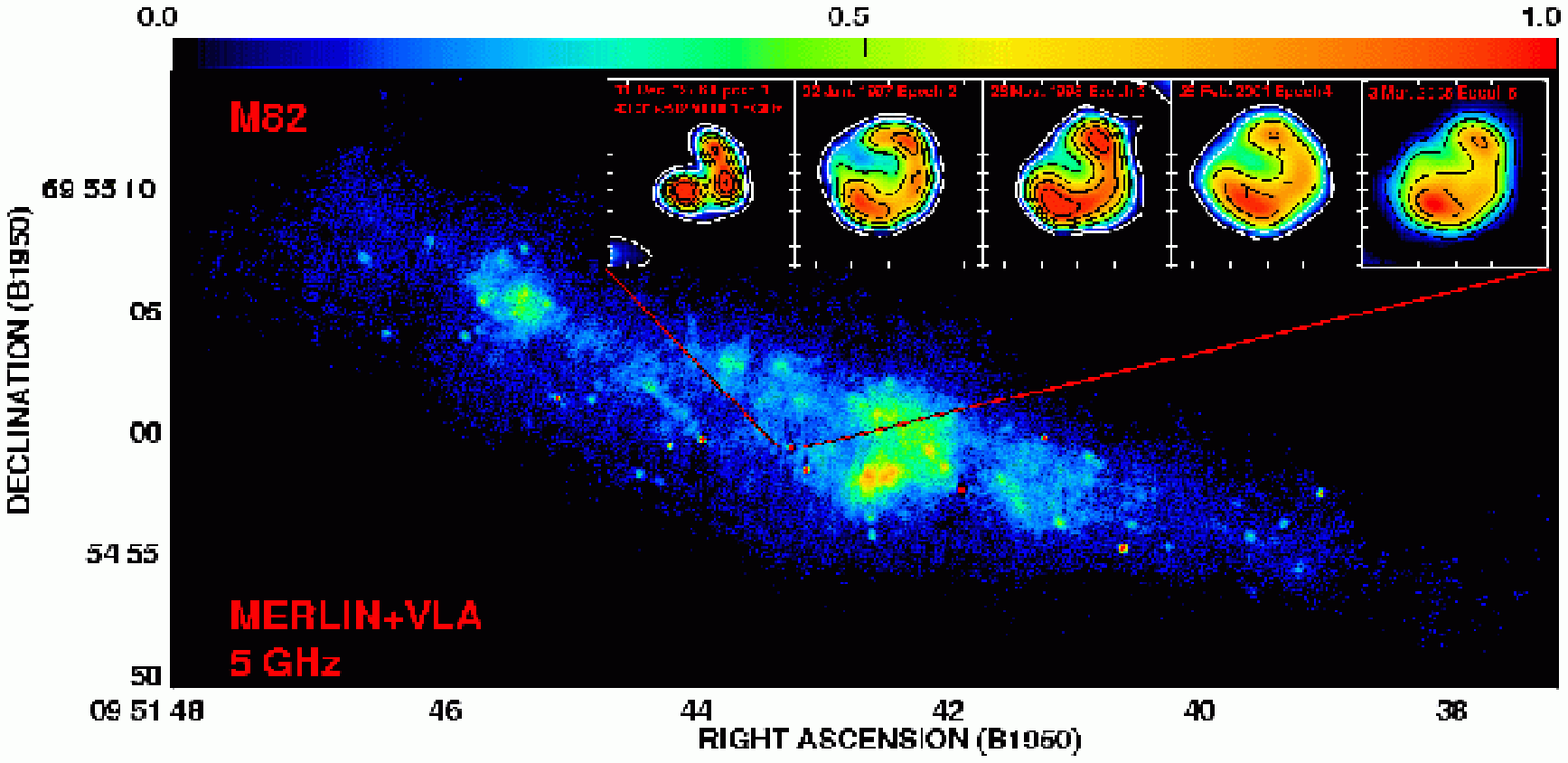,angle=90,width=13cm,clip=}
\vspace*{-0.5cm}
\end{center}
\caption{\small {\it Top Left:} M82 at 327\,MHz with a angular resolution
of $\sim$2 arcsec, observed with the VLA A+B configuration and including
the VLBA antenna at Pie Town \citep{nog03}. {\it Top Right:} A MERLIN
image of M82 at 408\,MHz with an angular resolution of 0.5 arcsec. {\it
Bottom:} A combined MERLIN plus VLA 5\,GHz image of M82. This image is
convolved with a synthesised beam of 100\,milliarcsec. Insert images show
the expansion of the RSNe 43.31+592, as observed with VLBI over the last 20
years.}
\label{M82fig}
\end{figure}

At 1.6\,GHz, e-MERLIN will be able to observe nearby starburst galaxies at
angular resolutions of $\sim$0.2$''$, with $\mu$Jy sensitivities.  LOFAR
will provide similarly sensitive observations with (when international
baselines are included) sub-arcsecond angular resolution, but, crucially,
at previously unexplored low radio frequencies. The direct complementarity
of these two instruments will provide a unique and powerful probe of the
ionised interstellar medium in galaxies. For example, the nearby
(3.2\,Mpc) prototypical starburst galaxy M82 contains $\sim$50 compact
radio sources embedded within a diffuse radio halo (see
Figure~\ref{M82fig}). Currently the deepest, high-resolution, low radio
frequency (327\,MHz) observations made using the VLA, and including the
VLBA antenna at Pie Town, are just able to begin to image the clumpy
nature of the ionised foreground material within the centre of this
starburst \citep[Figure~\ref{M82fig};][]{nog03}. However, these
observations only achieve angular resolutions of $\sim$2 arcsec, which is
not adequate to resolve the ionised gas in fine-detail. At somewhat higher
resolution ($\sim$0.5 arcsec), but with lower sensitivity, MERLIN 408\,MHz
observations of M82 \citep{wil97} were able to separate several of the
compact radio supernova remnants from the diffuse background and hence
allow the free-free absorbing column to be measured against these sources
(Figure~\ref{M82fig}). However, these observations had insufficient
brightness temperature sensitivity to detect all but the brightest
regions of extended emission. These MERLIN and VLA studies represent the
best high-resolution, low-frequency observations that can be achieved with
today's radio telescopes. LOFAR will revolutionise such observations by
providing a large leap in the sensitivity and angular resolution
achievable at the lowest radio frequencies.

LOFAR will be able to image the radio emission of starburst galaxies like
M82 in great detail between 30 and 230\,MHz.  When such observations are
coupled with higher frequency observations, from instruments such as
e-MERLIN, they will enable the free-free absorption of the foreground
ionised gas within these galaxies to be mapped against the background
radio continuum on linear scales of a few parsecs, as well as allowing the
radio spectra of the individual compact radio sources to be determined.

\subsubsection{Jet-powered radio nebulae around extragalactic microquasars}

Relativistic jets in AGN are well-known for inflating lobes filled with a
magnetised relativistic plasma emitting radio synchrotron radiation. X-ray
binaries producing jet flows, so-called microquasars, should in principle
create similar structures when their jets interact with the ISM. A few
microquasars in the Galaxy seem to contain such radio synchrotron lobes
\citep{rod92,mir93,cor02}, which have been interpreted as structures very
similar to the lobes of radio galaxies \citep{beg80}. However, the
conditions in the ISM for the formation of radio synchrotron lobes around
microquasar jets are not always favourable \citep{hei02}. Another method
of detecting the presence of jet-powered lobes is to look for the radio
bremsstrahlung emitted by the shock-compressed, and at least partially
ionised, ISM around the lobe surface. This technique was successful in
detecting the lobe inflated by one of the jets of Cyg X-1
\citep{gal05}. Detection of radio lobes can place important constraints on
the time-averaged energy transport rate of microquasar jets, which are
otherwise not obtainable. In the case of Cyg X-1 it is now understood,
thanks to the discovery of the jet-powered lobe, that the energy
transported by the jets is at least comparable to the energy radiated away
by the material in the accretion disc \citep{gal05}.

Detecting radio lobes in the Galaxy is complicated by the expected large
size of these structures, combined with the large number of confusing
radio sources, e.g., \HII\ regions, in the Galactic plane. The detection
of radio lobes around microquasars in other galaxies, where these problems
are reduced, may be more straightforward. Microquasars are likely to be
associated with regions actively forming stars. Hence, even in other
galaxies, their radio lobes are expected to be located close to confusing
\HII\ regions and supernova remnants. The position of microquasars in
other galaxies can be obtained from X-ray observations. To then identify
the radio lobes as connected to the microquasar, good spatial resolution
is needed. In the case of lobes emitting sufficient synchrotron radiation,
the identification is simplified due to the characteristic spectrum. This,
however, requires good spatial resolution at more than a single
frequency. To date, at least one example of a non-thermal radio source
with properties consistent with an interpretation as a jet-powered radio
lobe has been associated with an ultra-luminous X-ray source in the galaxy
NGC 5408 \citep{sor06}.

The typical properties of the ISM and the expected jet powers of
microquasars infer an expansion speed of the radio lobe of roughly
100\,km\,s$^{-1}$ \citep{kai04}. X-ray binaries containing an accreting
black hole are expected to have typical ages of $10^6$\,years, thus
implying a typical size of the radio lobe of around 100\,pc. Placing such
a source into a nearby galaxy within a distance of 10\,Mpc then suggests a
typical angular size of the lobe of roughly 1$''$. At the highest
observing frequency of LOFAR (240\,MHz) resolving such a radio lobe would
require a baseline of at least 260\,km. Ideally, the lobe should be
resolved at lower frequencies as well, which would require somewhat longer
baselines. Clearly LOFAR stations located in the UK would allow such
observations, while they are impossible with LOFAR restricted to the
Netherlands. Taking the radio nebula in NGC 5408 \citep{sor06} as a guide,
sensitivity is not an issue for these observations as the source has a
flux density of roughly 0.3\,mJy at 4.8\,GHz, suggesting a comfortable
6\,mJy at 240\,MHz (assuming a spectral index of -1, as measured at GHz
frequencies).
 
\subsection{Cosmic Magnetism}

Magnetic fields fill interstellar and intracluster space, and play a
vitally important role in many aspects of astrophysics, from the onset of
star formation to the evolution of galaxies and galaxy clusters. Despite
their importance, little is still known about the origin, structure and
evolution of magnetic fields. When were the first magnetic fields
generated in the Universe? Are they primordial, or was their generation
associated with early structure formation? How did magnetic fields evolve
as galaxies evolve? These are all unanswered questions.

Radio emission offers the best probe of astrophysical magnetic fields. The
intrinsic polarisation of a radio source yields information about the
orientation and degree of ordering of the magnetic field. Faraday rotation
of the polarisation vector as the radio wave passes through the magnetised
medium between the radio source and the observer gives a view of the
magnetic field along the line of sight. However, magnetic field strengths
are typically weak, so only the nearest or brightest objects have so far
been studied. LOFAR's high sensitivity will permit these studies to be
extended into much weaker field regimes, such as galaxy haloes, galaxy
clusters, and the intergalactic medium. LOFAR's multichannel
spectropolarimetric capabilities will be essential in this aim, as they
will enable accurate measurements of rotation measures and intrinsic
polarisation position angles in a single observation in a single (up to
32MHz) frequency band.

Polarisation information in the survey data will allow: Faraday tomography
of the interstellar medium in the Milky Way and in the disks and central
regions of nearby galaxies; studies of the extensions of galaxies into
their halos or intergalactic space, due to the effects of interactions and
galaxy winds; tracing of the full extent of magnetised halos in galaxy
clusters; study of the origin of magnetic fields in the intracluster
medium of galaxy clusters; measuring the magnetic fields in galaxies out
to $z \sim 2$. The UK has considerable expertise in radio polarimetry,
both on galactic and extragalactic scales. In addition to the scientific
results that LOFAR will produce, studying the weak magnetic fields that
LOFAR will see will enable UK researchers to develop the technical and
scientific analysis tools that will be essential for polarisation studies
with the SKA, which will be able to probe to significantly deeper depths.
\clearpage

\section{LOFAR-UK and Radio Transients}
\label{transients}

The very wide field of view of LOFAR, particularly in the low band, makes
it ideal for the discovery and monitoring of variable radio sources. This
has been identified by ASTRON as a key science area for LOFAR, in the form
of the Transients Key Project \citep{fen06}. The potential temporal scales
are from microseconds to years.

The UK has a strong history and interest in variable radio sources,
primarily focussed around high-energy astrophysical phenomena associated
with relativistic objects such as neutron stars and black holes. The UK
can also bring considerable additional benefits to the transients project,
in particular due their access to the RoboNet Telescopes (the Liverpool
and Faulkes telescope) which are able to acquire transient sources such as
Gamma Ray Bursts within one minute of an alert being generated. RoboNet is
coordinated by the Liverpool John Moores University (LJMU), which is part
of the LOFAR-UK consortium. LJMU has considerable access to these
telescopes, which is offered for optical follow-up of any LOFAR transients
as part of LOFAR-UK's ``in-kind" contribution. LOFAR-UK anticipates a clear
and productive working relationship with the Dutch LOFAR Transients Key
Project team.

\subsection{X-ray binaries / microquasars}

X-ray binaries are binary systems in which accretion of material from a
more or less `normal' companion star onto a collapsed relativistic object
(neutron star or black hole) results in an enormous release of
gravitational potential energy in the form of both radiation and powerful
relativistic jets. These jets are ubiquitously associated with radio
emission, resulting from shock-accelerated leptons spiralling in magnetic
fields and producing synchrotron emission. Clear patterns of behaviour in
the radio band have been linked to changes in the luminosity and `state'
(geometry and radiative efficiency of the accretion flow) of the accretor
\citep[e.g.,][]{fen04}. Studying such objects not only provides us with
valuable insight into the physics of accretion and jet formation, but has
been shown to be scaleable to accretion onto Active Galactic Nuclei
\citep[e.g.,][]{mch06}.

Observations of X-ray binaries with LOFAR can proceed in two ways:

\begin{description}
\item[(a)] discovery of new X-ray binary transients in the radio band.
\item[(b)] daily monitoring of known sources (including those discovered by
LOFAR itself), providing a unique resource.
\end{description}

Procedures for the detection and identification of new X-ray binary
transients with LOFAR are currently under development by the LOFAR
transients project team. LOFAR will be able to both scan a large fraction
of the sky daily {\em and} localise new transients with arcsec accuracy,
by a stepwise ramping up of baselines from core to full array, and
frequency from 30 to 240 MHz. It may well become the most productive
source for the discovery of new transients, following the likely demise of
the all-sky monitor (ASM) on-board the RXTE mission within two years.

\begin{figure}[!t]
\begin{center}
\psfig{file=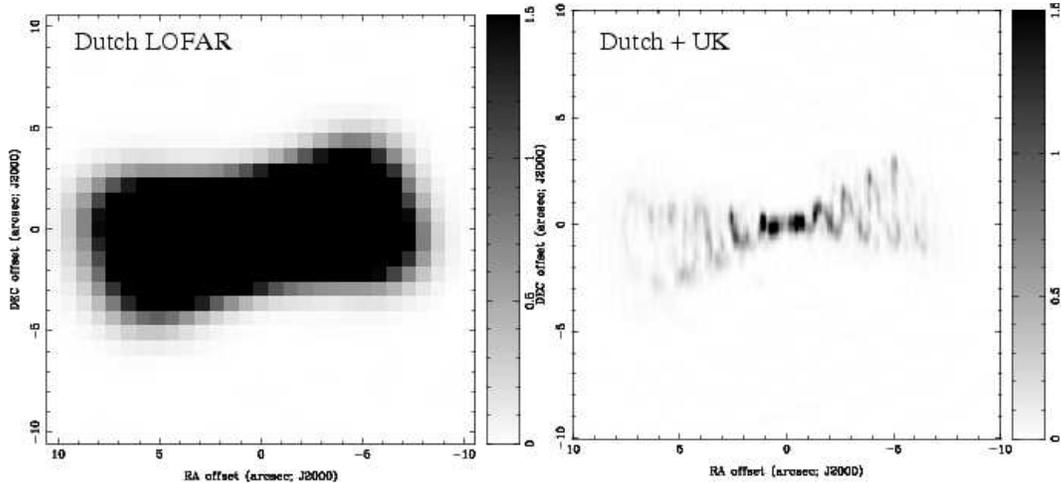,width=14cm,clip=}
\end{center}
\caption{\label{jetsourcefig} \small A simulated image of the SS433 jet,
based on a 4hr exposure with a Netherlands-only LOFAR (left panel), and
with the addition of UK baselines (right panel). Credit: Mark Hill and
Faye Cashman.}
\end{figure}

The international baselines provided by the UK will be primarily of use in
imaging the transient radio structures associated with relativistic
ejection events (cf. Figure~\ref{jetsourcefig}). Such events have been
mapped on angular scales from milli-arcsec (with GHz VLBI, and beyond the
capabilities of even an Europe-wide LOFAR) to several arcmin. Regularly
tracking the ejecta to large angular scales will allow the measurement of
deceleration and shock acceleration as the jet and ISM interact and
exchange energy.

UK expertise in the variable radio counterparts of X-ray binaries (also
known as `microquasars') includes phenomenological models of disc-jet
coupling, and theory and modelling of outbursts and jet
formation. Observational support for UK involvement will include
observations of new transients with e-MERLIN and optical to X--ray follow
up with RoboNet and SWIFT.

\subsection{AGN outbursts and variability}

LOFAR will open up a new window for studying radio-emission from AGN in
the time domain.  Whilst at low frequencies the radio emission from most
AGN is dominated by large-scale steep-spectrum lobes, which will be very
steady in the LOFAR band on timescales of decades or longer, some AGN are
highly variable on much shorter timescales (sometimes less than days).  In
some cases this is attributable to scintillation and/or relativistic
beaming, but in other cases it is clear that variability is physically
associated with the accretion/jet process itself. It is now thought that
AGN jets play a crucial role in regulating the growth of massive galaxies
and galaxy clusters, and variability is an essential tool to probe these
jets close to their origin.

Using two bands in the $\sim$hundred MHz range, LOFAR can cleanly separate
the wavelength-dependent interstellar scintillation effects from intrinsic
jet variability.  LOFAR's sub-mJy sensitivity on day time-scales will
enable, for the first time, the study of the variability of faint AGN jets
known to exist in `radio-quiet' AGN such as Seyfert galaxies, radio-quiet
quasars and low-luminosity AGN (LLAGN), as well as hundreds of more
distant and powerful radio-loud objects.  X-ray studies of AGN have led to
the identification of characteristic time-scales of variability, which
scale with black hole mass and inversely with accretion rate
\citep{mch06}.  With LOFAR it will be determined if similar scalings hold
in the radio for thousands of AGN, allowing definitive tests of how jet
sizes scale with mass and accretion rate; this will provide the key
physical parameters for models of how jets affect the galactic and
intergalactic environments.  LOFAR will also be sensitive to variability
in the relativistically beamed jets of hundreds of blazars to large
look-back times. This will determine whether jet variability evolves with
redshift, as might be expected if the black holes are still growing, or
their environment is changing.

As well as studying variability in the known AGN population, LOFAR will be
sensitive to transient extragalactic sources which are impossible to find
using conventional radio telescopes.  For example, it has long been
suspected that the `quiescent' supermassive black holes (SMBH) harboured
by every galaxy should occasionally flare up as they tidally disrupt, and
then accrete, material from a star that has wandered too close.  These
events are sufficiently rare that, to date, only a handful of candidate
events have been seen (in the X-ray band), in otherwise `normal' galaxies.
Given that accreting SMBH (i.e. AGN) are known to emit in the radio in the
form of jets, it is likely that LOFAR will be sensitive to these stellar
tidal disruption events.  Tidal disruption rates are a strong function of
black hole mass, and the stellar environment: measurement of these event
rates will provide constraints on models for the environments close to
SMBH, as well as the SMBH mass function in the local Universe.  The events
themselves will be of special astrophysical interest, revealing how the
accretion of the stellar remnants unfolds.  The flaring emission should
last for at most a few years, but can rise from zero within weeks, making
LOFAR ideally suited to detecting these events compared to more sparsely
sampled surveys.  Given the expected frequency of stellar tidal disruption
events in the local Universe, and assuming the same scaling of radio and
X-ray luminosity as seen in normal AGN \citep[e.g.][]{mer03}, it is
expected that LOFAR might detect a few tens of these events each
year. This would revolutionise this whole area of research.

\subsection{Gamma-ray bursts}

Gamma-ray bursts (GRBs) are amongst the most luminous events in the
Universe. They are short flashes of gamma rays, coming from random
directions, that instantaneously outshine every other gamma-ray source in
the sky including the Sun. They are currently detected at the rate of
about one per day, by all-sky monitors on orbiting satellites. They are
often followed by ``afterglow" emission at longer wavelengths, from
X--rays through to radio wavelengths.

The phenomenologically rich radio afterglows of GRBs arise from
synchrotron emission from electrons accelerated in shocks, a consequence
of the collision between an ultra-relativistic outflow from the GRB event
-- the ``fireball" -- and the external medium into which it expands.  As
such, it is a crucial indicator of the physical parameters of the blast
wave, providing constraints on the energetics of the explosion itself and
the density and structure of the circumburst medium with which it
interacts \citep[e.g.,][]{wij99}.  Despite synchrotron self-absorption
limiting the low-frequency visibility of the afterglow at early times, a
combination of factors nevertheless work in LOFAR's favour as an
instrument for detecting and following-up GRBs.

\begin{figure}[!h]
\centerline{
\psfig{file=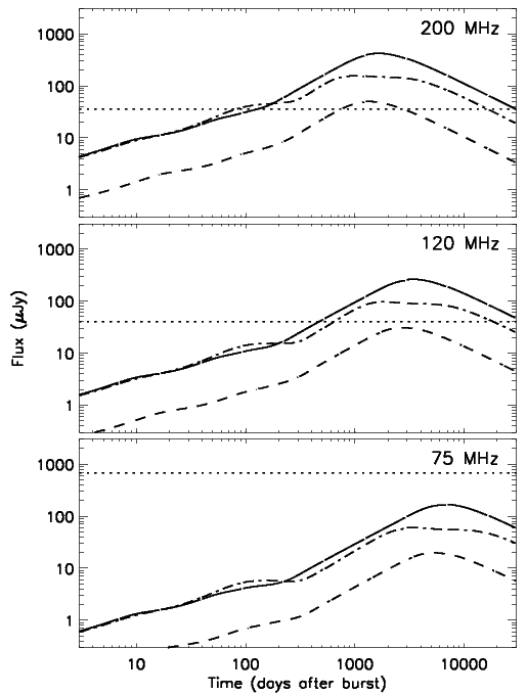,width=7cm,clip=}
}
\caption{\label{grbfig} \small The predicted light curve of the $z=0.16$
GRB 030329 at three frequencies within the LOFAR observing range
\citep[from][]{vdh07}. The solid and dash-dot line correspond to two
different models for the afterglow emission, whilst the dashed line gives
the corresponding light curve if the GRB were located at $z=1$. The
horizontal dotted line shows the limiting sensitivity of LOFAR in a four
hour integration.}
\end{figure}

Firstly, the afterglow evolves on a much longer timescale at lower
frequencies than in the optical and X-ray regimes. Therefore, not only is
one afforded a ``slow-motion" view of events such as the ``prompt
emission" -- an early ($\sim$1--2 day) flare due to a reverse shock
\citep[e.g.,][]{kul99} whose optical equivalent occurs too rapidly
($\sim$1 minute) for systematic study -- but also the evolution can be
tracked out to much later times than at high frequencies.  As the fireball
expands, the afterglow spectral peak shifts to longer wavelengths at later
times. Thus the radio afterglow does not immediately exhibit the power-law
decay seen in the optical and X-rays, but rises to a peak and declines
slowly thereafter.  Indeed, at the lowest frequencies, the light curve
does not peak until months or years after the burst. This means that LOFAR
will be able to track the brightest afterglows, such as GRB030329
\citep[Figure~\ref{grbfig};][]{vdh07}, for years after they have ceased to
become visible in the optical.

Secondly, the radio component is relatively insensitive to the geometry of
the relativistic fireball \citep{fra00}.  It is thus possible to detect
the more isotropic radio emission from GRBs whose high energy emission is
confined to a beam that, in the majority of cases, lies out of our line of
sight. Models predict that the true rate could outnumber the observed
(beamed) rate by a factor $\sim$50--100 \citep[e.g.,][]{gue05}. The
all-sky monitoring capability of LOFAR thus provides an unbiased census of
GRB beaming statistics, and a measure of their true rate.

There is also an intriguing possibility that the rate of low luminosity
(and hence low redshift) bursts could be much larger (of order 100 times
larger) than that of classical cosmological GRBs \citep{pia06,sod06}. Such
a low-z burst would not be seen in gamma-rays but, even if the GRB itself
is not seen (and consequently no X-ray or optical afterglow is found), it
is possible to detect the radio emission from the afterglow since, at late
times, the radio emission is isotropic.  Emission line spectra of the host
galaxies will be within reach of 8m class telescope.  This would confirm
the low redshift nature of these bursts and establish whether or not they
represent a class distinct from cosmological GRBs.

Bright, nearby bursts such as GRB030329 ($z$=0.1685) are expected to peak,
at 100--200MHz, 5--10 years after the event itself, and should be
detectable in this frequency range with LOFAR (Figure~\ref{grbfig}).  At
increasing redshift, an ``inverse $k$-correction", analogous to that
encountered for sub-mm galaxies, compensates to some extent for luminosity
distance dimming, as one probes a higher rest-frequency and an earlier
time in the light curve \citep{cia00}.  Thus, although events of the
magnitude of GRB030329 are rare, the lack of dependence of radio emission
on beaming geometry, combined with their straightforward detectability out
to $z\sim1$ (Figure~\ref{grbfig}) and the protracted evolution at low
frequency, make it feasible that LOFAR could detect as many as 20 such
bursts per year.  It will also be possible to obtain accurate localisation
of past bursts that have, in the LOFAR bands, yet to reach their
peak. This will enable the study of afterglows that were missed in the
optical, and therefore allow a statistical assessment of whether the
faintness of these ``dark bursts" was intrinsic, or due merely to survey
incompleteness.  The long baselines provided by UK LOFAR stations would be
particularly important in this respect, to provide high positional
accuracy for the bursts.

The progenitors of short-duration GRBs are also predicted to give rise to
radio flares, detection of which would be a potential test of progenitor
models.  The majority of short bursts are thought to result from the
coalescence of either neutron-star-neutron-star or neutron-star-black-hole
systems. Low frequency radio flares, which may be detectable by LOFAR, are
predicted to arise from such systems in two ways: (1) currents induced by
the motion, prior to coalescence, of one neutron star through the magnetic
field due to the other \citep{han01}; (2) variations in the surface
currents associated with the relativistic magnetised wind flowing from the
binary, which, though peaking at $\sim$1MHz, may have a high-frequency
tail that is detectable in the low-frequency LOFAR bands \citep{uso00}.
UK researchers have developed very advanced numerical simulations of the
merging of these binary systems of compact stars \citep[e.g.,][]{ruf01},
and detailed LOFAR observations will enable testing and refinement of
these models.

\subsection{Pulsars and related phenomena}

Pulsars are steep spectrum objects whose pulsed flux density usually peaks
in the 100--200\,MHz range. With an unprecedented sensitivity in exactly
this frequency range, LOFAR will make an excellent instrument for studies
of pulsars and their use as galactic probes. Moreover, as the emission
process of pulsars is still only poorly understood, the observed radiation
properties, as well as simply the presence or absence of low-frequency
emission from pulsars, will serve as powerful constraints on models of the
pulse emission process.

How many pulsars LOFAR will discover depends on the low-end of the pulsar
luminosity function. There are indications that the luminosity function
turns over in the range 0.3--1\,mJy\,kpc$^2$, but at the same time it is
likely that existing (high-frequency) surveys are already incomplete at
the 10\,mJy\,kpc$^2$ level. The extremely high sensitivity of LOFAR will
be able to measure the low-end of the pulsar luminosity function
significantly better than any previous survey.  Understanding this
function is vital for our knowledge of how many pulsars there are, which
in turn constrains the massive star population and supernova rate in the
Galaxy.

\begin{figure}[!bt]
\centerline{
\psfig{file=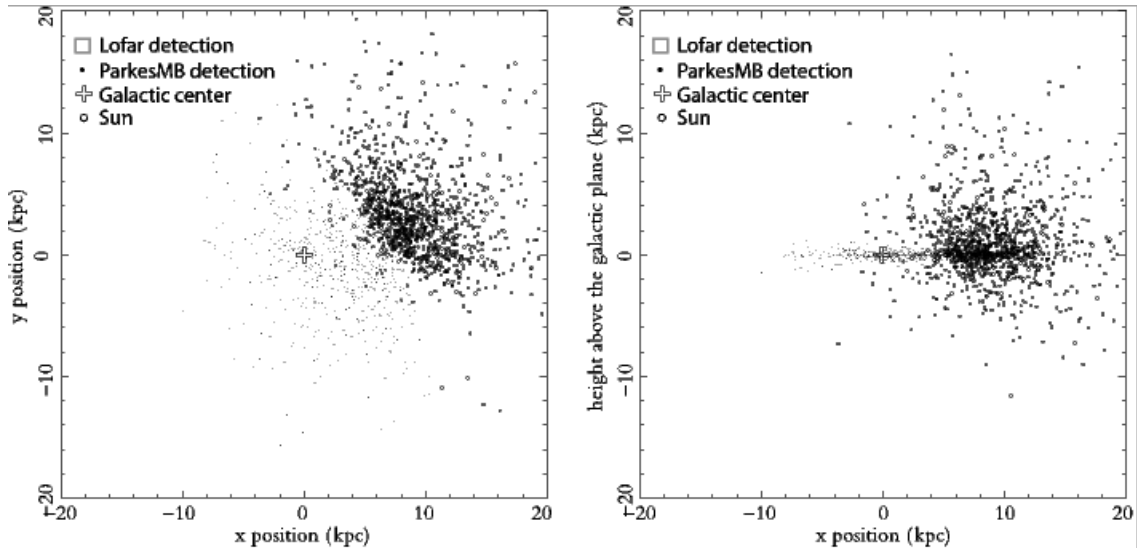,width=15cm,clip=}
}
\caption{\label{pulfig2}  \small Simulated detections of pulsars in the LOFAR
surveys (compared to Parkes Multibeam) for 1-hour LOFAR pointings. The
left panel shows these projected on the Galactic plane, whilst the right
panel shows them projected on the plane through the Galactic centre and
sun, perpendicular to the disk \citep[from][]{sta07}.}
\end{figure}

Recent pulsar surveys have moved to higher and higher frequencies to
escape the deleterious effects of dispersion due to the passage of the
pulses through the interstellar medium. However, the natural frequency
decimation of LOFAR, combined with the significant processing power
available, means that these effects can be limited. This means that
scattering in the interstellar medium will be the limiting factor in
determining the distance to which pulsars are seen with LOFAR. Even with
the most conservative estimates, one finds that a pulsar survey with LOFAR
of the entire Galaxy (i.e., those parts visible from Northern Europe) will
find $\approx$1500 new pulsars, almost doubling the total number of
pulsars known (see Figure~\ref{pulfig2}). This large addition to the known
population will allow us to better test the period distribution of
pulsars, which is an important ingredient in understanding supernova
physics and the physics of neutron stars. LOFAR will also provide an
improved spatial distribution of pulsars in the Galaxy -- this is in
particular true in the Northern hemisphere -- and thus at distances well
above and below the Galactic plane.

\begin{figure}[!tb]
\centerline{
\psfig{file=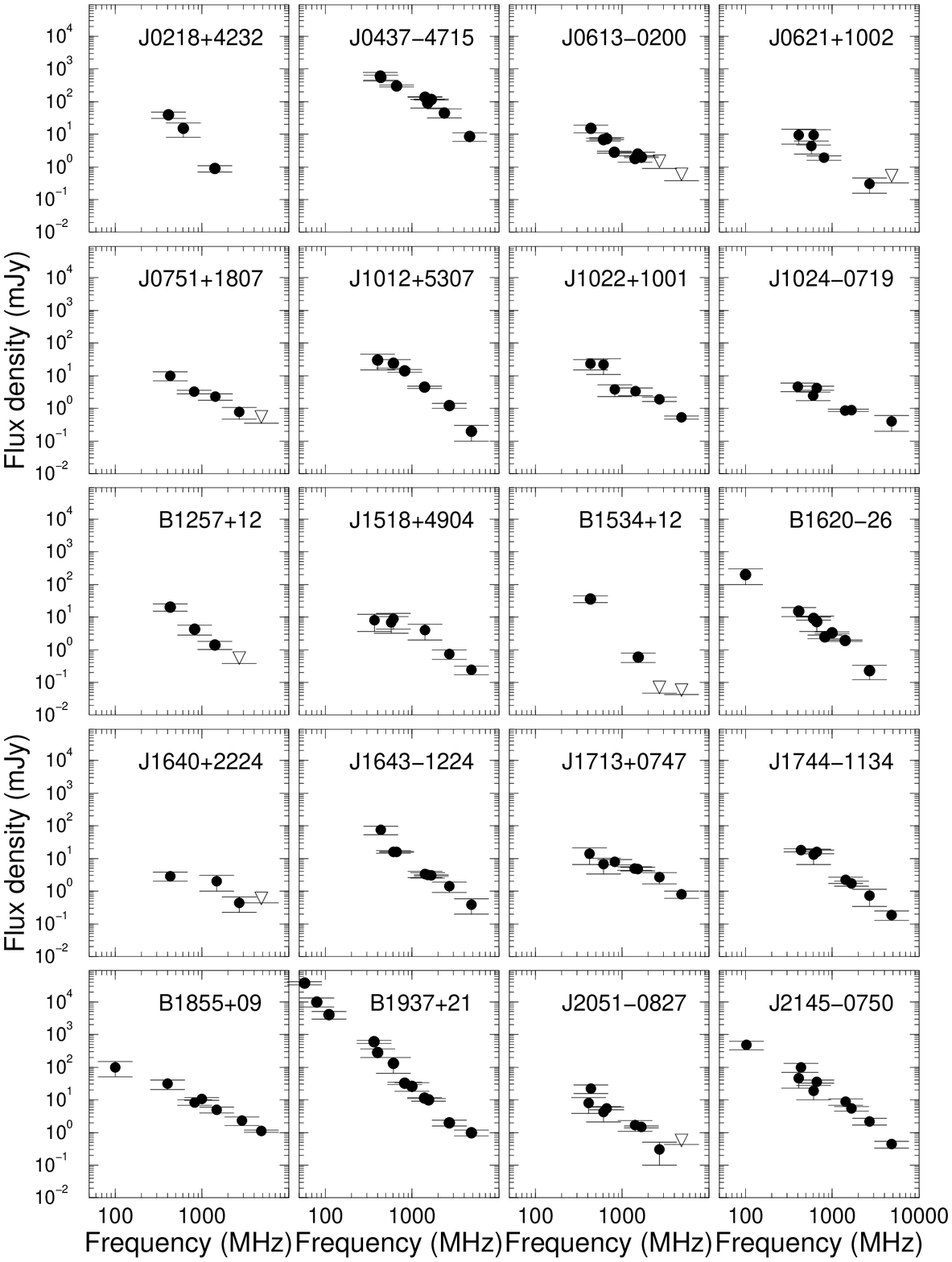,width=16cm,clip=}
}
\caption{\label{pulfig1} \small Example flux density spectra of
millisecond pulsars \citep[from][]{kra99}, demonstrating that the flux
density spectrum of many millisecond pulsars seems to continue rising to
the lowest frequencies.}
\end{figure}

The apparent detection of a few pulsars only at frequencies near or below
100 MHz, and the existence of pulsars like B0943+10 (which have
flux-density spectra with a spectral index steeper than -4) and
millisecond pulsars (which have steep spectra which do not turnover even
down to frequencies of 30 MHz; see Figure~\ref{pulfig1}) suggests that
there is a large number of pulsars that are detectable only at low
frequencies. This behaviour could be either intrinsic to the emission
mechanism of pulsars or due to geometrical effects. The latter reason
seems more likely, given that the pulsar emission cone width increases at
lower frequencies. Indeed, the increase in the pulsar emission cone size
could become quite significant at low frequencies, leading to a large
increase in the `beaming fraction' of the illuminated sky near 100
MHz. Thus, pulsars which are not beamed towards us at higher frequencies
may well be detectable at low frequencies. If there is a significant
population of such sources then LOFAR will be in a unique position to
detect them. Detection of these pulsars would provide an improved
understanding of the emission process and a clearer understanding of the
total population of radio pulsars.  Moreover, the shear number of pulsars
that will be discovered by LOFAR provides for the possibility that the
sample will include some exotic systems, like double neutron stars, double
pulsars, or pulsars with planets.

\begin{figure}[!tb]
\centerline{\psfig{file=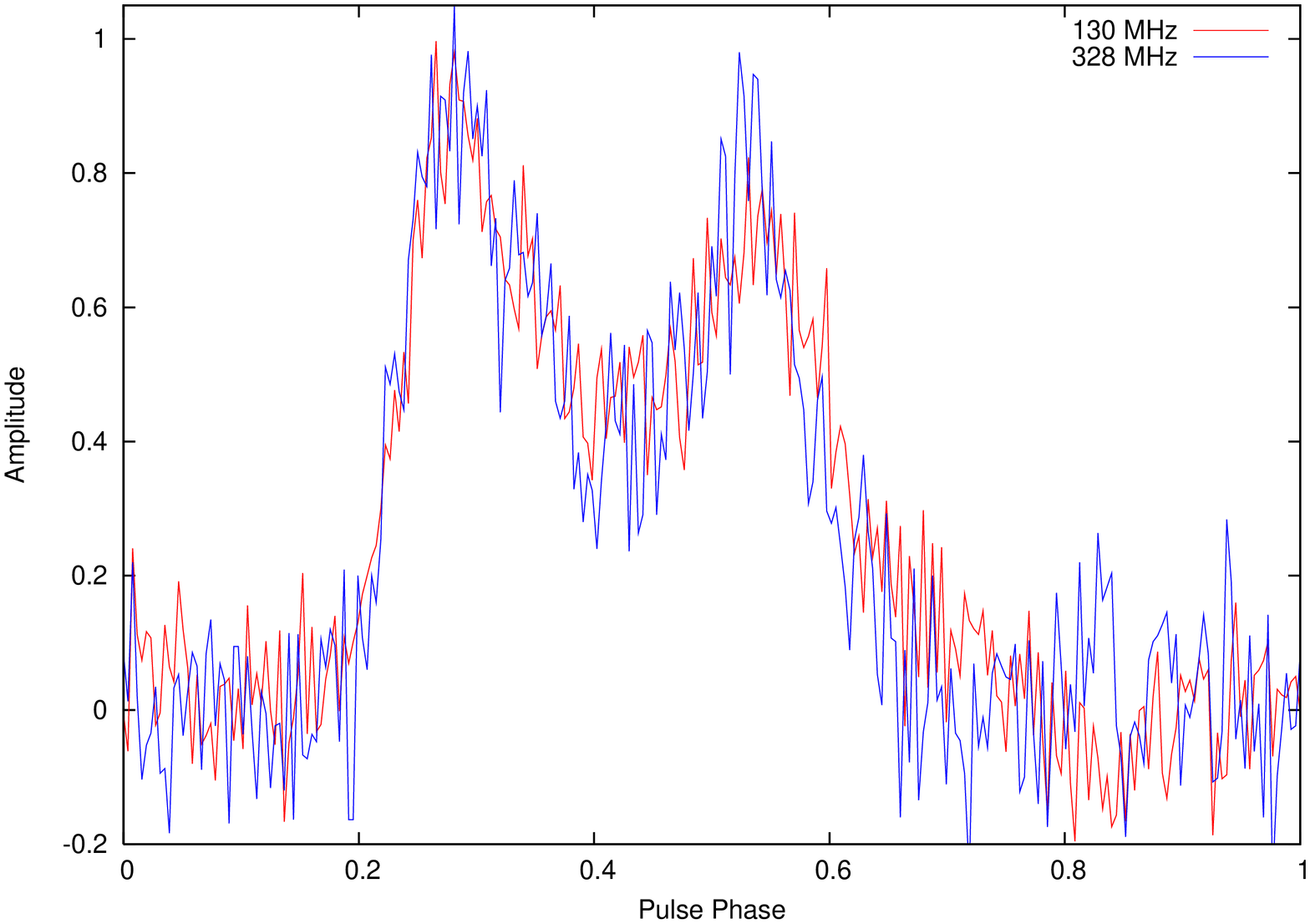,angle=-90,width=10cm,clip=}}
\caption{\label{pulseprof} \small The low frequency pulse profile of a
millisecond pulsar as observed with Westerbork, demonstrating that low
frequency observations can be comparable to, or even better than, high
frequency observations.}
\end{figure}

The average pulse profiles of radio pulsars are made up of the
superposition of a large number of sub-pulses. These sub-pulses are
believed to have a very strong link to the emission process. In some
pulsars these sub-pulses are observed to drift in an organised fashion
through the pulse window and this is thought to be due to properties
within the pulsar magnetosphere. Some pulsars show significant frequency
evolution in the properties of these drifting sub-pulses and observations
at low-frequencies, which probe a very different sight line across the
emission, can be used to reconstruct the distribution of emission within
the magnetosphere. Greater constraints can be obtained by simultaneous
observations at different frequencies across the LOFAR bands, and in
combination with high frequency facilities. This will reveal how the
``radius-to-frequency mapping'' (i.e., the dependence of emission height
on radio frequency) manifests itself in the drifting sub-pulses, and will
determine whether the drifting is more or less organised at the low
frequencies. There is also some evidence that both pulse intensities and
drifting are less stable at low frequencies. The greatly increased sample
of pulsars for which single pulse studies with LOFAR are possible will
help to confirm or refute this evidence.

Single pulses from radio pulsars offer the best view of the emission
process. The fact that LOFAR will enable the detection of single pulses
from so many more sources is a great step forward. Low frequencies are
particularly interesting because micro-structure (quasi-periodic emission
seen in some pulsars, with periods and widths of the order of
microseconds) and sub-pulses tend to be stronger at low frequencies
(cf. Figure~\ref{pulseprof}), where the density imbalance and plasma
dynamics are expected to be the most noticeable. Observations such as
these have potentially high payoff, because they come closest to the
timescales and predictions that can be made by theoretical models of the
emission region. The high time resolution requirement of these
observations will limit the sources because of interstellar scattering,
but there are still many tens of sources for which studies can be made.

Pulsars are excellent probes of the ionised component of the interstellar
medium through scintillation, dispersion measure, and Faraday rotation
studies.  Scintillation studies have been revolutionised in the last five
years by the discovery of faint halos of scattered light extending out to
10--50 times the width of the core of the scattered image.  This, in turn,
gives a wide-angle view of the scattering medium with milli-arcsecond
resolution, and the illuminated patch scans rapidly across the scattering
material because of the high pulsar space velocity.  Some of the most
interesting effects are visible at low frequencies, and LOFAR's
combination of sensitivity, frequency coverage, and signal flexibility are
an excellent match to this new science.  As such, LOFAR will also make
important contributions to traditional dispersion measure and rotation
measure determinations.  By almost doubling the number of known pulsars,
the proposed survey will add a dense grid of new sight lines through the
Galaxy to those that already exist. Combining the dispersion measures of
this new sample with those already known will improve our global model of
the distribution of the ionised ISM. Putting this together with the
rotation measures of a large fraction of the new and known pulsars will
place important constraints on the overall magnetic field structure of the
Milky Way, which is still not well characterised. At LOFAR frequencies it
is also possible to determine the very small rotation measures of the
nearby population of pulsars, providing an unprecedented tool for studying
the local magnetic field structure. The large number of new sight lines
will also allow statistical studies of small-scale fluctuations of
electron density and magnetic field variations.

\subsubsection{Extragalactic pulsars}

\begin{figure}
\vspace*{2cm}

{\Large{\bf See attached JPG file; 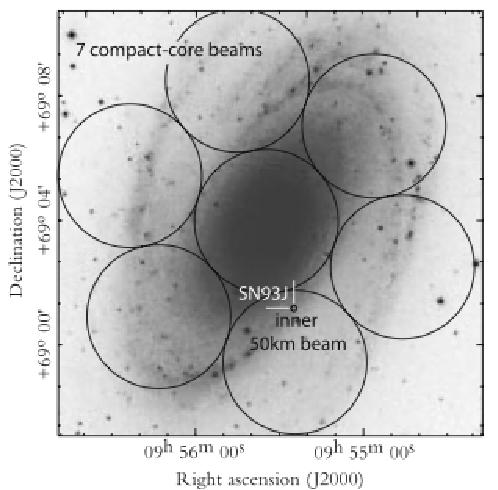}}
\vspace*{2cm}

\caption{ \small Coherently formed beams from the compact core and the
inner 50-km of LOFAR, projected on the core of M81, the second best
candidate galaxy for an extragalactic survey and the home of supernova
1993J.
\label{M33}}
\end{figure}

Due to its sensitivity, LOFAR can be the first telescope to find
extragalactic pulsars, besides those in the Magellanic Clouds
(cf. Figure~\ref{M33}).  If spiral and irregular galaxies (which will host
young, bright, Crab-like pulsars) are observed face-on and located away
from the Galactic disk, the scatter broadening will be relatively low and
a LOFAR survey will have excellent sensitivity for pulsars of all spin
periods. There are at least 20 galaxies for which LOFAR will have good
sensitivity to their pulsar population: for a relatively close galaxy like
M33, LOFAR could detect all pulsars more luminous than ~50Jy kpc$^2$,
above which level the Milky Way hosts 10 pulsars. Complementary to this
normal pulsar emission, some pulsars show ultra-bright `giant pulses' that
could be visible in even more remote galaxies \citep[e.g.][]{mcl03}.

A survey for extragalactic pulsars would investigate whether the bright
end of the pulsar distribution in other galaxies differs from that in our
galaxy, and how the pulsar distribution depends upon galaxy type and star
formation history. Extragalactic pulsars can also help the understanding
of the missing baryon problem, the history of massive star formation in
these galaxies and also, if sufficient numbers can be found, they can be
used to probe the turbulent intergalactic medium.

\subsection{Extrasolar planets}

Extrasolar planets is one of the most dynamic and exciting areas of
astrophysics, engaging both astronomers and the general public. New
discoveries as to the structure and formation of extrasolar planets are
being made at a rapid rate, and LOFAR will make important advances in this
field, particularly concerning the magnetic field and magnetosphere of
extrasolar planets.

As of March 2007, there are over 200 known extrasolar planets, with orbital
periods ranging from just over 1 day up to several years. The derived
planetary masses range from around 6 Earth masses up to the brown dwarf
limit. A significant number of these planets also have high orbital
eccentricities, and it is clear that many of the extrasolar planetary
systems are rather different from the solar system.

LOFAR is expected to detect the magnetospheres of extrasolar planets, and
this will offer a unique chance to investigate the magnetic field
strengths of extrasolar planets, planetary rotation (which may be very
difficult to access via any other means), and also possibly the presence
of moons orbiting the planet. This will be done based on extrapolations
from the solar system, where 5 planets (the Earth and the 4 giant planets)
have been detected at low frequencies, either from the ground or from
space-based observations.  The radio emission from the solar system
planets is coupled to the solar wind, and the radio flux is proportional
to the amount of solar wind incident on the planetary magnetosphere.
Consequently, the level of radio emission depends on planetary parameters,
such as the magnetic moment and the orbital period (which determines how
much solar wind is incident on the magnetosphere), and also stellar
parameters (such as the solar wind mass-loss rate and wind velocity).

The low frequency radio emission is due to electron cyclotron maser
emission, with an upper cut-off frequency determined by the magnetic field
strength close to the surface of the planet.  For Jupiter, emission is
observed extending up to around 40 MHz, and the Jovian low frequency
emission was one of the early surprises of radio astronomy.  For the other
planets the emission is at frequencies below the ionospheric cut-off
frequency for the Earth, and these have only been detected from space. The
frequency of emission from extrasolar planets is expected to be related to
planetary mass, with higher mass planets having emission extending up to
higher frequencies. Further to this, the electron cyclotron maser may also
play a role in the poorly understood radio emission from brown dwarfs
\citep{hal06}, where the masses are higher and the emission is at GHz
frequencies.

The level of radio flux can be easily estimated by using the measured flux
from the solar system objects (most notably Jupiter) and using parameters
appropriate for the extrasolar planetary systems \citep[see, for
example,][]{laz04,ste05}. The brightest exoplanets are expected to have
fluxes in the range of a few mJy, with a substantial number having fluxes
accessible with LOFAR. The highest level of emission is expected to come
from massive planets, orbiting at short periods around stars with strong
stellar winds (which translates to younger stars). The very short period
planets will be tidally locked and this may affect the level of radio
emission (and this would be one aspect where LOFAR will make a
contribution).

In addition to studying known extrasolar planets, there is also the aspect
of serendipitous detection of extrasolar planets. The emission should be
primarily in the range of 10-150\,MHz, be broad-band in nature, and also
show strong bursts. There have been several searches for extrasolar
planets at 150MHz (using the GMRT) and the VLA (74MHz) but no detections
have been made \citep{far02b,geo07}. One possible problem is
distinguishing between emission from the star and from the planet. This
will be achievable because the planetary emission is expected to be highly
polarised, while the stellar emission will not be.

The initial goal of using LOFAR in the study of extrasolar planets will be
to obtain a number of detections of the known extrasolar planets. Using
these, the systematics of low frequency radio emission from planets
outside our solar system can be investigated.  Once a number of
measurements of each planet have been obtained then, by analogy with
Jupiter, it will be possible to investigate the presence of periodicities,
which are indicators of planetary rotation and the presence of moons (Io
in the case of Jupiter). Whether these can be disentangled without
additional information may be challenging.  The wide spectral band of
LOFAR will be important in understanding the magnetospheres of extrasolar
planets. Identifying the high frequency emission cut-off will determine
the magnetic field strength and magnetic moment of extrasolar planets
(assuming the spectral characteristics are similar to those of solar
system objects), which are not accessible by other means.

In summary, it is confidently expected that LOFAR will detect a number of
extrasolar planets, opening up a new field of study. These detections will
enable extrasolar planets to be investigated in ways that are simply not
possible at other wavelengths. In addition, the low frequency radio
emission from extrasolar planets is likely to throw up a number of
surprises.

\subsection{Search for Extraterrestrial Intelligence}

It is generally believed that communications from an extraterrestrial
intelligence will come in two forms. The first may be a civilisation
more advanced than ours, that has the technology and power to broadcast
signals across the Galaxy specifically for others to detect (a
``beacon"). The second are technologically younger civilisations, like
ours, that are just beginning to use advanced communications which are
leaking into space for others to eavesdrop on. As outlined in
\citet{pen04}, radio frequencies are the most natural place to look
for both types of signal and, therefore, a majority of past, present
and future searches for extraterrestrial intelligence are in the radio
(microwave) range of the electromagnetic spectrum.

The most ambitious search for extraterrestrial beacons was the Phoenix
Project, which ran for nearly ten years (from 1995 to 2004), and observed
800 stars (out to 240 light years) with Arecibo, Parkes and the Green Bank
Telescopes (1.2 - 3 GHz ranges). Alternatively, the BETA project used a
26m radio telescope to perform an all-sky, narrow-band, microwave search
for extraterrestrial beacons in the so-called ``Water Hole" from 1400 to
1720 MHz (a radio--quiet region between the hydrogen line and the
strongest hydroxyl line). In both cases, no unusual radio signals were
detected.

The future for the Search for Extraterrestrial Intelligence (SETI) is to
push beyond our nearest stars, and survey a much larger volume of our
Galaxy. We already know from the detection of hundreds of extra--solar
planets, that other planetary systems exist relatively nearby, e.g., the
nearest known extra-solar planet is only 10.5 light years away around a
sun-like star called Epsilon Eridani. The number of such systems will
increase rapidly with new missions like Kepler, which is predicted to find
up to 640 inner-orbit planets by monitoring $10^5$ nearby main--sequence
stars; $\sim35$ of these will be Earth-like planets in the habitable zone
\citep[see][]{per06a}. Therefore, it is possible that life does exist
relatively nearby.

The next generation of SETI experiments will soon be underway with the
full Allen Telescope Array (ATA; 350 antennas) becoming operational in
2008. The ATA plans to survey $10^6$ stars out to 900 light years for
extraterrestrial signals within the range of 1 to 10 GHz. Beyond the ATA,
the Square Kilometre Array (SKA) should provide the sensitivity to explore
up to $10^{8}$ stars looking for powerful radio signals similar to those
presently used by humans \citep[see][]{tar04,pen04}.

Recently, \citet[LZ07]{loe06} have raised the issue of looking for
extraterrestrial intelligence in the 100\,MHz regime of the radio
spectrum, rather than at GHz frequencies around the 21cm hydrogen line (as
with ATA and the Phoenix Project above). The rationale for this idea is
two-fold. First, humans communicate mostly in the 100\,MHz regime, e.g.,
through television and radio. Second, the search for the ``Epoch of
Reionisation" (EoR) has pushed the next generation of radio observatories
(MWA-LFD, PAST, LOFAR) towards large surveys in the 80--300\,MHz range, to
detect redshifted 21cm emission from the first galaxies at $z\sim 6$ to
15. LZ07 therefore note that a SETI project could ``piggy-back" on such
EoR surveys.

In their paper, LZ07 focus on military radar as one of the most
powerful sources of radio ``leakage" into space from Earth. The radar
employed by the US Ballistic Missile Defence System (BMDS) can
generate isotropic radiation with a total power of $2\times10^9$W, or
two orders of magnitude higher if beamed (LZ07). Likewise,
``over-the-horizon" radar, that bounces signals off the ionosphere, can
reach similar power output \citep{tar04}. Using such signals as a
blueprint for possible extraterrestrial radio emission, LZ07 predict
that LOFAR could detect civilisations like ours out to a distance of
$\sim70$pc, containing up to $\sim 10^5$ stars.

The challenge would be finding such a signal in the LOFAR data-stream, as
it would appear as a faint spectral line whose frequency would not
coincide with any known atomic or molecular transition. Furthermore, the
line would be Doppler shifted by both the orbital (around the star) and
rotational (spin on axis) motions, as well as a possible modulation in the
strength of the line if the emission was not uniformly spread over the
entire surface of the planet. The emission may also be switched on and
off, or be beamed ``lighthouse" fashion. The detection of such an unusual
signal requires new algorithms and extensive computational resources, as
the analysis must run on the raw data-streams with as much frequency
resolution as possible.

In summary, LOFAR has the sensitivity to detect human-like radio signals
out to the nearest Sun--like planet systems. Therefore, LOFAR (alongside
ATA and other 21cm surveys) will provide the first real opportunity to
detect serendipitous radio emission from the communications of advanced
extraterrestrial civilisations. Even if no signal is detected, a LOFAR
SETI initiative would provide important training for the SKA, especially
in the development of the required computational techniques and resources.

\subsection{Exploration of the Unknown}

The challenges outlined in the LOFAR science case are today's problems:
will they still be the outstanding problems that will confront astronomers
in the next decade and beyond? If history is any example, the excitement
of LOFAR will not be in the old questions which are answered, but by the
new questions that will be raised by the new types of observations it will
permit.

Most of the phenomena observed today using telescopes across the
electromagnetic spectrum were unknown a few decades ago, and to an amazing
extent were discovered by scientists using increasingly powerful
instruments and following their curiosity when they found the unexpected.
Examples include non-thermal radiation, radio galaxies, quasars, pulsars,
gravitational lensing, cosmic evolution, extra-solar planetary systems,
cosmic masers, molecular clouds, dark matter, and the cosmic microwave
background.  These discoveries have changed the whole face of astronomy in
fundamental ways.  Some discoveries came about as a result of more
sensitivity, others from better spatial or temporal resolution, still
others by observing in a new wavelength band or even from misguided
theory.  Many involved recognising a new phenomenon and being able to
distinguish it from a spurious instrumental response.

How can one plan for discovery? Despite the apparent capriciousness of the
aim, history tells us that a basic requirement is to carry out systematic
work with one or more observing capabilities (sensitivity; spatial,
temporal or spectral coverage; spatial, temporal or spectral resolution)
having at least an order of magnitude improvement over what has been
achieved before. LOFAR will greatly enlarge known parameter space as a
result of: i) a much greater sensitivity at low radio frequencies; ii) a
very large instantaneous field-of-view; iii) multiple, independently
steerable, beams. The sensitivity and sky coverage advances combine to
make a large volume of space accessible -- and hence the chances of
finding intrinsically rare objects in large scale surveys will be much
enhanced. LOFAR is therefore well-placed to make serendipitous
discoveries.

LOFAR's design means that it will be a highly multiplexed instrument
involving multi-beaming on a massive scale as compared with current
telescopes. The multiple beams provide users with a highly flexible and
responsive instrument and will inevitably lead to changes in observing
style. The challenge to LOFAR is to develop a philosophy of operations and
data archival which allows individuals, small groups, and larger
communities the freedom to innovate, and encourages users to explore
completely new ways of collecting, reducing and analysing data: in other
words, to allow for discovery as well as explanation. Some possible
considerations for LOFAR planners are:

\begin{itemize}
\item Award some time to successful groups or collaborations on the basis
of their past record -- a `rolling time allocation grant' which can be
sustained or closed down on the basis of performance integrated over
several years.

\item Allow high-risk or unproven new-style observations. The availability
of independent beams will help to make this feasible without compromising
conventional observing programs.

\item Maintain technical expertise in the community. In the lead up to
LOFAR and then the SKA, innovative research and development is
re-invigorating the world-wide community and involving a new generation of
engineers and students. However, when first LOFAR and then SKA are
completed, it is vital to allow a cadre of technical people world-wide to
gain continuous access to parts of the system for continuous
experimentation.
\end{itemize}
\clearpage

\section{Ultra-High Energy Cosmic Rays and Neutrinos with LOFAR-UK}

Ultra-high energy cosmic rays (UHECRs) are predominantly protons and heavy
nuclei with energies far exceeding those in terrestrial particle
accelerators. The mechanism by which these particles are accelerated to
such high energies is not known, but candidates include the most extreme
processes in the Universe, such as Active Galactic Nuclei (AGN) and Gamma
Ray Bursts.  They could even arise from the decay of unknown massive
particles, often predicted in Grand-Unified Theories (GUTs), or from
topological defects arising during the period of inflation after the Big
Bang.

Although the spectrum of charged cosmic rays has been measured over 14
orders of magnitude \citep{abb05,man05,shi06}, no diffuse cosmic neutrino
spectrum has yet been observed. Unlike cosmic rays, neutrinos are
undeflected by magnetic fields and so will point back to their source.
They also travel cosmological distances unattenuated.  A multi-messenger
approach that includes data from cosmic rays, neutrinos and gamma rays,
will be necessary to further our understanding of the origin of UHECRs and
the mechanisms that drive their acceleration.  In addition, the
interactions of neutrinos with energies above $10^{17}$ eV, in any
detection medium, occur at centre-of-mass energies that exceed those seen
in the high energy particle accelerators on Earth. These interactions thus
offer the opportunity to probe new particle physics phenomena at energies
approaching the GUT scale, for example, the presence of extra spatial
dimensions, which would enhance the neutrino-nucleon interaction cross-section.

UHECRs produce a particle shower when they impact the atmosphere, and
currently the most important methods for detecting such showers have been
observations of the fluorescence light emitted by the air showers, or
directly detecting the shower particles that reach the surface of the
Earth. However, the UHECR-initiated air showers also produce a radio
signature. When the secondary charged particles in the air shower are
deflected in the Earth's magnetic field, they produce radiation that is
emitted in the forward direction, and is coherent for wavelengths that are
larger than the size of the shower front, i.e., for frequencies less than
100\,MHz. These radio signatures from cosmic rays were first detected over
forty years ago \citep[see][for an historical overview]{wee01}. The
unprecedented size of the LOFAR array allows this technique to be applied
to neutrino detection for the first time. Air showers from neutrinos can
be distinguished from showers from other particles if the neutrino path is
highly inclined with respect to the down-going direction. At those angles,
the atmosphere filters electromagnetic energy from cosmic rays showers
produced at high altitudes, while neutrinos can penetrate deeper into the
atmosphere and produce a detectable cascade at lower altitudes. The high
effective area provided by LOFAR's antennae would offer a greatly improved
sensitivity to the radio signature produced by air showers from both
ultra-high energy cosmic rays and neutrinos \citep{fal04}.

Additionally, LOFAR has a unique opportunity to detect ultra-high energy
neutrinos through a different radio signature that they would produce in
their interactions with the moon. This radio Cerenkov signature was first
proposed by Gurgen Askaryan. When a neutrino interacts in matter and
produces a shower, that shower will develop a charge asymmetry, at the
level of approximately 20\%. This can be viewed as a ball of negative
charge travelling faster than the speed of light in that medium, which
results in Cerenkov radiation. For wavelengths longer than a transverse
size of the shower, of order 10~cm, the signal is coherent.  That
coherence occurs in the radio microwave region.  At the same frequencies,
there exist media that occur naturally in large volumes and are
transparent to this signal.  Ice, salt and sand are three such media and
past, present and future experiments that use this technique all use one
of those media. The GLUE experiment was the first to search for ultra-high
energy neutrinos by viewing the sandy surface of the moon (the regolith)
with a radio antenna \citep{gor03}. LOFAR, with its antennas pointed at
the moon, would be sensitive to the same signature.

LOFAR offers the opportunity to extend the search for cosmogenic neutrinos
into a higher energy regime than is possible with existing and proposed
detectors. The UCL HEP LOFAR group (Connolly, Lancaster, Nichol, Waters)
has expertise in the simulation of neutrino interactions and the
associated radio signal, through their participation in the ANITA radio
high-energy neutrino experiment, which is expected to be the first
experiment to find evidence for UHE neutrinos in the energy region below
LOFAR. This expertise will be applied to the neutrino-moon interactions
that LOFAR has sensitivity to, with a view to probing predictions of the
neutrino flux to $10^{22}$~eV and identifying any sources, should a flux
be observed.

\clearpage

\section{Solar and Heliospheric Physics with LOFAR-UK}
\label{solar}

The Sun is a powerful emitter at radio wavelengths, not only during
intense bursts of activity related to phenomena such as solar flares and
coronal mass ejections (CMEs), but also during times when it is considered
quiet at other wavelengths. The radio domain provides a particularly
sensitive diagnostic tool for accelerated particles on the Sun because
even weak disturbances, such as the tiny events thought to be related to
coronal heating, can give rise to detectable radio signatures via coherent
plasma emission. Radio emission from energetic and dynamic phenomena such
as solar flares and CMEs is of particular interest, as they are major
drivers of space weather and can affect the Earth's space environment.
Flares and CMEs are also challenging physical phenomena to be understood
in their own right.

In the solar context, LOFAR will open up an unexplored window of radio
observations, that is particularly useful for studying particle
acceleration and large-scale dynamics.  It will enable the study of
accelerated particles, from very weak to very energetic events and,
especially when combined with multi-wavelength observations from other
terrestrial- and space-based observatories, allow key topics to be
addressed. With LOFAR, dynamic processes involved in solar flares and CMEs
can be studied from their origins on the Sun as they propagate out through
the solar atmosphere. Using radio interplanetary scintillation
measurements, CMEs can be tracked, and the solar wind and interplanetary
magnetic field conditions analysed, from the outer solar corona to the
Earth and other planets.  Furthermore, radio observations with LOFAR will
provide powerful diagnostics of the Earth's ionosphere and
magnetically-linked regions of its magnetosphere. Uniquely, in the radio
domain, LOFAR will enable much of the Sun-Earth system to be probed with a
single instrument facility.

LOFAR's native angular resolution is a few arcseconds, and this, in
combination with its enhanced sensitivity and time resolution, makes for a
solar radio instrument far superior to any previous.  Prime benefits of
LOFAR to solar physics will be its increased sensitivity and its ability
to perform radio imaging spectroscopy, scanning rapidly in many
frequencies, and following dynamic sources as they propagate through the
corona on (sub)second timescales (for comparison, the Nan\c{c}ay
Radioheliograph has 5 channels between 150 MHz and 450 MHz, and occasional
VLA solar imaging is also done in only 3 or 4 channels).  A major
advantage of LOFAR will be the opportunity to examine phenomena for which
radio signatures occur preferentially at low wavelengths, such as coronal
shock waves and radio noise storms.

Sources of radio waves close to the plasma frequency in the solar corona
are broadened by plasma and wave turbulent scattering \citep{bas04}. For
example, sources at double the local plasma frequency, near 200\,MHz,
should have a size of around 100$''$.  While this means that it may not
always be possible to exploit the full spatial resolution capabilities of
LOFAR for solar studies, it does allow the study not only of the source
structure (for sources emitting at higher than the local plasma frequency)
but also of the effects of propagation and refraction of radio waves in
the solar corona.

Solar and heliospheric physics stands to benefit greatly from LOFAR's
increased capabilities. The UK has large, internationally-leading, solar
and heliospheric physics communities that are active in theory, modelling
and data analysis from gamma-rays to radio wavelengths, of phenomena
occurring below the solar photosphere, through the solar atmosphere, into
the solar wind and on to the atmospheres and surfaces of the planets.
These communities host considerable expertise in radio methods and are
well-placed to take advantage of LOFAR to study the genesis of solar
disturbances and their effect on the Earth and the other inner planets.
Radio studies will naturally be augmented by multi-wavelength observations
of the dynamic corona available from other STFC-supported missions such as
STEREO and Hinode (Solar-B).

The programmes of study included in this science case cover a wide range
of solar physics and Sun-Earth connections, from eruptive activity and
coronal shock waves on the Sun, through to the propagation and evolution
of structures in the solar wind and the signatures of disturbances in the
magnetosphere and ionosphere of the Earth, as well as the connection
between cosmic rays and atmospheric electricity. Below the main science
case for solar physics, heliospheric physics and terrestrial/planetary
physics are outlined.

\subsection{Solar Flares}

Solar flares are complex, extremely energetic and often dynamic phenomena.
Among other things they produce immense bursts of energetic electrons, the
source and acceleration of which is not well-understood. The primary
diagnostics for these accelerated electrons are hard X-rays and radio
waves.  In the metric and decametric regime, their emission is dominated
by coherent radiation at the local plasma frequency and/or its
harmonics. This is mode-converted plasma radiation from unstable
distributions, such as non-thermal beams of particles.  These beams,
moving at $\sim$0.2--0.5c outwards into interplanetary space and, more
rarely, sunwards, produce emission drifting rapidly in frequency: type III
bursts, reverse-drift type III bursts and bi-directional beams.  There are
many other, less explored, types of radio emission, which may be closely
related to the main flare energy release, such as radio spike bursts and
pulsations \citep[e.g.,][]{isl94}.  Together, these give unique insight
into the production of non-thermal distributions in the corona.

With LOFAR we can also study flare electrons in large-scale magnetic
structures that are difficult to see at other wavelengths. U-bursts, which
are sometimes observed below 200 MHz \citep[e.g.,][]{bas98}, are produced
by electrons travelling up and then down along large magnetic loops.
Sometimes such large loops are visible in X-ray and EUV images of the
solar corona \citep[e.g.,][]{kha00,poh01,fou05}. LOFAR is also expected to
see incoherent gyro-synchrotron of thermal or mildly-relativistic
electrons in magnetic fields of up to 10G, corresponding to coronal loops
on the scale of a solar radius. Imaging with LOFAR will reveal the
properties and development of large-scale, low density closed coronal
loops: how they form, erupt and reform, and their relation with active
regions, flarings and the CME structures. The incoherent radiation will
provide estimates of the energy distribution and numbers of particles in
these structures.

\begin{figure}[!t]
\centering
\includegraphics[width=11cm]{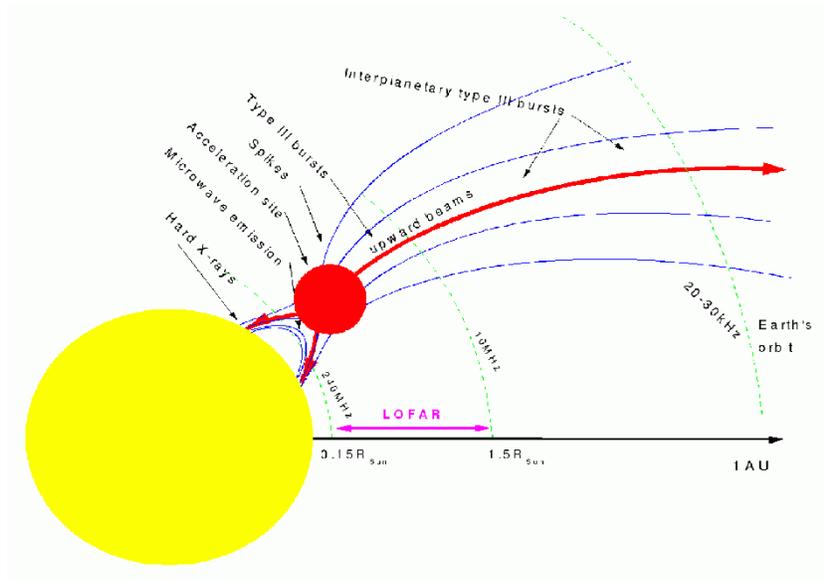}
\caption{\small The approximate location of different types of solar radio
emission, in the context of the flare magnetic geometry}
\label{solarscale}
\end{figure}

The LOFAR frequency range corresponds to plasma densities of $10^7$ to $7
\times 10^8$\,cm$^{-3}$, present between about 1.15 to 2.5 solar radii
(cf. Figure~\ref{solarscale}). In some flares, the {\it source} of the
electrons which generate type III emission will lie in this range.  This
is particularly the case for events where magnetic reconnection is
believed to occur high above the flare kernel, say at the top of large
loop system, as proposed in the break-out model \citep{ant98,ant99}. In
many cases the type III emission from upward-propagating electrons can be
studied, for example, in comparison with the downward-propagating
electrons which give rise to hard X-ray emission. Spatial and spectral
observations of type III sources will address the following questions:

\begin{itemize}
\item What is the starting location (i.e., the all-important acceleration
site) and propagation path of electrons producing type III and other
coherent bursts?

\item How do the burst spectra change as a function of time and position?
This is a diagnostic of local plasma conditions, e.g., the presence of low
frequency turbulence.
\end{itemize}

Another coherent radio signature, radio spike bursts, are believed to be
produced by accelerated electrons at, or very close to, the energy release
site. \citet{kha06} found, using Nan\c{c}ay Radioheliograph data, that the
spikes appeared high in the corona apparently in response to compression
of magnetic structures by an ongoing CME. This suggests a direct link
between MHD disturbances and electron acceleration, which can be studied
further with LOFAR.

\subsection{Coronal Mass Ejections}

\begin{figure}[!tb]
\centering
\includegraphics[width=10cm]{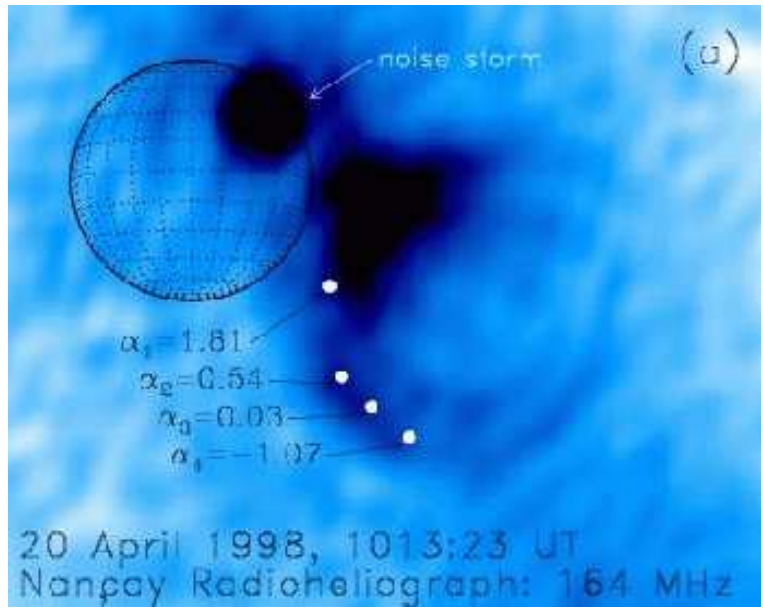}
\caption{\small The radio signature of a coronal mass ejection
from the Nan\c{c}ay Radioheliograph \citep{bas01}.}
\label{nancay}
\end{figure}

Coronal mass ejections (CMEs) are important for the production of
interplanetary (IP) shock waves and particle acceleration, and are one of
the most significant space weather phenomena affecting the Earth's
environment. CMEs are studied to determine the physical processes that
precede and cause them, their intrinsic properties and their consequences,
as well as how to identify the most geo-effective events. CMEs have been
imaged at radio frequencies \citep[e.g.,][cf. Figure~\ref{nancay}]{bas01}
at heights below two or three solar radii. In these cases, the emission
appears to be thermal and non-thermal gyro-synchrotron emission, providing
immediate information on the electron energy distribution and the magnetic
field strength in a CME. However, with comparatively few events analysed
so far, little is known of their origins or consequences.

With LOFAR, combined with imaging and spectroscopy at other wavelengths,
the following questions will be addressed:

\begin{itemize}
\item What are the trajectories of CMEs, and what are their effects on the
surrounding corona?
\item What can be learned about the physical structure (density, field
strength) and driver of the CMEs?
\item What are the relations between CMEs and flares?
\item Is the eruption of hot large loop structures in the solar corona a
major component of the later CME?
\item What is the relation between CMEs and shock waves seen in the
corona and in interplanetary space?
\end{itemize}

\subsection{Coronal shock waves}

A familiar radio signature of a shock wave in a metric to decametric
dynamic spectrogram is a broad spectral feature drifting slowly to lower
frequencies. This so-called type II radio burst is believed to be due to
plasma emission from electrons accelerated at an outward-propagating shock
wave \citep{wil72}.

Radio observations from LOFAR will enable the coronal type II bursts to be
followed and the physical properties (density, magnetic field strength) of
these shocks to be determined. Studies of coronal shock waves will seek to
understand the physical origin of coronal type II radio bursts and their
relation to interplanetary type II bursts and shock waves.  Regarding the
origin of the shock waves, it is aimed to determine whether the shock
waves are launched by, e.g., a piston of moving matter or a pressure
pulse.  LOFAR will also reveal the association of these bursts with flares
or CMEs, and determine whether they are directly--driven or
freely--propagating shocks.

Studying shock waves on the Sun is important not only for understanding
them in their own right, but also for understanding their effects on other
solar phenomena. It is known that large-scale shock waves may disturb
filaments \citep{mor64}, initiate eruptions (including CMEs), and trigger
`sympathetic' flares, all at remote locations. \citet{kha00} identified a
new class of coronal mass ejections associated with the eruption of
large-scale trans-equatorial loops attributed to the passage of a shock
wave. In addition, shocks are probably also the drivers of oscillations
observed in coronal loops \citep{hud04,bal05}.  Studies of coronal shock
waves will seek to determine their propagation characteristics and their
effects on the surrounding corona and other solar structures.

It should be noted that type II radio bursts preferentially occur at low
wavelengths (typically $<$150\,MHz) and, consequently, comparatively few
type II bursts have been studied to date.  The potential for discovery in
the study of coronal type II radio bursts and large-scale coronal shock
waves with LOFAR is very high.

The investigation of both large- and small-scale shock waves in radio data
will augment the pre-eminent UK effort in the detection and analysis of
MHD waves at X-ray, EUV and optical wavelengths. Since shock waves are
ubiquitous in astrophysical plasmas, advances in understanding solar
coronal shock waves and their role in particle acceleration will also be
of interest to the wider astrophysics community.

\begin{figure}[!t]
\centering
\vspace*{2cm}

{\Large{\bf See attached JPG file; M33.jpg}}
\vspace*{2cm}

\caption{\small The radio signature of a shock wave propagating through
the solar corona \citep[from][]{pic05}.}
\end{figure}

\subsection{Non-flaring active region energy release}

The corona of an active region evolves continuously under the action of
sub-photospheric driving, and it is believed to sustain a permanent
population of non-thermal electrons \citep[as has been observed for active
dwarf stars;][]{kun87}. Small micro-flares occur every few minutes when
there are active regions on the disk \citep{lin84}, many of them producing
non-thermal particles \citep{lin01}, and the ubiquitous presence of
non-flare-associated type III bursts may also reveal streams of enhanced
density associated with multi-million degree coronal jets
\citep[e.g.,][]{aur94,kun95}. The quasi-continuous occurrence of even
smaller events has been proposed as the mechanism that maintains the hot
corona \citep{par88}.  High brightness temperature radio noise storms are
a signature of accelerated electrons lasting for hours or even days, in
correlation with identifiable events in active region evolution
\citep{ben00a}.  These are a major source of solar radio emission at low
frequencies, outside of the times of flares and CMEs. The origins of the
noise storms, in particular how the accelerated population of electrons is
produced and maintained, are not yet understood.  Noise storms, like type
II radio bursts, preferentially occur at low frequencies and so LOFAR will
be particularly useful in their study. Small numbers of quiet--time
non--thermal electrons, hard to observe or invisible in other energy
ranges (optical, X-ray), should be sufficiently bright via coherent
emission to be imaged with LOFAR. Such studies will address fundamental
issues in solar coronal physics: the occurrence of non-thermal electron
distributions in non-flaring active regions; the partitioning of energy
between heating and particle acceleration; the role of nano-flares in
coronal heating and in the `gradual' evolution of active regions.

\begin{figure}[!h]
\centering
\vspace*{2cm}

{\Large{\bf See attached JPG file; NS\_941018\_1.jpg}}
\vspace*{2cm}

\caption{\small An example of a solar radio noise storm on 18 October 1994
as seen in a radio spectrogram (courtesy the Astrophysikalisches Institut
Potsdam).  The numerous short-duration, narrow-band bursts are noise storm
emission and indicate continued particle acceleration.}
\end{figure}

\subsection{Radar mapping of the solar corona, and plasma turbulence}

Using LOFAR as a receiver for future active experiments on the solar
corona could open a new era of radar studies of the Sun, such as is
already used to probe the Earth's ionosphere. The solar corona is a unique
example of a turbulent plasma medium yet to be fully explored.
Electromagnetic emission is effectively scattered by electron plasma
waves, whereas absorption at the local plasma frequency is strongly
determined by the anisotropy of the turbulence spectrum \citep{kho06}.
Early active experiments \citep{jam70} show that the radar signal
reflected from the corona can vary by a few orders of magnitude, implying
a highly anisotropic medium. The radar echoes at 38 MHz imply ever-present
compressional waves in the corona, which may be associated with coronal
heating. LOFAR will answer crucial questions about Langmuir wave spectra
in the solar corona, inaccessible by in-situ measurements. Such an
approach in the ionosphere successfully identifies the parameters of
Langmuir turbulence \citep{kon05}, and similar efforts may be possible
with the solar atmosphere. In addition, passive observations of sub-second
radio emission provides a diagnostic tool to measure the magnetic field
and main plasma parameters \citep{ste04}. There is growing international
interest in active radar studies of the Sun (e.g., the LOFAR Outrigger In
Scandinavia (LOIS): {\it www.lois-space.net}).

\subsection{Radio scintillation observations of the 3D solar wind}

The solar wind is the medium by which solar disturbances, such as coronal
mass ejections and fast solar wind stream interaction regions, are
conveyed to the Earth and the other planets. It is also representative of
the stellar winds surrounding other mid-life cool stars and interacting
with any planets they might possess. The plasma processes operating in the
solar wind are the same as those found in the atmospheres of other
Sun-like stars and are relevant to other astrophysical processes, while
the evolution of the large-scale structure of the solar wind is
fundamental in determining the manner in which it couples to planetary
environments in the solar system. Current instruments -- and currently
planned instruments -- can only give a partial view of solar wind
structure. LOFAR has the potential to provide a high-resolution view of
density and velocity structures in the solar wind from inside the orbit of
Venus to beyond Earth orbit.  The science questions hoped to be addressed
with LOFAR are:

\begin{itemize}
\item How do structures within coronal mass ejections (CMEs) interact, and
how does this affect the interaction between the CME and the background
solar wind?
\item How do the interaction regions between CMEs and stream-stream
co-rotating interaction regions (CIRs) develop, and how important is this
in determining the geo-effectiveness of CMEs?
\item How does large-scale structure in the solar wind evolve in the inner
solar system?
\item How do small-scale structures (turbulence) in the solar wind evolve
with increasing distance from the Sun, and how rapidly is energy
transported between scales?
\end{itemize}

To achieve this, LOFAR will be used as a multi-beam interplanetary
scintillation (IPS) telescope, measuring the small variations in apparent
intensity of astronomical radio sources arising from scattering of the
signal by small-scale density irregularities in the solar wind. As the
irregularities are being carried out from the Sun by the solar wind, IPS
measurements provide information on both (relative) solar wind density and
solar wind outflow speed, as well as providing estimates of other solar
wind parameters \citep{col96}. The technique of IPS, using modulation of
signals from astronomical radio sources by the solar wind, has been used
to study the density and outflow velocity structure of the solar wind for
over forty years \citep[e.g.,][]{hew64,arm72,koj77}.  In the last ten
years there have been considerable advances in the technique, allowing it
to realise its full potential as a tool for untangling the evolving
structure of the solar wind \citep[e.g.,][]{col96,bre99,bre06}.

\begin{figure}[!h]
\centering
\includegraphics[width=10cm]{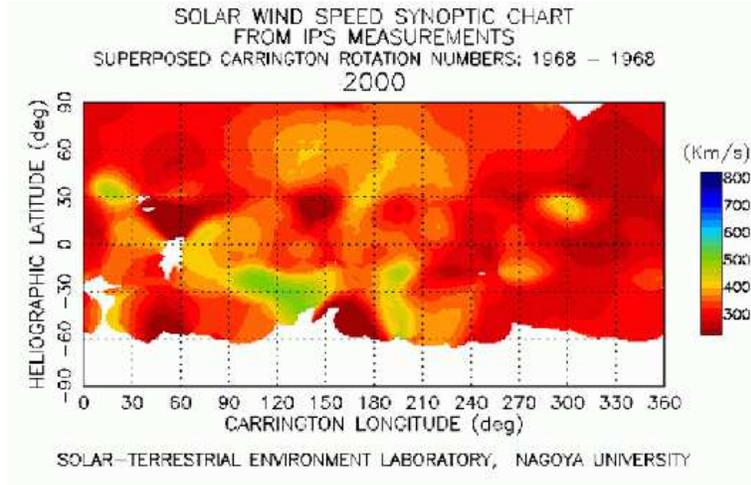}
\caption{\label{solwindfig} \small An overview map of solar wind velocity
structure over one solar rotation near solar maximum (credit: M. Kojima)}
\end{figure}

LOFAR possesses the features necessary to provide a view of the solar
wind with sufficient resolution to image in detail the passage of CMEs
and the development of stream-stream interactions, and thereby to gain
the full scientific benefit from the next generation of space
missions, such as the Solar Orbiter. These observations require large
numbers of density and velocity measurements, with good velocity
resolution \citep{bre03}, over a large number of radio sources across
the inner heliosphere each day. LOFAR has the collecting area and
sampling rate (100 Hz minimum) to detect the many weak, and
weakly-scintillating (at the 1-10\% level), radio sources necessary to
do this. The construction of UK stations as part of the LOFAR
programme will provide the long baselines needed for accurate
measurements of solar wind speed. This combination of sufficient
spatial resolution to image the internal structure of solar wind
disturbances and accurate measurements of their velocity would
represent a huge advance on the capabilities of any current system,
and will provide entirely new information on the solar wind.

\begin{figure}[!h]
\begin{tabular}{cc}
\psfig{file=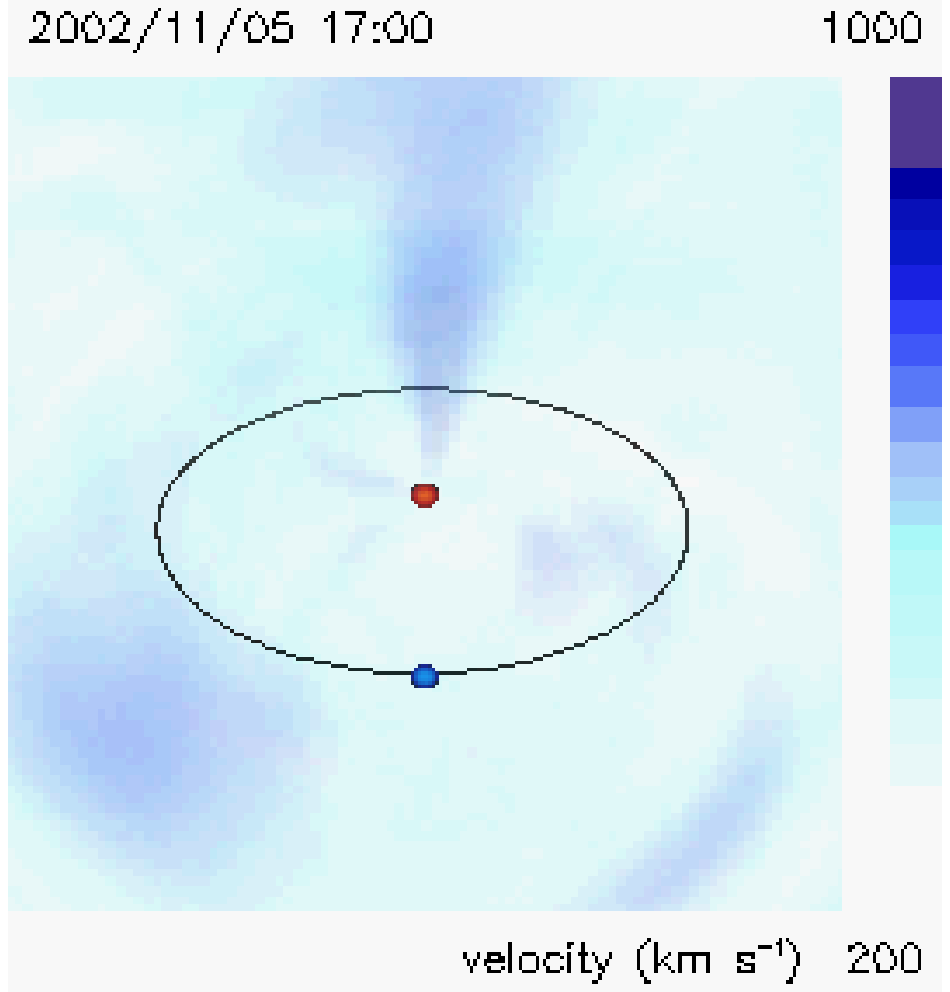,width=8cm,clip=}
&
\psfig{file=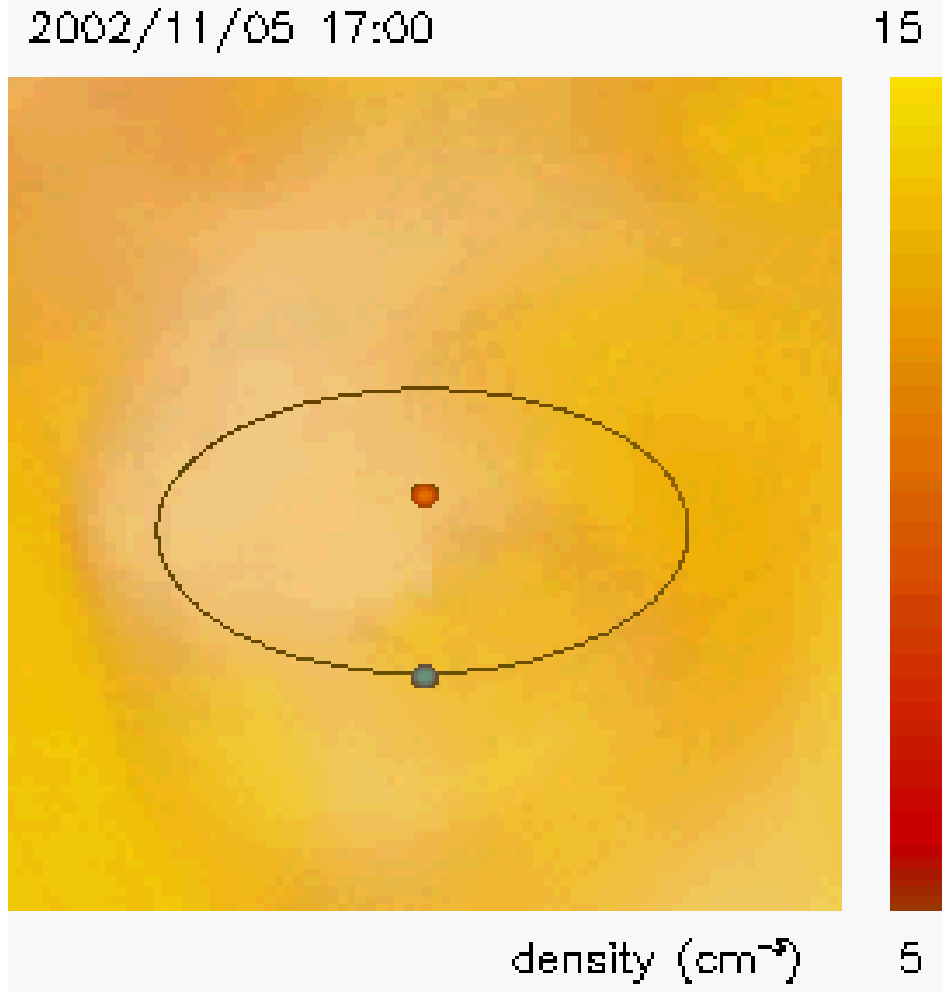,width=8cm,clip=}
\end{tabular}
\caption{\small Tomographic reconstructions of 3D solar wind velocity
(left) and density (right) structure, using Nagoya STELab IPS data
\citep[from][]{hic01}.}
\end{figure}

LOFAR will be able to provide measurements of solar wind density and
velocity from $\sim$80 solar radii out to beyond 230 solar radii, spanning
the distance range from within the orbit of the ESA Solar Orbiter mission
to beyond Earth orbit, and covering the regions in which the structure of
the solar wind seen in the outer regions of the corona is transformed into
that sweeping across the orbit of the Earth. It will be able to provide
information on the large-scale 3-dimensional structure of the inner solar
wind, both to place high-resolution coronal and in-situ measurements in
their broader context (how do structures elsewhere relate to those
observed in detail?) and to study the evolution of the solar wind in
velocity and density with latitude, longitude, distance from the Sun and
phase of the solar cycle. Specifically, LOFAR will:

\begin{itemize}
\item provide information on the large-scale structure of the solar wind,
and particularly on the interaction regions between streams of fast and
slow solar wind and between coronal mass ejections and the background
solar wind (a large number of observations per day are required for
this). The high sensitivity of IPS measurements to electron density,
together with their ability to measure velocity, will provide an
overlapping dataset which will be the ideal complement to white-light
coronal density measurements of the inner heliosphere;
\item provide context for high-resolution IPS (EISCAT, MERLIN, VLBA) and
in-situ measurements, by revealing the large-scale structure of the solar
wind at interplanetary distances. The results will be especially valuable
in providing context for Solar Orbiter measurements;
\item provide information for 3D tomographic reconstruction of the solar
wind;
\item provide density and velocity information for solar wind and
heliospheric modelling.
\end{itemize}

The multiple-frequency capabilities of LOFAR will also make it possible to
probe differences in behaviour between different scales of irregularities,
building on the recent work of \citet{fal06}. The IPS mode for LOFAR would
be capable of providing information suitable for operational space weather
monitoring, although this would not be the primary purpose of these
observations.

A great advantage of LOFAR as an IPS instrument is its multi-beam
capability, which will enable routine IPS observations to be run without
contention for telescope time. This will make synoptic observations of the
solar wind practical. The large number of observations required to image
density structures would be made by using the whole collecting area of the
telescope to observe scintillation of many weak sources. Velocities could
be estimated from those source-observations with high signal-to-noise,
using the method developed by \citet{woa92}, with velocity calibration
achieved by using widely-separated LOFAR station groups to observe a more
limited number of strong sources simultaneously, allowing use of a 2-site
cross-correlation approach to velocity estimation
\citep[e.g.,][]{col96,bre99,bre06}. LOFAR will provide a unique
opportunity to study the large-scale structure of the interplanetary solar
wind -- a field of science where the UK is already one of the world
leaders.

\subsection{Riometric Observations of the terrestrial space environment}

The Earth's ionosphere and magnetosphere form a coupled plasma system,
strongly influenced by the external drivers of the solar wind and coronal
mass ejections. While it is carrying out astronomical observations, LOFAR
will also be able to diagnose conditions in the ionosphere, thus providing
large-scale, high-resolution measurements of the magnetically-linked inner
magnetosphere, specifically the so-called `slot region' between the
Earth's inner and outer radiation belts. Studying this crucial zone, not
currently covered by any existing instrumentation, will advance, in
particular, our understanding of how relativistic particles respond -- are
accelerated and lost -- as geomagnetic and electric fields vary.

The science questions that will be addressed are:

\begin{itemize}
\item What is the morphology of the F-region plasma at latitudes
equatorwards (``sub-auroral'') of the auroral oval during storms and
sub-storms?
\item What is the extent of relativistic electron loss to the ionosphere
across the outer radiation belt?
\item What is the morphology of the sub-auroral ionosphere during
sub-storms?
\item What is the cause, and the spatial and temporal structure, of
electron precipitation from the slot region during geomagnetic storms?
\item By how much does solar radio emission degrade absorption
measurements in the auroral regions?
\end{itemize}

As radio signals pass through the ionosphere they are attenuated, primarily
by increased electron density. Although this is a hindrance for
astronomical observations, it is this loss of signal that provides a
mechanism for observing the ionosphere. By utilising the loss of signal,
LOFAR can act as a large-scale riometer (relative ionospheric opacity
meter) at the same time as it is making astronomical observations. At
30\,MHz the beam width is 5.5$^{\circ}$, corresponding to a spatial
resolution of $\sim$8km in the ionospheric D-region. This is on a par with
the detail provided by the new ARIES (Advanced Rio-Imager Experiment in
Scandinavia) system recently deployed at Ramfjord, Norway. However, the
latitude of LOFAR means that it will probe a different, lower-latitude,
part of the ionosphere from ARIES, and one in which the attenuation is
expected to be due to very high energy electrons leaking from the
radiation belt, or to very low energy electrons from the Earth's
plasmasphere -- the region of cold, dense plasma trapped on field lines
which rotate with the Earth.

The essence of using LOFAR as a riometer is to extract the raw signal from
each beam before any corrections are applied to compensate for ionospheric
attenuation. This can then be compared to a quiet day signal. The
technique will work best for frequencies in the 20--60\,MHz band; at
higher frequencies the smaller absorption values will be hard to
distinguish from the noise level. This is especially important for
monitoring the subtle increases and decreases in absorption in the F
region, associated with, e.g., the loading of the F-region when the
Earth's plasmasphere is eroded during geomagnetic storms, or the decreases
due to nightly ionospheric flows. The longitudinal coverage of LOFAR will
provide large-scale images of the ionosphere with high temporal and
spatial resolution. This is something that satellites are currently not
able to do -- for example, 4 point measurements in 90 minutes is typical
for a polar orbiting satellite.

Since riometers are essentially sensitive radio receivers, they suffer
when the sun is emitting additional radiation at their operating
frequency, and when high-energy solar protons produce so-called polar-cap
absorption events that mask the true ionospheric absorption. LOFAR will
provide a measure of the solar radio emission when the sub-auroral
ionosphere is inactive, so that corrections can be made to riometer data
from other, high-latitude, instruments.

\subsection{LOFAR as an ionospheric sensor}

The mid-latitude ionosphere is far from being understood and is the key
region showing a dramatic response to solar storms. Newly-discovered
ionospheric phenomena include storm enhance density (SED) and the
associated sub-auroral polarisation stream (SAPS). GPS ionospheric
monitoring \citep{fos02,yin04} has aided the understanding of the
phenomena over the US and Europe. However, only very few storms over the
last solar maximum have been studied. One problem is the lack of
mid-latitude instrumentation -- many ionospheric instruments have been
decommissioned and there is increasing reliance on data from satellites
that are not guaranteed to be in the right location at the right time.

There are two key areas where LOFAR can aid in modern ionospheric
studies. The first is that the unprecedented sensitivity of LOFAR to radio
signal delay will allow measurements of path delay to less than 1 cm. The
same types of techniques that have already been developed to use GPS
signals to image wide regions of the mid-latitude ionosphere are directly
applicable to LOFAR signals, resulting in thousands of simultaneous and
accurate measurements to be used in multi-instrument data
assimilations. Simultaneous with the ionospheric response, the solar radio
bursts will also be detected, allowing high-temporal and spatial
resolution studies of solar--terrestrial storms. In addition, the
capability to observe at high resolution across a wide-area will expand
the observable spectrum of ionospheric waves.

The second area, an even more novel use of LOFAR, is as a passive
mid-latitude backscatter radar, using signals-of-opportunity as a system
to monitor the expansion of the polar cap during highly-disturbed
space--weather events and to allow the density and velocity of the plasma
over Europe to be monitored continuously.  This would provide a
substantial European contribution to support the science goals proposed in
US satellite missions such as `Living with a Star Ionosphere-Thermosphere
Storm Probes'.

\subsection[Radio from Lightning Flashes and Cosmic Rays]{Radio from Lightning Flashes and Cosmic Rays: Natural Radio
Waves in the Earth's Atmosphere}

Naturally occurring lightning flashes and cosmic rays transmit bursts
of invisible electromagnetic waves into the Earth's atmosphere,
similar to radio broadcasting stations. It was recently suggested that
lightning flashes may be initiated by energetic cosmic rays
\citep{gur05}.  The physical properties of lightning flashes and cosmic
rays can be inferred from simultaneous measurements of electromagnetic
waves with LOFAR and extremely low frequency radio wave magnetometers.

The proposed study of radio waves from lightning flashes and cosmic rays
is based on the comparison of simultaneous electromagnetic wave recordings
from lightning flashes and cosmic rays. The experimental observation of
radio waves builds on the expertise with remote sensing at the University
of Bath \citep{ful04}, while the observation of radio bursts from
ultra-high energy cosmic ray air showers (UHECRs) is based on the
expertise of ASTRON \citep{fal05}. The naturally occurring electromagnetic
waves will be observed with radio wave antennae at Extremely-Low
Frequencies (ELF), from 10\,Hz to 4\,kHz, in collaboration with the British
Geological Survey (BGS) in Edinburgh. The BGS operates a geophysical
observatory at Eskdalemuir in Scotland, which is remotely located, away
from human activity, and hence ideally suited to record naturally
occurring radio waves. The observation of cosmic rays can be done with
LOFAR. 

\subsection{Other science areas}

There is considerable interest within the UK solar, heliospheric (and
solar system) science communities in other opportunities for exploiting
LOFAR. These include studies of small-scale structure in the terrestrial
ionosphere using the technique of ionospheric scintillation, and active
radar studies of space plasma processes, with LOFAR acting as a network of
receiving arrays.
\clearpage

\section{LOFAR-UK as a stand-alone array}

There will be times when some or all beams from the international LOFAR
stations will not be correlated with those of the Dutch LOFAR stations,
either because they are not required (e.g. when EoR observations are
carried out using only the Dutch core stations), or because of limitations
of the correlator. In these circumstances, beams from individual LOFAR-UK
stations can be used independently, or may be correlated with the other
international stations (and any unused Dutch stations) as a
sparsely-sampled array. There are a number of valuable science goals that
could be carried out in this way.

\subsection{Pulsar observations with individual LOFAR stations}

The individual LOFAR stations have sufficient sensitivity to be able to do
very interesting monitoring observations of some known pulsars and other
transient radio emitting neutron stars.  They could be used to perform
regular timing observations of a few dozen radio pulsars which are known
to glitch.  One of the keys to understanding the equation of state of the
super dense material of a neutron star is seeing how often neutron stars
suffer starquake-like glitches and, when they do occur, how the rotation
rate recovers from the glitch. The frequent monitoring that would be
possible with individual LOFAR stations would allow more glitches to be
caught, and for them to be sampled more regularly. Presently less than a
handful of pulsars are monitored frequently enough to be able to study
their glitches in sufficient detail.  These timing observations could also
be used to form timing ephemerides of the brightest pulsars which could be
used by high energy satellites like GLAST to look for gamma-ray emission.

Recent discoveries have highlighted two new interesting classes of
radio-emitting neutron stars. These are the RRATs and the intermittent
pulsars. RRATs are characterised by infrequent bright single pulses
(typically a few per hour). Monitoring of these sources may greatly
improve the number of pulses detected and thus allow a better
determination of the nature of the sources.  Moreover by implementing a
trigger algorithm on the station data, piggy-back searches could also be
performed for more of the short duration pulses that characterise these
sources, thus potentially greatly increasing the number known. The wide
field-of-view of a LOFAR station will greatly facilitate this
search. These observations would also be sensitive to short duration
pulses like the one recently discovered in the direction of the Magellanic
Clouds but which is believed to be at more than a Mpc away. The origin of
this single very bright burst is at present unknown but there are
suggestions that there could be hundreds per day.  Their brightness means
that they could be detected by an individual LOFAR station.

The intermittent pulsars are the other new class of sources which could be
studied with a LOFAR station. They are pulsars which are seen to emit
radio waves from anything from a few days to years and then turn off for
similarly long periods.  Intriguingly during the off periods they are seen
to exhibit different spin properties to when they are on, thus providing a
unique probe of the emission mechanism itself. To better understand the
phenomenon, better statistics are needed on the repetition timescale of
the source turning on and off. It is also of interest to know what happens
exactly at the moment when the on-off or off-on transition happens. The
monitoring capabilities of a LOFAR station are again ideally suited to
this task.

\subsection{Solar observations with individual LOFAR stations}

As discussed in Section~\ref{solar}, one of LOFAR's Key Science Projects
is concerned with observing and studying the properties of the Sun.
However, such observations will only be carried out for a subset of the
time. Although the angular resolution of an individual LOFAR station will
be low, it will be able to operate as a very sensitive, high temporal
resolution and high frequency resolution solar radio spectrograph, greatly
increasing the capabilities for solar activity monitoring.  Spectrographic
observations are essential to make sense of the solar radio emission in
the full-LOFAR Solar imaging observations, because a variety of radio
bursts have been classified according to their form in spectrograms and
these, in turn, have been related to various physical phenomena. Moreover,
several very interesting fine structures have been seen in recent
spectrogram data.  LOFAR acting as a very sensitive spectrograph will very
likely reveal much more information on the detailed spectral behaviour of
solar radio emissions, and thus on the details of the particle
accelerations and plasma process involved.

Continual solar spectroscopic monitoring would also allow
burst-trigger-mode observations, whereby the entire LOFAR array could be
triggered to observe the Sun at relatively short notice if an interesting
burst is detected in the radio spectrograms. Furthermore, relatively crude
centroiding images from the LOFAR-UK stations would provide useful
information (in the absence of better radio imaging) for large-scale
phenomena such as CMEs and shock waves.

\subsection{Heliospheric Physics with individual LOFAR stations}

The UK has considerable expertise in indirect imaging of the
interplanetary medium using scintillation techniques.  These methods have
proven effective in monitoring the solar wind, providing alerts for the
possible onset of geomagnetic activity and determining the connections
between solar activity and interplanetary weather. In addition these data
have direct relevance to satellite and spacecraft operations, the
commercial operation of high-latitude electric power grids and radio
communication.  Near real-time monitoring of the solar wind has
significant potential in all these fields but has been hampered by the
lack of suitable radio telescopes on the ground, capable of monitoring
several hundred compact extragalactic radio sources per day and measuring
their scintillation levels. However, each LOFAR station has the
sensitivity, collecting area and bandwidth to perform these stand-alone
scintillation measurements, necessary for solar wind imaging. Indeed, the
beam-forming capabilities of a single LOFAR station are far superior to
any low frequency array currently in operation worldwide and offer the
opportunity to exploit this technique to the full for the first time.

Measurements of density structures do not need interferometric baselines,
so this task could be performed by a single UK LOFAR station. With more
than one station, intensity correlation analyses will reveal the motion of
the scintillation pattern over the ground and therefore the velocity of
the solar wind. Long-baseline intensity correlation observations combining
data from UK station(s) with measurements from non-core LOFAR sites in
Germany and France would provide high-precision measurements of solar wind
speed and direction. LOFAR will offer a unique combination of imaging of
density structures in the solar wind together with long-baseline,
high-precision velocity measurements, and will provide an entirely new
view of the evolving interplanetary medium. Again, these observations can
be performed without relaying data back to the LOFAR data processing
facility in Groningen.

\subsection{Ionospheric diagnostics with individual LOFAR stations}

As LOFAR operates at low radio frequencies, its performance will be
sensitive to conditions in the Earth's ionosphere. In order to ensure
optimal performance from LOFAR, adequate ionospheric diagnostics will be
required. The UK has a record of expertise in ionospheric radio science,
and this will be drawn on to map ionospheric structures above LOFAR in
near real-time. This would provide the information necessary to correct
for ionospheric effects, even during periods of geomagnetic disturbance.
This capacity could be developed with a single UK LOFAR station.

\subsection{Correlating E-LOFAR stations for early long-baseline surveys}

The ({\it u-v})-plane coverage of an international array of 8-10 European
LOFAR stations will be sufficient to make reasonable images at
1$^{\prime\prime}$ resolution.  Although these observations will not have
the sensitivity of the full array, they should be able to observe every
source in the 1.4-GHz FIRST survey \citep{bec95} to a flux limit of 1~mJy
at 1.4-GHz, in just a few beam-months. The correlation requirement would
be relatively modest, despite the high resolution, because the FIRST
survey tells us exactly where the sources are; only a small area around
each source need be correlated, resulting in greatly reduced requirements
on channel width and integration time. Ionospheric calibration would be
done by observing only in fields around bright ($>$200-mJy) point-source
calibrators. The amount of sky that can be covered depends critically on
the ionospheric stability at the time of observations, but for typical
isoplanactic patch sizes we expect 30--50\% coverage should be possible.

There would be a number of key advantages in such a programme. First, the
entire FIRST survey could be examined for gravitational lenses. These have
typical dimensions of $<2^{\prime\prime}$ (so require E-LOFAR's
resolution), and are of importance for studying mass distributions in
galaxies and galaxy evolution, as discussed in Section~\ref{lenssec}. The
numbers of lenses which could be detected would depend on assumptions
about source structure, but it is notable that the CLASS survey
\citep{mye03} discovered 22 lenses out of the currently known $\sim$150 by
a radio survey of flat-spectrum radio sources to the 30-mJy level. A
1$^{\prime\prime}$--resolution survey at this depth would also push into
new parameter space for the study of radio source populations.

Second, it is almost certainly imperative for the planning of a full-scale
survey with the whole E-LOFAR array to have a pilot survey on long
baselines. Such a survey, in addition to producing scientific output,
would allow optimisation of observing strategies given the calibration
problems produced by the ionosphere. This will be crucial, because the
full E-LOFAR survey will in any case require increased
computing/correlator resources, both because it will have more long
baselines and also because it will be a blind survey, without the
advantage of prior knowledge of source positions. Considerable experience
with the pilot survey will therefore be necessary before this is carried
out.
\clearpage

\section{Technical Case}
\label{technical}

\subsection{The locations of UK LOFAR stations}

\begin{figure}
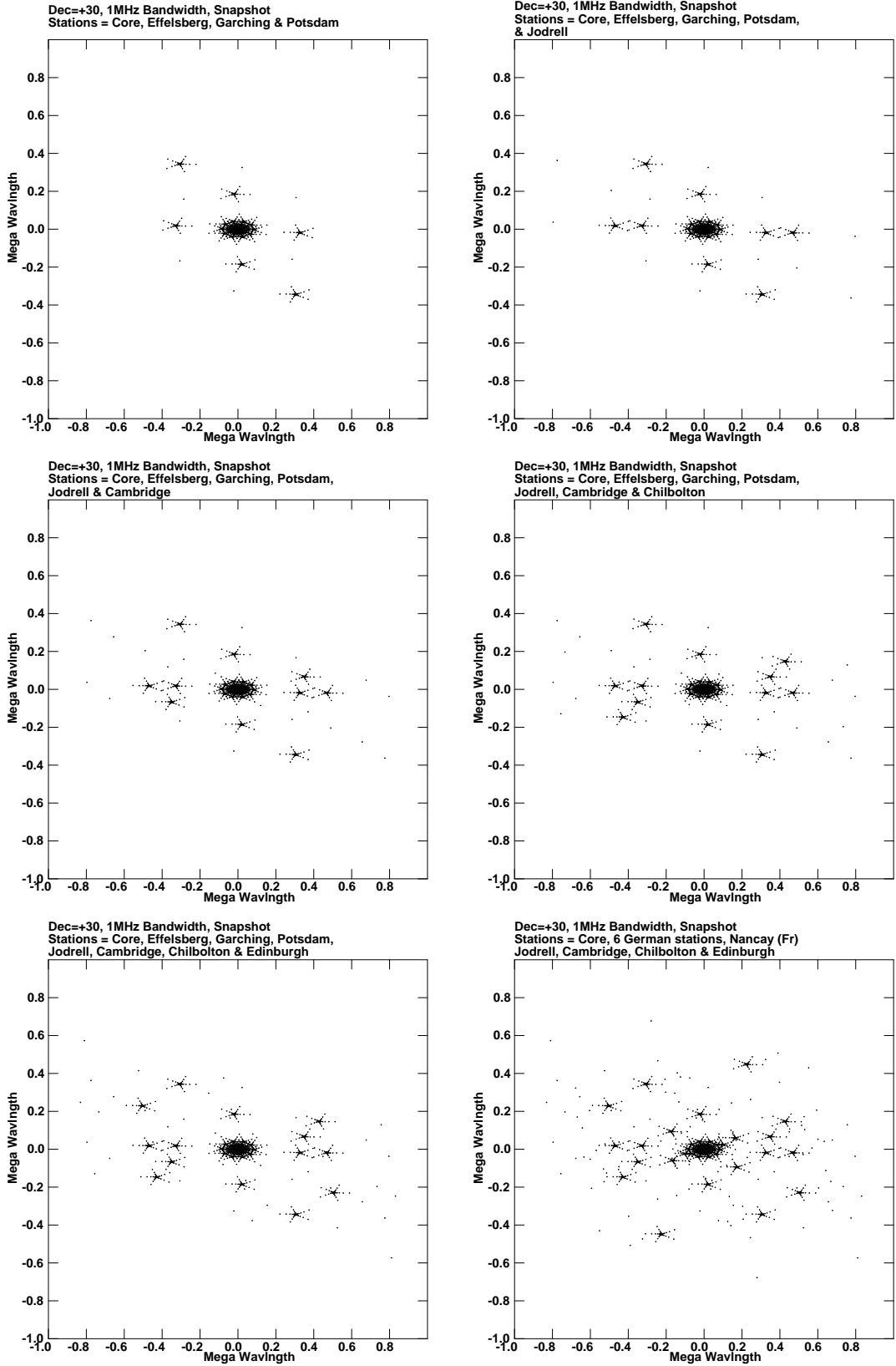

\begin{center}
\begin{tabular}{cc}
\psfig{file=Snapshot-dec30-1.ps,width=70mm} 
& 
\psfig{file=Snapshot-dec30-2.ps,width=70mm}
\\
\psfig{file=Snapshot-dec30-3.ps,width=70mm} 
& 
\psfig{file=Snapshot-dec30-4.ps,width=70mm}
\\
\psfig{file=Snapshot-dec30-5.ps,width=70mm} 
& 
\psfig{file=Snapshot-dec30-6.ps,width=70mm}
\\
\end{tabular}
\end{center}
\caption{\small The simulated {\it u-v} coverage for snapshot observations
at 230\,MHz (other frequencies only change the axis scalings), with 1\,MHz
bandwidth, of a source at 30$^{\circ}$ declination. The top-left plot
shows the {\it u-v} coverage provided by the Dutch LOFAR plus three
fully-funded German stations. The subsequent four plots show the
improvement in {\it u-v} coverage when various subsets of UK stations are
added. The lower-right plot finally shows the {\it u-v} coverage produced
from 6 German, 1 French and 4 UK stations.}
\label{uvsnap}
\end{figure}

\begin{figure}
\begin{center}
\begin{tabular}{cc}
\psfig{file=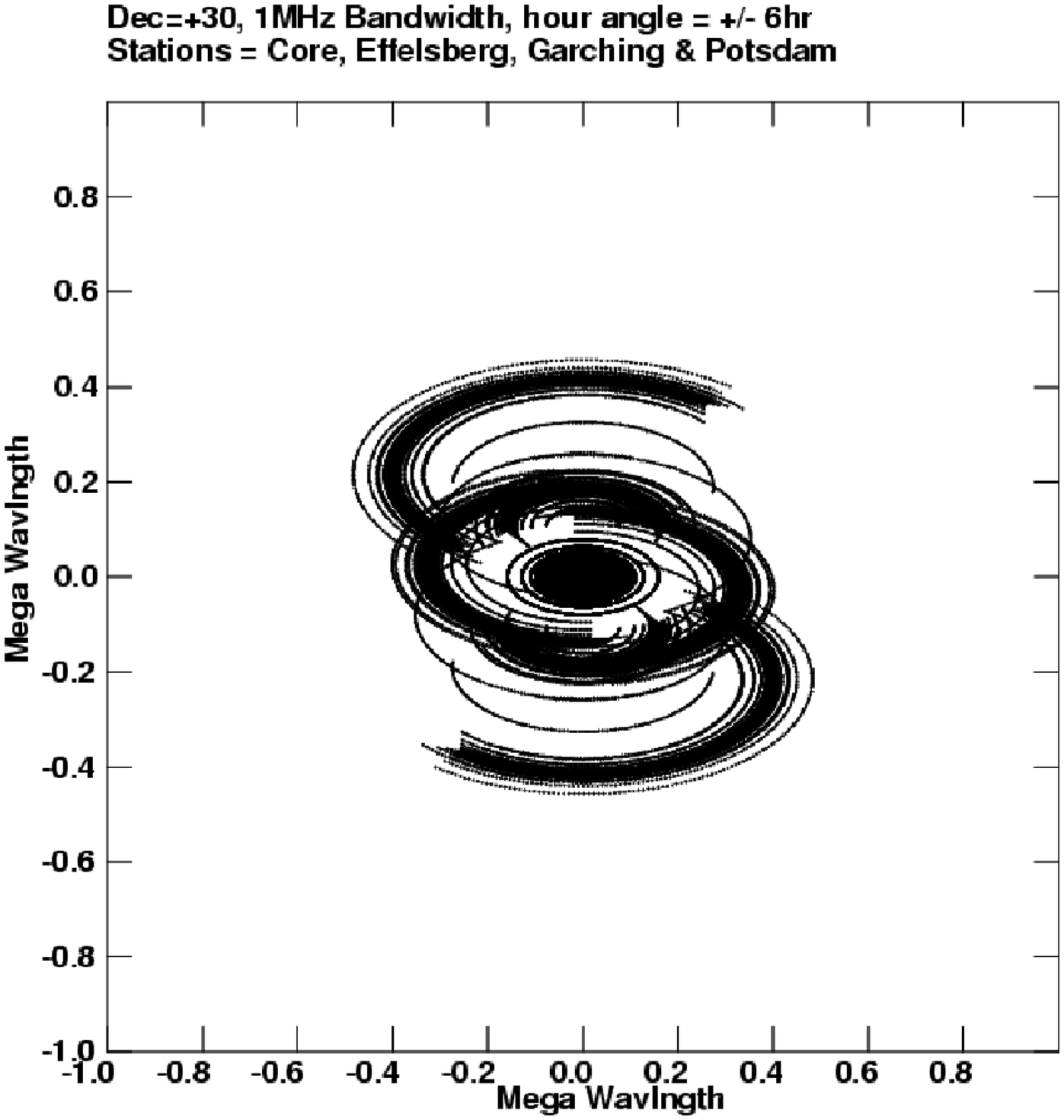,width=72mm} 
& 
\psfig{file=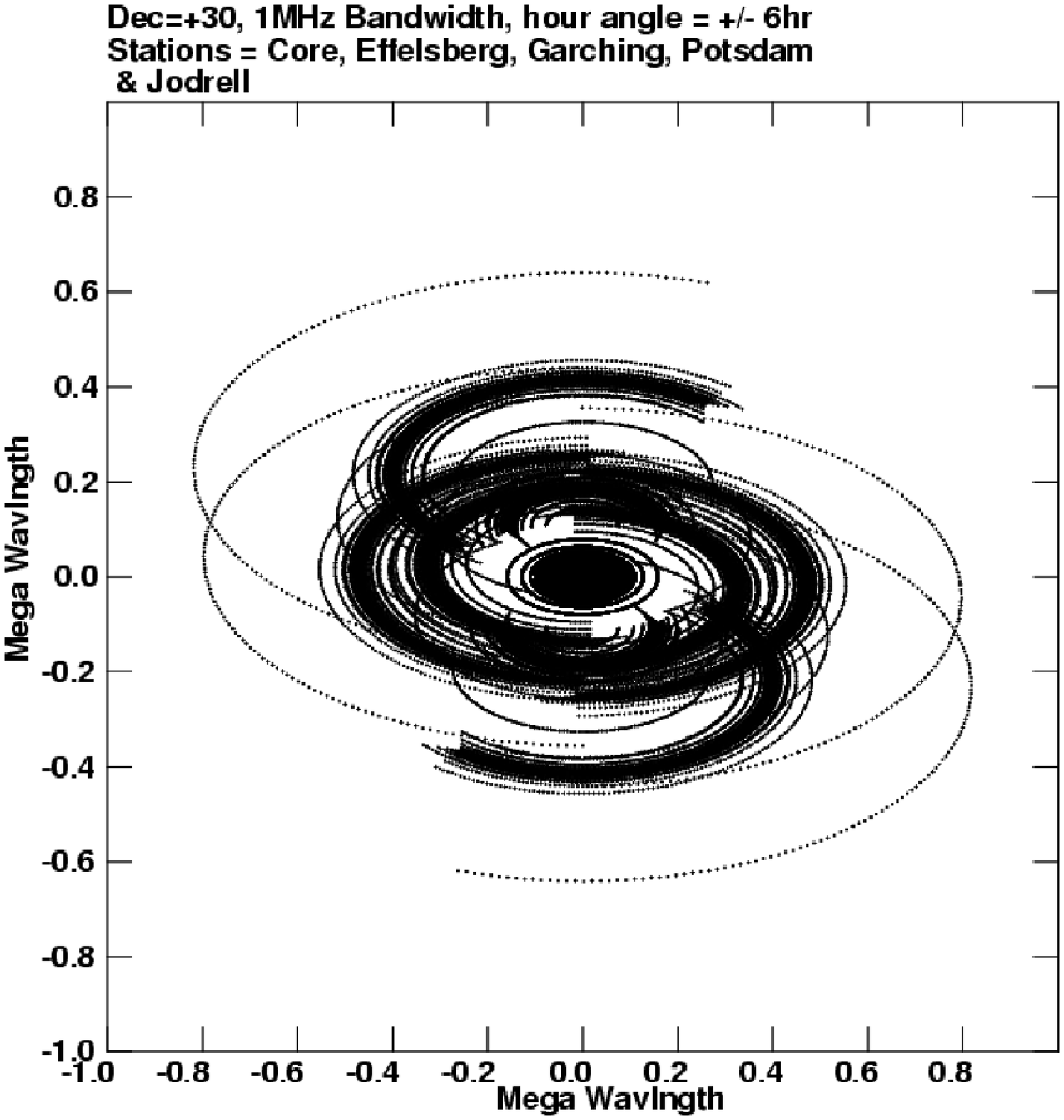,width=72mm}
\\
\psfig{file=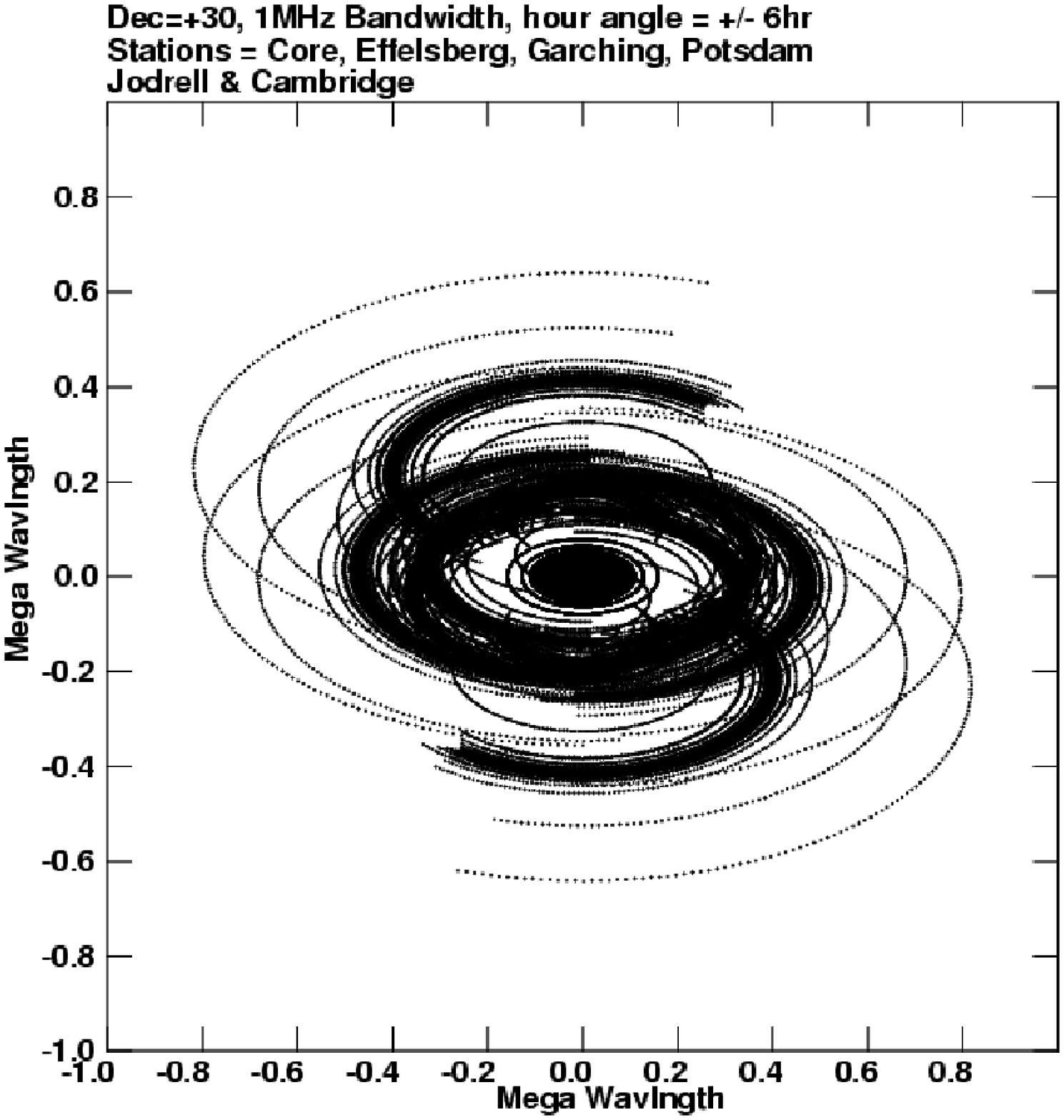,width=72mm} 
& 
\psfig{file=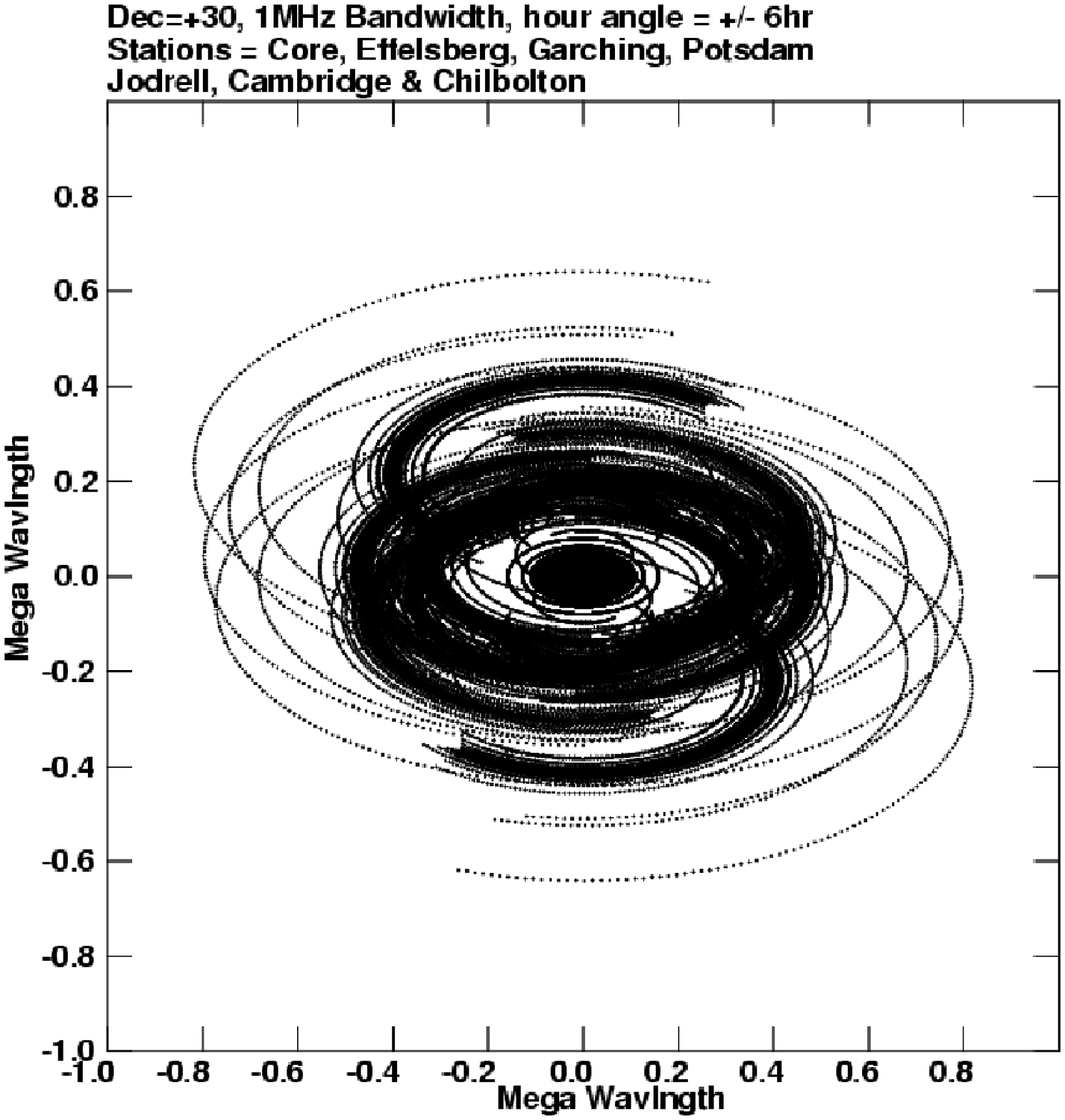,width=72mm}
\\
\psfig{file=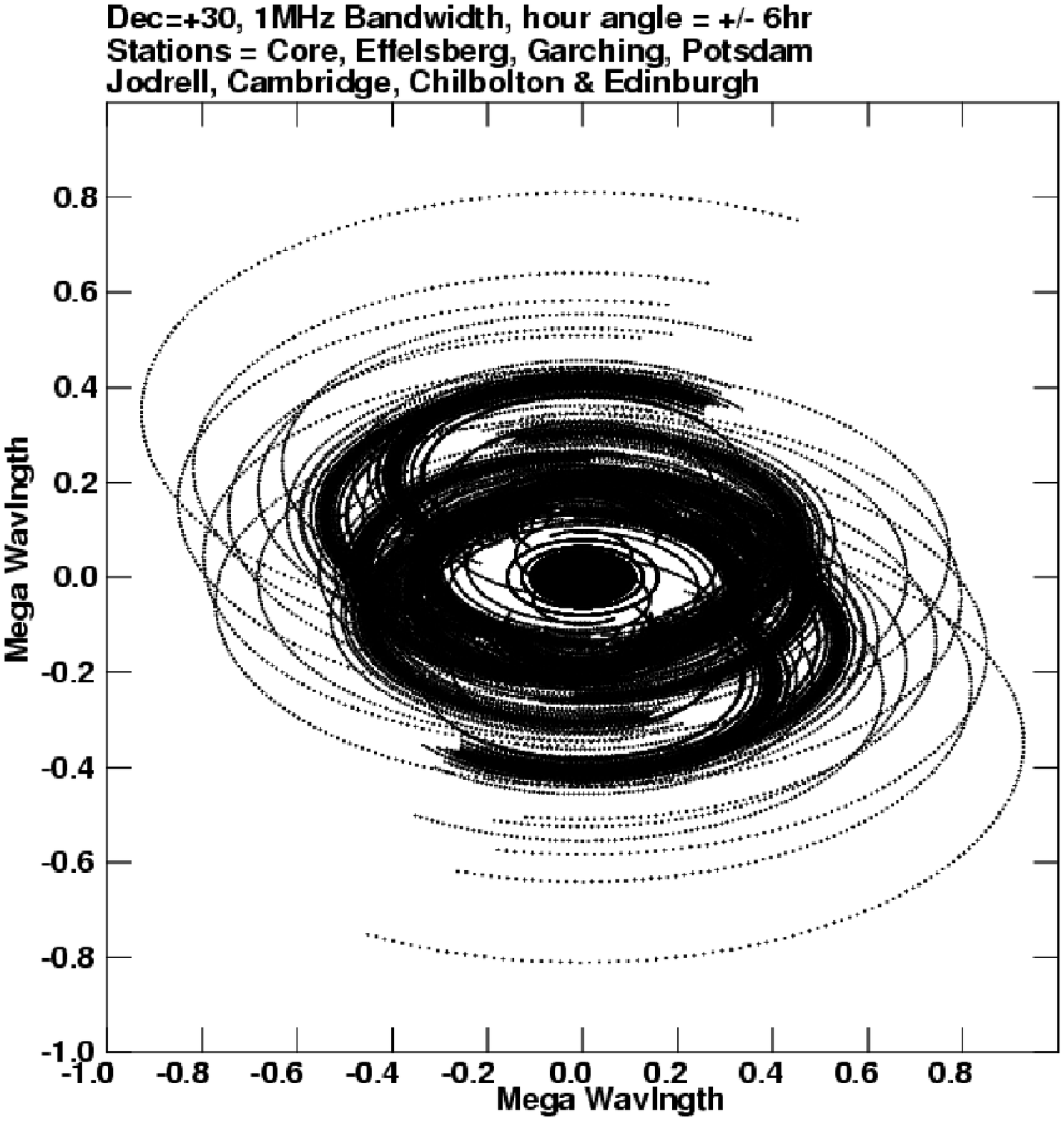,width=72mm} 
& 
\psfig{file=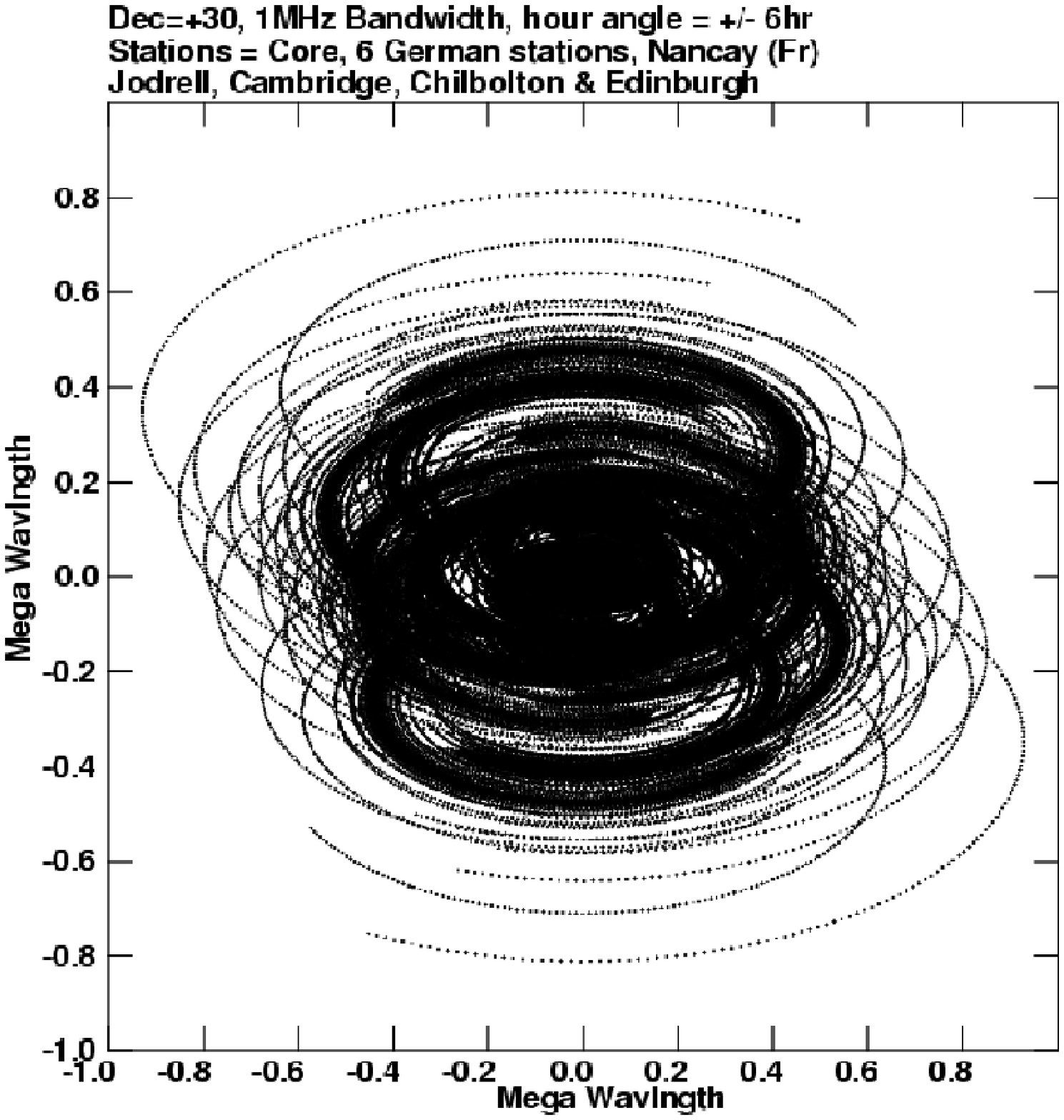,width=72mm}
\\
\end{tabular}
\end{center}
\caption{\small The simulated {\it u-v} tracks for full 12\,hr (hour angle
 $\pm 6$) synthesis observations of a source at 30$^{\circ}$ declination,
 at 230\,MHz with 1\,MHz bandwidth. The various combinations of LOFAR
 stations are as described in Figure~\ref{uvsnap}.}
\label{uvfull}
\end{figure}

Scientific simulations, plus considerations of practicalities, have led to
the conclusion that the optimal approach for LOFAR-UK would be the
deployment of four UK LOFAR stations, subtending a range of angles and
baseline lengths relative to the Dutch core. Leading candidates for the
sites of these stations are {\it Jodrell Bank}, {\it Lord's Bridge}, {\it
Chilbolton}, and {\it Edinburgh}. The rationale for the choice of these
sites is discussed below. Numerous simulations have been carried out of
the {\it u-v} coverage that would be provided by various subsets of these
sites, for sources of different declinations in both snapshot and full
12-hr synthesis observations. A full suite of these simulations is
available at {\it www.jb.man.ac.uk/$\sim$rbeswick/LOFAR/}; two
illustrative sets of examples are shown in Figures~\ref{uvsnap}
and~\ref{uvfull}. It is clear that the proposed UK stations will: (i)
increase the maximum baselines in the {\it u-v} plane, leading to an
increase in the angular resolution of the array by a factor of $\sim 2$
compared to LOFAR with just Dutch and German stations (and a factor of
several compared to Dutch-only LOFAR); (ii) allow much more uniform
coverage of intermediate baselines, providing smoother and more symmetric
beam-shapes.

The Jodrell Bank and Lord's Bridge sites are natural choices for UK LOFAR
stations, for a number of reasons. These are locations where, as discussed
in Section~\ref{context}, there is a long history of radio astronomical
facilities, and thus extensive local technical expertise. Land has been
donated to the LOFAR-UK project at these sites by the University of
Manchester and the University of Cambridge, respectively. The proposed
stations are also close to e-MERLIN fibres, thus minimising the additional
cost of fibre connections. The sites have recently been RFI surveyed by
the LOFAR team, and initial findings show that in both cases the average
interference environment is better than an average LOFAR site in the
Netherlands.

The Chilbolton Observatory site, close to STFC Rutherford Appleton
Laboratory, also offers excellent potential to host a LOFAR-UK
station. RFI surveying of this site by the LOFAR team has demonstrated
that this is also better than typical LOFAR sites in the Netherlands. In
addition, the majority of facilities required to support the establishment
of a LOFAR station are present. The only notable short-coming is the lack
of a connection to dark fibre. Connection to dark fibre is thought to be a
more cost effective method of providing data transport than using the
Chilbolton Observatory's connection to the BT fibre network. The South
East England Development Agency (SEEDA) are very enthusiastic about the
location of a LOFAR station in this part of the country and may provide
financial support (see Section~\ref{consortsec}) to assist in data
connectivity.

A Scottish LOFAR station has the clear benefit of providing both the
longest possible UK baselines, and also a different subtended angle
relative to the main LOFAR core, providing improved coverage of the {\it
u-v} plane. There is also the possibility that funding for a Scottish
station could be provided directly from the Scottish Funding Council. The
currently-favoured site for a Scottish LOFAR station is about 10\,km south
of Edinburgh, close to the Edinburgh Technopole Science Park. A 10\,Gb/s
fibre connection is available from this science park, connecting directly
into the SuperJanet 5 fibre network.  Investigations of precise site
locations, and the cost of installing the final 1-2\,km fibre connection,
are still in progress. The nearby location of the Astronomy Technology
Centre may be beneficial in terms of technical support.

Other potential sites for LOFAR stations within the UK are still under
consideration. In particular, the possibility of siting a LOFAR station in
Central Wales, as part of a larger funding bid to the Welsh Assembly for
fast-fibre connectivity of that region, is currently being investigated.
There have also been informal discussions with the Regional Development
Agency and other funding sources in the North-East of England related to
the construction of a LOFAR station there.
 
\subsection{Data transport}

The anticipated data rate from a LOFAR station outside the central core is
2 Gb/s, based on a total bandwidth of 32 MHz, which can be divided in up
to 8 beams, in each of 2 polarisations, with 12-bit sampling, and allowing
up to 512 Mb/s for formatting and framing overheads along with control and
monitoring data. For LOFAR within the Netherlands, the data will be
transmitted from each station across a dedicated Wide Area Network using
Gigabit Ethernet protocols to the central processor. Connecting stations
from outside this dedicated WAN will require additional work and
resources.

For remote stations in the UK, we expect that the LOFAR station data
can be transmitted across the SuperJanet 5 (SJ5) network, which has
trunk data rates and access points at 10 Gb/s, to interconnect with
the Geant network (the EC-funded network which connects the various
European research networks) for transmission to Amsterdam. Subsequent
transmission to the LOFAR central processing facility will be handled
by Surfnet in the Netherlands. A similar route is currently being used
for regular e-VLBI experiments, where data from radio telescopes at
Jodrell Bank, Cambridge (via radio link to Jodrell Bank), Torun
(Poland), Medicina (Italy), Westerbork (Netherlands) and Onsala
(Sweden) are transmitted at data rates of up to 512 Mb/s to the EVN
data processor at JIVE (located in Dwingeloo, in the Netherlands) for
real-time correlation there. Data are transmitted using standard IP
protocols across national research networks and Geant.  Scientists and
engineers at Jodrell Bank Observatory, the MERLIN/VLBI National
Facility and the High Energy Physics group at the University of
Manchester have played important roles in the development of e-VLBI
and are engaged in a number of research projects to develop the
technique and capabilities still further. Currently, data transmission
from Jodrell Bank also makes use of the UKLight connection from
Manchester to Amsterdam, provided by Janet(UK). The last two years
have seen a rapid development of this e-VLBI capability, with
sustained data rates climbing from 32 Mb/s to 512 Mb/s, and an
EC-funded project EXPReS is currently tasked with making 1 Gb/s
real-time operations a routine procedure for the European VLBI
Network. Sustaining continuous transmission across production networks
at these rates is a real challenge and a very active area of research
in terms of processing and throughput bottlenecks, data protocols and
switching and routing techniques.

Issues for data transport from LOFAR-UK stations include the initial
connection from the station to the SJ5 network, and consideration and
optimisation of the data rate and management of the data flow across SJ5
and Geant to Amsterdam. For the Jodrell Bank site, a 2.5 Gb/s connection
to Manchester already exists, funded through e-MERLIN and a PPARC grant
for e-VLBI, with operational costs currently met by MERLIN. Equipment is
currently being installed to upgrade this connection to 13 Gb/s. For the
Cambridge site, there may be options to make use of the e-MERLIN dark
fibre connection to connect with national trunk fibre networks and then
SJ5. Other sites will require fibre rental or new optical fibre cables to
be installed over distances of up to a few kilometres.

The nature of the data flow from a LOFAR-UK station needs to be considered
in detail.  While the 12-bit sampling depth is a specification for the
LOFAR core and stations in the Netherlands, and may be appropriate for the
combination of stations where there are considerable correlated
interfering signals present, it may be possible to reduce this sampling
depth at remote stations where the interfering signals are much more
independent. In addition, it may be possible to increase the amount of
local signal processing in order to reduce the transmitted data rate. In
the end, this comes down to issues of cost: for the LOFAR core, it is
cheaper to transmit larger volumes of data and do the processing centrally,
while for the most distant stations it may be more cost effective to do
more processing locally and reduce the data rate.  As data transmission
becomes cheaper and easier, it will be possible to increase the data rate,
number of beams and sampling depth, even for the remote stations.

The management of the data flows within the UK, and the interconnections
with Geant and Surfnet, will require close interaction with Janet(UK),
Dante, Surfnet and LOFAR staff.

\subsection{Long-baseline calibration}

The calibration and imaging of wide fields of view at high resolution and
at low frequencies is one of the key technical challenges facing LOFAR.
Substantial effort is being invested by the LOFAR team to study, develop,
implement and test calibration and imaging strategies, algorithms and
software. Extending LOFAR to long baselines, especially with isolated
stations, makes some of these issues more challenging. Solving these
issues will not be the first priority of the Dutch LOFAR team and so it
will be important for the UK to contribute effort in this area.

The principal issue for LOFAR calibration will be the number of sources
per station beam. Even for the `Minimum Ionosphere Models' developed by
Noordam, this number is likely to be 20-30. Longer baselines will resolve
a significant fraction of these sources, reducing the number of
calibration points per field of view and hence the accuracy with which the
ionosphere can be characterised above those stations. Shorter baselines
benefit from the coherence of the ionospheric fluctuations on length
scales of 100km; longer baselines are likely to require denser calibration
observations on the sky in both time and space. Keeping track of the whole
number of turns of phase delay introduced by the ionosphere for remote
long baseline stations may also be more problematic.

The LOFAR calibration strategy involves solving for instrumental and
ionospheric parameters at the locations of individual bright sources and
then subtracting these accurately from the data. The brightest sources in
the sky (Category 1, roughly the A-team of radio astronomy: Cyg A, Virgo
A, etc) are subtracted wherever they are in the sky. The next brightest
sources (Category 2) within the beam are used to calibrate and are then
subtracted; their solutions are then interpolated over the whole field of
view.

Longer baselines require the data to be correlated in narrower frequency
channels and with shorter integration times. In general, to keep the field
of view constant, the number of sample points has to increase as the
square of the baseline length. The increase in correlator output data rate
has consequences for calibration, and in particular the (discrete Fourier
transform) subtraction of category 1 and 2 sources.  The data processing
for imaging also increases with the correlator output rate and the size of
the image in pixels. It will be worthwhile investigating how to optimise
the calibration and data processing techniques for long baselines to the
UK and other international partners.

\subsection{Operational requirements for Solar and Heliospheric observations with LOFAR}

Solar and Heliospheric Physics are not part of the Dutch Key Science plan
for LOFAR, and so we discuss here the observational requirements for this
aspect of the LOFAR-UK science case. There are clear differences between
the operational requirements of these two sets of studies: solar
observations (flares, CMEs and coronal shocks, quiescent energy release
and solar radar studies of plasma turbulence) will require dedicated
observing modes with beams tracking the Sun, whereas the majority of the
heliospheric applications require many radio sources across a wide area of
the sky to be observed near-simultaneously. Further details are provided
below.

Since Solar and Heliospheric Physics is not one of the Dutch Key Science
Projects, the European partners will also be responsible for archiving the
LOFAR data for this project.  We are aware that the Astrophysical Institut
Potsdam (AIP) intends to host a LOFAR Solar Data Archive.  The University
of Glasgow is also exploring the possibility of supporting the archiving
of the solar data from LOFAR, in close discussion with AIP.
Considerations for the support of the Heliospheric data are also underway.
 
\subsubsection{Solar observational requirements}

Observing the Sun with LOFAR will differ in technique from observations of
other astronomical objects.  Solar observations with LOFAR require
observations taken in four ``modes": spectrograph-mode, snapshot-mode,
burst-trigger-mode, and campaign-mode, which are described below.  Since
the interest is in observing and understanding phenomena which are
frequency- and time-dependent on the Sun, a spectrographic capability for
LOFAR is required to be available alongside the LOFAR radio imaging
observations.

For spectrograph mode, it is proposed to use one, or a few, ground
stations to have LOFAR operate as a very sensitive and high temporal
resolution solar radio spectrograph.  Spectrographic observations are
essential to make sense of solar radio emission during imaging
observations in burst-trigger-mode and campaign-mode.  The reason for this
is because a variety of radio bursts have been classified according to
their form in spectrograms and these, in turn, have been related to
various phenomena.  Moreover, several very interesting fine structures
have been seen in recent spectrogram data.  LOFAR acting as a very
sensitive spectrograph will very likely reveal much more information on
the detailed spectral behaviour of solar radio emissions.

Snapshot-mode solar observations with LOFAR are imaging observations
taken, ideally, at several wavelengths, regularly throughout daylight
hours every day.  The cadence of the images is to be negotiated, but an
initial suggestion is one set of images every three minutes. These provide
`snapshots' of solar phenomena which will be useful for general and
long-term studies.

In addition to snapshot-mode, burst-trigger-mode observations are
required.  That is, at certain times, LOFAR would be available to respond
to requests to observe the Sun at relatively short notice if bursts are
detected.  Since the majority of bursts of interest occur infrequently,
this approach will be more effective for the solar science than simply
allocating time for high resolution observations during which no
interesting phenomena may occur.  One way in which this mode could be
envisaged to operate is as follows. Radio spectrographs from the
Astrophysical Institut Potsdam (or from an individual LOFAR station)
detect the onset of an interesting radio burst in the LOFAR frequency
range.  This information is then rapidly communicated to LOFAR, and at the
same time a request is made for LOFAR to observe this burst with the
highest possible temporal and spatial resolution.  The typical total
duration of most solar radio bursts of interest will be a few minutes or a
few tens of minutes. Ideally the images should be taken with sub-second
(millisecond?) cadence and, preferably, in at least two (ideally more)
frequencies.  This mode will allow us to adequately observe highly dynamic
important solar phenomena.

The other mode of observations that is proposed is campaign-mode.  In the
solar community, multi-wavelength, multi-instrument, internationally
co-ordinated observations are sometimes carried out. These observations
are planned typically weeks to months in advance.  In would be highly
beneficial for LOFAR to participate in such campaign observations. These
will involve higher cadence observations than in snapshot mode, and most
likely a number of frequencies.  The specific time intervals and cadence
will be specified well in advance of the observations, to allow
co-ordination between all instruments. This mode may be used to study, for
example, the evolution and development of a solar active region.

Most solar images (e.g., during snapshot mode) will require observations
in a relatively small region of the sky, several degrees across, but in
some situations (e.g., detailed study of CMEs) larger fields of view might
occasionally be required.

\subsubsection{Heliospheric observation requirements}

Radio scintillation observations of the solar wind, riometric studies of
magnetosphere-ionosphere coupling and ionospheric studies all require
observations of many weak radio sources over the course of a day, with
both the amplitude and phase of the signals sampled rapidly. In the case
of radio scintillation observations, the total power received from each
radio source must be sampled at 100\,Hz across a bandwidth of several MHz,
centred near the upper limit of LOFAR frequencies, with the source
observed for around 15 minutes in order to build up a well-defined
spectrum. The demands of riometric and ionospheric observations would be
slightly lower, as regards sampling rate, but once again large numbers of
sources would need to be observed.
 
In order to successfully image solar wind density and velocity structures,
a radio scintillation experiment running on LOFAR would need to be able to
record scintillation patterns for a large number of astronomical sources
lying within about 40 degrees in the sky from the Sun. Most of these
sources would be relatively weak, so as a guideline it can be assumed that
point-like sources with flux densities of a few Jy at $\sim$200\,MHz would
need to be observed (the choice of a frequency band near the upper limit
of LOFAR capabilities is to optimise the observations to study the solar
wind inside the orbit of the Earth, and to reduce the effects of
ionospheric scintillation). As the variation in power due to scintillation
is of the order of a few per cent of the total flux strength, the
observations would be required to have rms noise levels of a few mJy in
order accurately measure the scintillation. As the scintillation patterns
vary only slowly with frequency, a wide bandwidth is advisable for
successful scintillation observations -- of the order of a few MHz (most
current IPS systems use bandwidths of 10 MHz or more). The collecting area
of the LOFAR core should be amply capable of fulfilling these
requirements, provided that wide-bandwidth measurement is possible.
 
Interplanetary scintillation is characterised by rapid variations in
signal power, on timescales of less than 0.1s to more than 10s. Estimates
of solar wind speed are particularly sensitive to the high-frequency limit
of the spectrum, so it is necessary to sample rapidly: sampling should be
at 100\,Hz at least, and it is generally desirable to sample more rapidly
and subsequently integrate to reduce the effects of stochastic noise. For
the weaker sources it will take about 15 minutes to build up a
well-defined power spectrum for the scintillating flux, and this is what
sets the required duration of each source-observation.  As many sources
need to be observed to build up an image of solar wind structures, the use
of multiple beams is highly desirable, but the wide bandwidth and rapid
sampling are essential to successful scintillation observations.  The best
science value would be obtained by running radio scintillation experiments
on LOFAR in campaign mode, choosing one or two intervals a year and
attempting to construct maps of solar wind density and velocity over a
solar rotation (27 days) -- these intervals would be selected to offer the
best complementarity with other (mainly space-based) observations. A
baseline for usage might be 150 sources observed every other day
(preferably every day during intervals of higher solar activity) over 27
days, no more than once a year. The requirement to observe a source for 15
minutes to build up a good spectrum of scintillations thus indicates a
need for at least 4 beams to observe all of the sources within a 12 hour
period, and ideally more beams for a shorter period to increase
simultaneity.
 
\clearpage

\section{The LOFAR-UK Consortium}
\label{consortsec}

\subsection{Consortium members and management}

The LOFAR-UK consortium comprises UK university departments and research
institutes who plan to work with funding agencies, educational
organisations, industry, and the main LOFAR team, to operate a number of
LOFAR stations within the UK. It will also initiate and coordinate
LOFAR-based and LOFAR-related scientific research, both within the UK and
in wider collaborations involving the whole LOFAR team. It plans to
coordinate the exploitation of UK skills and resources to maximise the
UK's participation in, and scientific return from, LOFAR.

In late 2006, fourteen institutions signed a Memorandum of Understanding
which formalised the LOFAR-UK consortium, and led to the formation of a
Management Council (MC); a fifteenth has recently also signed up. The
institutions, and their representatives on the Management Council, are
listed in Table~\ref{mctab}. Rob Fender (Southampton) has been elected as
the overall Project Leader of LOFAR-UK, and Steve Rawlings (Oxford) as the
Deputy Project Leader. These Project Leaders are responsible for
representing the interests of the whole LOFAR-UK consortium, both
internally within the consortium, and externally through negotiations with
the main LOFAR collaboration and other international partners, and with
funding agencies. The MC has additionally appointed three project
coordinators to oversee various aspects of the LOFAR-UK effort.  Philip
Best (Edinburgh) has been appointed as Science Coordinator, with
responsibility for coordinating the UK White Paper and the UK science
effort. Rob Beswick (Manchester) is the Technical Coordinator, with
responsibility for managing the technical effort, such as the issues of
data transport (liaising with Janet(UK)), long-baseline calibration, and
site testing (liaising with ASTRON).  Finally, Bob Nichol (Portsmouth) is
the LOFAR-UK e-Science coordinator, responsible for investigating novel
approaches to data archiving and interrogation, and GRID computing, as
well as maintenance of LOFAR-UK email lists and web pages.

\begin{table}[!bh]
\smallskip
\begin{center}
\begin{tabular}{ll}
\hline
Institution                        & LOFAR-UK representative\\
\hline
Aberystwyth University              & Andy Breen        \\
Cardiff University                  & Steve Eales       \\
Durham University                   & Alastair Edge     \\
Liverpool John Moores University    & Chris Simpson     \\
Open University                     & Glenn White        \\
STFC (RAL)                          & Brian Ellison     \\
University College London / MSSL    & Catherine Brocksopp   \\
University of Cambridge (Cavendish)~~~~~~ & Paul Alexander    \\
University of Edinburgh             & Philip Best       \\
University of Glasgow               & Graham Woan       \\
University of Hertfordshire         & Matt Jarvis      \\
University of Manchester            & Simon Garrington  \\
University of Oxford                & Steve Rawlings    \\
University of Portsmouth            & David Bacon       \\
University of Southampton           & Rob Fender        \\
\hline
\end{tabular}
\end{center}
\caption{ \small Participating institutions of LOFAR-UK, and their
  representatives on the Management Council.}
\label{mctab}
\end{table}

\subsection{Estimated costs and funding of LOFAR-UK}

The cost of purchasing all of the hardware associated with a LOFAR
station, and installing this on the ground, is currently estimated at
$\pounds$610k per station. This excludes the cost of acquiring and
preparing the land, and of providing a fibre connection for the
station. Running costs per station include: fibre rental to transport the
data to the Netherlands -- this differs from site to site, being upwards
of $\pounds$20k per annum; electricity costs at $\pounds$10k per annum;
technician support and other maintenance at upwards of $\pounds$10k per
annum; yet-to-be-finalised contributions towards central processing costs,
likely to be of order $\pounds$40k per annum.

Currently thirteen UK universities have pledged funds to the LOFAR-UK
project, totalling $\pounds$600k. In addition, the University of
Manchester, the University of Cambridge, and the STFC (RAL) have all
pledged land for LOFAR stations, with e-MERLIN fibre connections
already available at the Jodrell Bank and Lord's Bridge sites. Further
universities have expressed interest in joining the LOFAR-UK
consortium, and are currently attempting to raise funds. Bids are also
submitted or in preparation for the additional funding required to
reach the UK's target of four LOFAR stations. These include:

\begin{itemize} 
\item A bid for $\pounds$500k to the Scottish Funding Council, submitted
in July 2007, as part of a larger SUPA2 initiative. This would make a
substantial contribution towards the initial purchase, installation and
commissioning costs of a Scottish LOFAR station.

\item A bid to HEFCE as part of the SEPNET `ASCE' initiative, to include
more south-eastern Universities in the LOFAR initiative ($\pounds$150k),
provide installation support for the Chilbolton site ($\pounds$148k), and
provide computing support for the LOFAR-UK effort.

\item A bid for $\approx \pounds$100k, to be submitted to the South-East
England Development Agency (SEEDA), to support the operation of a station
at the Chilbolton site.

\item A bid fro approximately $\pounds$4 million to STFC, to be submitted
in January 2008, to support the remainder of the construction and first
three-year running costs of the four LOFAR-UK stations, together with
postdoctoral and FEC support for addressing the technical challenges
associated with long-baseline calibration and data transport, and for
building a UK LOFAR archive.
\end{itemize}

It is envisaged that funding of the stations beyond the initial 3-year
period will be drawn from European Union funding through the Framework
Programme 7, from STFC Telescope Operations, or through Rolling Grants of
the LOFAR-UK consortium members.

ASTRON has indicated that there will be two phases or purchasing of LOFAR
stations, in Winter 2007-8 and Winter 2008-9. LOFAR-UK already has
sufficient funding in place that it has proceeded with the purchase of a
single LOFAR station in the Winter 2007-8 round. The goal is to follow
this with the purchase of three more stations next year. On a longer
timescale, there are aspirations to further increase the number of UK
LOFAR stations, with other funding sources being investigated.

\end{document}